\documentclass[fleqn,usenatbib]{mnras}

\DeclareRobustCommand{\VAN}[3]{#2}
\let\VANthebibliography\thebibliography
\def\thebibliography{\DeclareRobustCommand{\VAN}[3]{##3}\VANthebibliography}
\pdfoutput=1

\usepackage{amssymb}	
\usepackage{graphicx}
\usepackage{dcolumn}
\usepackage{bm}
\usepackage{epsf}
\usepackage{verbatim}
\usepackage{amsmath}
\usepackage{amsfonts}
\usepackage{ifthen}
\usepackage[normalem]{ulem}
\usepackage{multirow}
\usepackage{tikz}

\usepackage{xcolor}
\definecolor{darkgreen}{rgb}{0.0,0.5,0.0}
\definecolor{Amber}{rgb}{1.0, 0.75, 0.0}
\definecolor{carrotorange}{rgb}{0.93, 0.57, 0.13}
\definecolor{cinnamon}{rgb}{0.82, 0.41, 0.12}

\newcommand{\ie}{\emph{i.e.,} }
\newcommand{\eg}{\emph{e.g.,} }

\newcommand{\be}{\begin{equation}}
\newcommand{\ee}{\end{equation}}
\newcommand{\bea}{\begin{equation*}}
\newcommand{\eea}{\end{equation*}}
\newcommand{\beqr}{\begin{eqnarray} \nonumber}
\newcommand{\eeqr}{\end{eqnarray}}
\newcommand{\beqrb}{\begin{eqnarray}}
\newcommand{\eeqrb}{\nonumber \end{eqnarray}}
\newcommand{\fin}{\mbox{ .}}
\newcommand{\coma}{\mbox{ ,}}

\newcommand{\cm}{\mbox{ cm}}
\newcommand{\sr}{\mbox{ sr}}
\newcommand{\se}{\mbox{ s}}

\newcommand{\Myr}{\mbox{ Myr}}

\newcommand{\erg}{\mbox{ erg}}

\newcommand{\pc}{\mbox{ pc}}
\newcommand{\kpc}{\mbox{ kpc}}

\newcommand{\keV}{\mbox{ keV}}

\newcommand{\K}{\mbox{ K}}

\newcommand{\gama}{$\gamma$}

\newcommand{\unit}[1]{\bm{\hat{#1}}}

\newcommand{\DrawFigs}[1]{#1}

\newcommand{\kmps}{km s$^{-1}$ }
\newcommand{\eff}{{\mbox{\tiny{eff}}}}
\newcommand{\cmz}{{\mbox{\tiny{cmz}}}}
\newcommand{\tot}{{\mbox{\tiny{tot}}}}
\newcommand{\scmz}{{\mbox{\tiny{scmz}}}}
\newcommand{\sh}{{\mbox{\tiny{sh}}}}
\newcommand{\sd}{{\mbox{\tiny{sd}}}}
\newcommand{\inj}{{j}}

\newcommand{\DM}{{\mbox{\tiny{DM}}}}
\newcommand{\tage}{{\mathbb{T}}}
\newcommand{\ah}{{\alpha}}
\newcommand{\MyfR}{{f_R}}
\newcommand{\Myfp}{{f_p}}

\newcommand{\Myfq}{{f_q}}
\newcommand{\MyfqInv}{{f_q^{-1}}}
\newcommand{\Myfm}{{f_m}}
\newcommand{\tr}{{s}}

\newcommand{\Mvir}{{M_{\mbox{\tiny vir}}}}
\newcommand{\rvir}{{r_{\mbox{\tiny vir}}}}
\newcommand{\cvir}{{c_{\mbox{\tiny vir}}}}
\newcommand{\modj}{{k}}

\newcommand{\tv}{{\tau_v}}
\newcommand{\tz}{{\tau_z}}
\newcommand{\tzInv}{{\tau_z^{-1}}}

\newcommand{\disk}{disc}
\newcommand{\Disk}{Disc}
\newcommand{\disks}{discs}

\defcitealias{SuEtAl10}{S10}
\defcitealias{FermiBubbles14}{F14}

\defcitealias{Keshetgurwich18}{KG18}
\defcitealias{Keshetgurwich17}{KG17}
\defcitealias{Sarkaretal15a}{S15b}
\defcitealias{Sarkaretal15b}{S15a}
\defcitealias{Irwinetal19}{I19}
\defcitealias{ZhangGuo2020}{ZG20}

\newcommand{\KS}{{\citetalias{Sarkaretal15a}}}

\title[FB shock-edge implications]{Fermi bubbles: the collimated outburst needed to explain forward-shock edges}

\author[Mondal et al.]{
Santanu Mondal,$^{1}$\thanks{E-mail: santanu@post.bgu.ac.il}
Uri Keshet,$^{1}$\thanks{E-mail: ukeshet@bgu.ac.il}
Kartick C. Sarkar,$^{2}$\thanks{E-mail: sarkar.kartick@mail.huji.ac.il}
and Ilya Gurwich$^{3}$
\\
$^{1}$ Physics Department, Ben-Gurion University of the Negev, POB 653, Be'er-Sheva 84105, Israel\\
$^{2}$Center for Astrophysics and Planetary Science, Racah Institute of Physics, The Hebrew University of Jerusalem, Israel\\
$^{3}$Department of Physics, NRCN, POB 9001, Beer-Sheva 84190, Israel\\
}

\date{Accepted XXX. Received YYY; in original form ZZZ}

\pubyear{2021}
\begin{document}
\label{firstpage}
\pagerange{\pageref{firstpage}--\pageref{lastpage}}
\maketitle

\begin{abstract}
The bipolar, nonthermal, high-latitude lobes known as the Fermi bubbles (FBs) are thought to originate from a massive energy release near the Galactic centre (GC).
We constrain the FB engine and the circumgalactic medium (CGM) by analytically and numerically modeling the FB edges as strong forward shocks, as inferred from recent observations.
A non-directed energy release produces shocks too spherical to account for observations even for a maximally massive Galactic {\disk}, critical CGM rotation, or injection effectively offset from the GC.
In contrast, collimated injection nearly perpendicular to the {\disk} can account for observations in both ballistic (free expansion) and slowdown regimes, as we show using a simple stratified evolution model verified by hydrodynamic simulations.
FBs still in their ballistic regime require injection (at $z\simeq100\pc$ heights in our model) with a half-opening angle $\theta\simeq4^\circ$, a normalized velocity $\beta_{-2}\equiv v/(0.01c)\gtrsim 0.4$, and an energy $E\gtrsim2\beta_{-2}^2\times 10^{55}\erg$, launched $\tage\simeq 3.3\beta_{-2}^{-1}\Myr$ ago, showing a distinctive low-pressure region behind the bubble head.
Slowing-down (mass accumulated) FBs require a faster injection, a thinner jet, a smaller $E/(\beta_{-2}\theta)^{2}$, and a comparable $\tage$, and follow a ballistic stage that must reach a height $z_{\tr}\gtrsim 5\kpc$.
\end{abstract}


\section{Introduction}
\label{sec:Intro}
The Fermi bubbles (FBs) are bipolar, non-thermal lobes, extending to latitudes $|b|\simeq 50^\circ\text{--}55^\circ$ above and below the Galactic centre (GC).
Their high-latitude signature was first revealed in \gama-rays, by the {\it Fermi} Gamma-ray Space Telescope \citep[][henceforth S10]{DoblerEtAl10, SuEtAl10}, but subsequently also in microwaves \citep{DoblerFinkbeiner2008,Dobler12, PlanckHaze13}
and in X-rays \citep[][hereafter KG18]{Keshetgurwich18}.
The edges of the FBs are distinct and easily traced \citep[][hereafter KG17]{Keshetgurwich17}, linking them to large \citep{Sofue00}, intermediate \citep{Blandhawthorncohen03} and small \citep{Baganoffetal03} scale \citep{Sofue00, Blandhawthorncohen03} nonthermal features, and thus identifying the FBs as arising from a major, $\geq 10^{55}$~erg event \citep[][and references therein]{VeilleuxEtAl05} near the GC.

The origin of the FBs is still uncertain, present models including a starburst \citep[][hereafter S15b]{CarrettiEtAl13,Lacki14,Sarkaretal15b}, an SMBH jet \citep{ChengEtAl11,Guo12,ZubovasNayakshin12}, SMBH accretion wind \citep{Mouetal14, Mouetal15}, or steady star formation \citep{Crocker12}.
Due to their dynamical, nonthermal nature, and the vast energy involved, an accurate interpretation of the FBs is important for understanding the energy budget, structure, and history of our Galaxy.

Correctly interpreting the FBs is important for understanding additional, possibly related phenomena, such as `chimneys' which may play a role in the transport of energy to the base of the FBs \citep{Pontietal2019}, symmetric lobes on a few $100\pc$ scales \citep{Heywoodetal2019}, and magellanic echoes \citep{Bland-Hawthornetal19}, where two bipolar ionization radiation cones are  associated with the GC took place $\sim$3.5~Myr ago.
Recent UV absorption survey of quasar sightlines passing through
the FBs have been used to map the kinematics of associated cold clouds.
The velocity of the clouds are symmetric in both hemispheres, decreases from $\sim 900$ \kmps at low latitude to $\sim 350$ \kmps at high latitudes \citep{Foxetal2015,Bordoloietal2017,Karimetal2018, Ashleyetal2020}.

In spite of their dramatic appearance in the \gama-ray sky, even the nature of the FB edges is not widely agreed upon. Previous models have interpreted the FB edge as either an outgoing shock \citep{FujitaEtAl13}, or a termination shock of a wind \citep{Lacki14}, or a discontinuity \citep[][\KS]{Crocker12}.
More recent studies provided evidence that the edges are forward shocks.
This evidence is based on the little variation of the \gama-ray spectrum with position along the edge, indicating a strong, Mach $M>5$ shock \citepalias{Keshetgurwich17},
on a faint X-ray shell inside the FB edges found by stacking {\it ROSAT} data and consistent with $M>4$ forward shocks \citepalias{Keshetgurwich18},
and on a combined analysis of the microwave and \gama-ray spectrum (Keshet \& Gurwich, in preparation).
These claims are, however, in tension with some other results \citep{KataokaEtAl13, CarrettiEtAl13,Kataokaetal2021}.

The identification of the FB edges as strong forward shocks simplifies, in a sense, their analysis, because their supersonic motion renders them less sensitive to details of the energy release and CGM properties than the alternative, discontinuity or reverse shock models.
We thus use hydrodynamic simulations to study which properties of the FB engine and of the CGM are consistent with the observed morphology of the FBs.
In particular, we use the edges identified by applying gradient filters to the {\it Fermi} data \citepalias{Keshetgurwich17}, and the projection of the toy model for the FBs in 3D \citepalias{Keshetgurwich18}.

The paper is organised as follows. In \autoref{sec:Method}, we describe the methodology and the numerical setup.
The analysis of putative FBs generated by an isotropic, non-directed injection of energy is presented in \autoref{nonjettedresults}, and shown to yield bubbles too wide to be consistent with the FBs.
In \autoref{jettedresults}, we study FBs generated by the collimated injection of both momentum and energy, presenting a stratified evolution model in \autoref{sec:ToyModels} and the corresponding simulations in \autoref{subsec:JettedSims}.
Our results are summarised and discussed in \autoref{sec:discussion}.
We present the Galactic model implemented in the simulations in \autoref{appendix:A}, and representative convergence tests in \autoref{appendix:B}.

\section{Method} \label{sec:Method}

We simulate a simple numerical model of the Galaxy, inject energy and in some cases also momentum rapidly near the Galactic centre, and examine the resulting, evolved FBs with an emphasis on the robustly-measured geometry of the expanding, forward shock.
The parameters of the Galactic model and of the injecting engine are varied, to quantify the dependence of the observed quantities upon the assumptions, and to identify the physical condition which could account for the observed FB edges.

The analysis is based on hydrodynamic simulations using the publicly available Eulerian grid code PLUTO-v4.0 \citep{Mignoneetal07}.
The simulations are axisymmetric and non-relativistic, modeling a viscous fluid with no heat conduction, with an ideal gas of polytropic index $\gamma=5/3$, without cosmic-rays, and with a mean particle mass $\mu_m m_p$, where $m_p$ is the proton mass and $\mu_m\simeq 0.6$. These assumptions are reasonable for the present problem, as we focus mainly on the geometry of the shock, at $\gtrsim$ kpc scales.
Spherical coordinates $(r,\theta,\phi)$ centered on the GC are used, with polar coordinate $\theta=\pi/2$ along the Galactic {\disk} and a frozen azimuthal coordinate $\phi$.
The computation box extends from $r_0$ to $15$ kpc in the radial direction and from $\theta=0$ to $\pi/2$ in the polar direction. The inner boundary in the radial direction is set to have the values corresponding to the hydrostatic equilibrium, except where we inject jet energy. The outer boundary in the radial direction is set to have zero gradient. In the polar direction, $\theta = 0$ and $\pi/2$ boundaries are set to be axisymmetric and reflective, respectively.
For convenience, we introduce also a cylindrical coordinate system $(R,\phi,z)$, centered on the GC with axial coordinate $z$ along $\theta=0$.

The Galactic model and its numerical setup follow the work of \citet[][hereafter S15a]{Sarkaretal15a}, as detailed in \autoref{appendix:A}. The model includes two gas components of the interstellar medium (ISM): a hot, $4\times10^4$~K, {\disk}-like component and a warm, $2\times10^6$~K, halo component. The model assumes that the gas components are in hydrostatic equilibrium with the gravity of dark matter, the {\disk}, and the bulge, and each component is initially isothermal. The warm component associated with the {\disk} rotates approximately with the stars; a rotation parameter
$f_h\leq 1$ is introduced to quantify the ratio between halo rotation and stellar rotation (see \autoref{appendix:A}).
The halo has a global baryon fraction $f_b=0.16$, consistent with the cosmic value of $\Omega_b$/$\Omega_m$.
The fixed parameters of our nominal Galactic model are provided in \autoref{table:fixedCGM}.
For these parameters, the mass density of the halo --- which plays an important role in the bubble evolution --- is distributed at large radii as $\rho_h\simeq 4.4\times 10^{-4}(r/10\kpc)^{-1.5}m_p\cm^{-3}$.
Galactic model parameters that are varied among simulations are provided in \autoref{Tab:NonDirected} for non-directed bubble simulations, and in \autoref{Tab:Jetted} for jetted bubble simulations.

\begin{table}
\scriptsize
\centering
\caption{\label{table:fixedCGM} Fixed CGM parameters. Additional CGM parameters varied among simulations are provided in \autoref{Tab:NonDirected} and \autoref{Tab:Jetted}.
}
\scalebox{1.2}{
\begin{tabular}{|c|c|c|}
\hline
\textbf{Parameter} &      \textbf{Definition}         &  \textbf{Value} \\
\hline
$\Mvir$ [$M_\odot$]               & Virial mass  & $1.2\times10^{12}$\\
$\rvir$ [kpc]               & Virial radius & 250 \\
$d$ [kpc]               & DM core radius & 6.0 \\
$r_s$ [kpc]              & Scale radius & 20.8 \\
$\cvir=\rvir/r_s$     & Concentration parameter & 12\\
$a_d$ [kpc]       &   {\Disk} scale length     & 3.0\\
$b_d$ [kpc]       &  {\Disk} scale height  & 0.4 \\
$a_b$ [kpc]       & Bulge scale radius  & 2.0\\
$\rho_{d0}$ [m$_p$ cm$^{-3}$] & {\Disk} mass density & 1.0\\
$\rho_{h0}$ [m$_p$ cm$^{-3}$]& Halo mass density  & 0.019\\
$T_d$ [K]& Disc temperature  & $4\times 10^4$\\
$T_h$ [K]& Halo temperature  & $2\times 10^6$\\
$T_{\cmz}$ [K]& CMZ temperature  & $10^3$\\
$f_{d}$   & Disc gas rotation   & 0.975\\
$f_{\rm cmz}$   & CMZ rotation   & 0.0\\
$r_{\max}$ [kpc]& Outer boundary of the box  & 15\\
$r_{\rm gd}$ [kpc]& Sun Galactocentric radius   & 8.5\\
\hline
\end{tabular}}
\end{table}

\begin{figure} 
\DrawFigs{
\centering{
\includegraphics[height=6.2truecm]{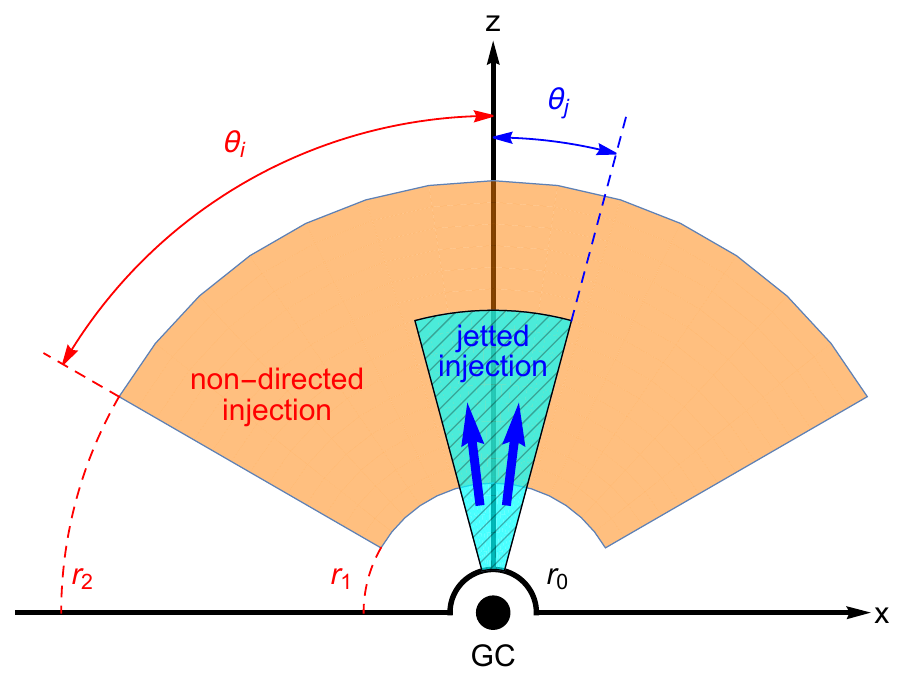}}
}
\caption{
Illustration of non-directed (orange shaded region) and jetted (hatched cyan, with arrows showing the momentum direction) injection is based on our simulated FBs.
\label{fig:illustration}}
\end{figure}

We consider two different modes of FB engines, operating over a short, $\ll\Myr$ timescale near the GC, as illustrated in \autoref{fig:illustration}: energy only (henceforth: non-directed) or both energy and momentum (henceforth: jetted) injection.
The non-directed mode can arise, for example, from starburst activity at some distance from the GC, or from a dissipated jet at some height above the GC, so we consider such injection both close to the GC (around $r_0$; henceforth GC injection) and at intermediate latitudes (around $r\sim$ kpc; IL injection).
Jetted emission can arise from a relativistic jet launched from the central black hole (CBH) in the GC, and is represented for simplicity as simultaneous energy and momentum in the $\theta=0$ direction, injected at $r_0$.
As we show below, jetted injection can lead to two different types of FBs: ballistic bubbles for high energies, and slowing-down, non-ballistic for low energies.

We detail the two (GC and IL) non-directed modes in \autoref{nonjettedinjection}, and the two (ballistic and slowdown) jetted modes in \autoref{jettedinjection}.
Shear in the flow, especially during the injection of slowdown jetted bubbles, can give rise to substantial Kelvin-Helmholtz instabilities (KHI), which are dealt with in \autoref{subsec:Viscosity}.
The simulation results are finally projected and compared to observations, in a method outlined in \autoref{subsec:projection}.
Convergence tests are provided in \autoref{appendix:B}.

\subsection{Non-directed injection}
\label{nonjettedinjection}

We begin with the simple, non-directed injection of energy with zero momentum, either from the vicinity of the CBH (GC),
or from intermediate latitudes (IL).
We assume that the injected energy is thermalised within a conical region (with half opening angle $\theta_i$) around the rotation axis of the Galaxy.
The injected energy is thus added to the internal energy
above the injection point, at some radial range $r_1<r<r_2$, as illustrated by the orange shaded region
in \autoref{fig:illustration}.
The internal energy is added at a constant rate
\begin{equation} \label{eq:GCInjectionErate}
\dot{e}_{\inj} = \frac{E_{\inj}}{V_{\inj} t_{\inj}}
\end{equation}
during a short time interval time
$t_{\inj}$, after which injection is turned off.
Here, both the total injected energy $E_{\inj}$ and the injection volume
\begin{equation}\label{eq:GCInjectionVolume}
    V_{\inj}=\frac{4\pi}{3} (r_2^3 - r_1^3)(1-\cos \theta_i)
\end{equation}
are defined for the full sky, summing both hemispheres.

With the above general setup, we explore the simulated appearance of the FBs for different central engines and Galactic structures, by varying the properties of energy injection and of the Galactic model, as detailed in \autoref{Tab:NonDirected}.
Such variations are essential because, as shown in \autoref{nonjettedresults}, extreme --- and even implausible --- choices of the Galactic model parameters are needed in order to produce simulated FBs of any resemblance to the observations.

In particular, we vary the mass of the Galactic {\disk} and of the bulge, which can modify the resulting bubble morphology, while keeping the virial mass of the galaxy fixed.
We also vary the rotation frequency of the Galactic components, which affect their structure and therefore also the appearance of the shock.
The geometrical parameters of the injection volume have little effect near the GC, but they do have some impact at high latitudes, where we consider in particular a narrow, $\theta_i\simeq 5^\circ$ cone representing a putative dissipated jet.

We find converged results with a minimal resolution of $1024\times256$ with a stretched\footnote{
See http://plutocode.ph.unito.it/files/userguide.pdf for details.}
grid along the $r$ and $\theta$ directions, with no need to introduce viscosity.
All of our production runs for the non-directed set-ups have this resolution.

\subsection{Jetted injection} \label{jettedinjection}

Next, consider jetted injection, introducing both energy and momentum in the $\theta\simeq 0$ direction, approximately perpendicular to the Galactic plane.
Injection into the computation domain is implemented by modifying the $r_0$, inner boundary within the half opening angle $\theta_j$ of the jet,
introducing a luminosity $L_{\inj}= E_{\inj}/t_{\inj}$ over a short time interval $t_{\inj}$, in the form of kinetic energy due to momentum in the $\unit{\bm{r}}$ direction.
This injection scheme is illustrated by the hatched cyan region in \autoref{fig:illustration}.
The quantities $E_{\inj}$ and $L_{\inj}$, like all extensive injection parameters above and below, pertain to the full sky, \ie to both hemispheres combined.

The energy injected is predominantly kinetic, so the injected mass is taken as
\begin{equation}\label{eq:JettedInjectionMdot}
\dot M_{\inj} = \frac{2L_{\inj}}{v_{\inj}^2} \coma
\end{equation}
where $v_{\inj}$ is the non-relativistic velocity of the injected gas; relativistic corrections are omitted.
The area of the injection boundary is given by
\begin{equation}\label{eq:JettedInjectionArea}
A_{\inj}=4\pi(1-\cos\theta_j)r_{0}^2 \coma
\end{equation}
so the mass density of the jet material injected at $r_0$ is given by
\begin{equation}\label{eq:JettedInjectionRhoDot}
  \rho_{\inj}=\frac{\dot M_{\inj}}{A_{\inj}v_{\inj}} \fin
\end{equation}
Kinetic energy is assumed to dominate at the base of the jet, so
the thermal energy flux $(\gamma -1)^{-1}{\gamma} P v_{\inj}$ is taken to be negligible with respect to the kinetic energy flux $(1/2)\rho_{\inj} v_{\inj}^3$ throughout the injection boundary.
Here, $P$ is the gas pressure.

The bubbles formed by jetted injection are not very sensitive to the Galactic model in the ballistic case, and depend mostly on the halo distribution for the slowdown case.
The minimum grid necessary for convergence is a $1024\times512$ uniform grid in the $r$ and $\theta$ directions for the ballistic case, and a $1536\times 768$ uniform grid for the slowdown case.


\subsection{Overcoming KHI}
\label{subsec:Viscosity}

The emergence of strong shear, in particular for jetted injection, gives rise to KHI that corrugate the discontinuities, thus leading to expensive simulations with a short time step and early dissipation of the kinetic energy. We avoid these instabilities, especially at the early stages of jet formation, because details of the early evolution are in any case poorly understood and not well-constrained by observations.

One modification we thus incorporate is smoothing of the injected velocity gradient, by switching from a step function at $\theta_j$ to a continuous, linear velocity profile
\begin{equation}\label{eq:ViscosityVelocity}
v(r=r_0,\theta<\theta_j;t<t_{\inj})=\left( 1-\frac{\theta}{\theta_j} \right)v_{\inj}\coma
\end{equation}
in which case the tabulated parameter $v_{\inj}$ refers to the maximal, axial velocity.
We keep $\rho_{\inj}$ uniform, so \autoref{eq:JettedInjectionRhoDot} is replaced by
\begin{equation}\label{eq:ViscosityRho}
  \rho_{\inj}=\frac{L_{\inj}}{2\pi  \left[1-6\frac{\theta_j-\sin(\theta_j)}{\theta_j^3}\right] r_0^2v_{\inj}^3} \fin
\end{equation}

More importantly, we incorporate viscosity, modeled for simplicity as the isotropic, dynamic viscosity of a nonmagnetized plasma \citep{Braginskii1958,Spitzer1962}
\begin{equation}\label{eq:ViscositySpitzer}
\mu\simeq 6\times10^3 \left(\frac{\ln\Lambda}{37}\right)^{-1} \left(\frac{T}{10^8 \text{K}}\right)^{5/2} \text{gm}~\text{cm}^{-1}~\text{s}^{-1},
\end{equation}
where T is the temperature in Kelvin and $\ln\Lambda$ is the Coulomb logarithm.
To prevent exceedingly small time steps, we cap the viscosity at $\mu_{\max}=10\mbox{ g}\cm^{-1}\se^{-1}$, which is sufficient for present purposes.
Indeed, KHI of wavelengths $\lambda$ are suppressed by viscosity \citep[\eg][]{Lamb1932,Chandrasekhar1961,Roedigeretal2013}
\begin{equation}
\mu \gtrsim  \frac{\rho \lambda \Delta{v}}{64 q_\rho^{1/2}} \coma
\end{equation}
where $\Delta{v}$ is the shear velocity and $q_\rho$ is the density contrast across the discontinuity.
If we replace $\lambda$ with the scale of the bubble at a given time (z), we will have the value necessary to insure KHI suppression at all scales.
At small scales, this upper limit keeps increasing as the bubble grows, but beyond $z\gtrsim 0.35kpc$, the density drops rapidly as the bubble grows. At $z\sim 0.35\kpc$ our simulation shows that $\rho_{d}$=0.3 cm$^{-3}$ and $\rho_h$=$2\times10^{-3}$ cm$^{-3}$ whereas, $\Delta{v}=0.1c$ is chosen for our nominal simulation, which gives, $\mu \sim 13$~gm cm$^{-1}$ s$^{-1}$ as the necessary threshold to suppress all KHI modes.

\subsection{Projection and comparison with observations}
\label{subsec:projection}

In order to compare the simulated bubble to FB observations, we project the simulation box for an observer at the solar position, taken as $(r,\theta,\phi)=(8.5\kpc,\pi/2,\pi)$.
We use the
Projection Analysis Software for Simulations \citepalias[{\fontfamily{qcr}\selectfont PASS};][]{Sarkaretal15b},
which computes the surface brightness
by integrating the emissivity along the line of sight (LOS),
\begin{equation}
F_X (l, b)=\frac{1}{4\pi} \int_{LOS} n^2 \varepsilon(T,Z) ds,
\end{equation}
where $n$ is the particle number density and
$T$ is the temperature.
The emissivity $\varepsilon(T,Z)$ is computed assuming a
metallicity $Z=0.2Z_\odot$
by interpolating the {\fontfamily{qcr}\selectfont MEKAL} from {\sc xspec} \citep{Meweetal1985,Meweetal1986,Liedahletal1995,Arnaud1985}
data;
we use the $2.0-10.0$ keV band, sufficient for our purposes.
The LOS is extended beyond the simulation box, out to $r = 100$ kpc, assuming steady-state equilibrium with the background gravitational field, extending the temperature and density profiles of our Galactic model.

Given the projected image of the simulation at a given time $t$, we identify the bubble edges by applying a gradient filter,
of $2^\circ$ width.
The maximal latitude $b$ of the projected bubble, and its maximal longitudinal half width $\Delta l$, are then measured as a function of $t$.
The age $\tage$ of the bubble is estimated by requiring $b(\tage)=52^\circ$, as inferred from FB observations.
The measured bubble edge at this time is compared to the edges inferred by applying a gradient filter to the \emph{Fermi} map \citepalias[$6^\circ$ width for southeast bubble;][]{Keshetgurwich17} and to the projection of a FB toy model \citepalias{Keshetgurwich18}.

\section{Non-directed injection} \label{nonjettedresults}

We simulate various non-directed injection models, as described in \autoref{nonjettedinjection}, and compare the bubbles as they reach $b\simeq 52^\circ$ with the observed FBs.
The parameters of select non-directed injection simulations, which are displayed in the following figures, are detailed in \autoref{Tab:NonDirected}.
The table also provides the central measured properties of each simulated bubble: its age $\tage$ and maximal half width $\Delta{l}$.

\begin{table*}
\scriptsize
\centering
\caption{\label{Tab:NonDirected}
Variable parameters for select non-directed injection simulations presented in figures  \ref{fig:base_GC_IL}--\ref{fig:IL_setups}.
}
\scalebox{1.2}{
\begin{tabular}{|l l|p{0.4cm} p{0.4cm}p{0.4cm}p{0.4cm}p{0.4cm} |p{0.4cm}p{0.4cm}p{0.4cm}p{0.4cm}p{0.4cm}|}
\hline
\textbf{Parameter} & \textbf{Definition}   & \multicolumn{5}{c|}{\textbf{GC Injection}}  & \multicolumn{5}{c|}{\textbf{IL Injection}}  \\
&  & S1    & S1a  & S1b & S1c & S1d   & S2    & S2a   & S2b   & S2c   & S2d \\
\hline
$M_d$ [$10^{10}M_\odot$] & {\Disk} mass & \multirow{1}{*}{6}    & \multirow{1}{*}{6}  & \multirow{1}{*}{12} & \multirow{1}{*}{6} & \multirow{1}{*}{12} & \multirow{1}{*}{6} & \multirow{1}{*}{6} & \multirow{1}{*}{12} & \multirow{1}{*}{6} & \multirow{1}{*}{12} \\
$M_b$ [$10^{10}M_\odot$] & Bulge mass &\multirow{1}{*}{2}  &\multirow{1}{*}{2}  &\multirow{1}{*}{2}  &\multirow{1}{*}{4}  &\multirow{1}{*}{4}   &\multirow{1}{*}{2}  &\multirow{1}{*}{2}  &\multirow{1}{*}{2}  &\multirow{1}{*}{4}  &\multirow{1}{*}{4}\\
$f_h$  &  CGM rotation    & \multirow{1}{*}{0.33}  & \multirow{1}{*}{0.9}  & \multirow{1}{*}{0.33} & \multirow{1}{*}{0.33} & \multirow{1}{*}{0.9} & \multirow{1}{*}{0.33} & \multirow{1}{*}{0.9}  & \multirow{1}{*}{0.33} & \multirow{1}{*}{0.33} & \multirow{1}{*}{0.9}\\
\hline
$E_{\inj}$[$10^{56}$erg] & Injected energy & \multicolumn{5}{c|}{\multirow{1}{*}{2}} & \multicolumn{5}{c|}{\multirow{1}{*}{2}} \\
$t_{\inj}$ [Myr] & Injection duration        &
\multicolumn{5}{c|}{\multirow{1}{*}{0.01}} & \multicolumn{5}{c|}{\multirow{1}{*}{0.01}} \\
$r_1$ [kpc] & Minimal injection radius &
\multicolumn{5}{c|}{\multirow{1}{*}{0.02}} & \multicolumn{5}{c|}{\multirow{1}{*}{1.0}} \\
$r_2$ [kpc] & Maximal injection radius &
\multicolumn{5}{c|}{\multirow{1}{*}{0.04}} & \multicolumn{5}{c|}{\multirow{1}{*}{2.0}} \\
$\theta_i$ & Injection opening angle  &
\multicolumn{5}{c|}{\multirow{1}{*}{$90^\circ$}} & \multicolumn{5}{c|}{\multirow{1}{*}{$5^\circ$}} \\
$r_0$ [kpc] & Inner boundary &
\multicolumn{5}{c|}{\multirow{1}{*}{0.02}} & \multicolumn{5}{c|}{\multirow{1}{*}{0.02}} \\
\hline
$\tage$ [Myr] & Bubble age (approximate) & \multirow{1}{*}{5} &\multirow{1}{*}{6} &\multirow{1}{*}{2.8} &\multirow{1}{*}{3.5}   &\multirow{1}{*}{2.8} &\multirow{1}{*}{4.3} &\multirow{1}{*}{5} &\multirow{1}{*}{2}   &\multirow{1}{*}{2.7}    &\multirow{1}{*}{1.5}\\
$\Delta l$  & Bubble half width (approximate) &\multirow{1}{*}{$37^\circ$}   &\multirow{1}{*}{$34^\circ$} &\multirow{1}{*}{$33^\circ$} &\multirow{1}{*}{$34^\circ$} &\multirow{1}{*}{$27^\circ$}  &\multirow{1}{*}{$36^\circ$}  &\multirow{1}{*}{$33^\circ$} &\multirow{1}{*}{$33^\circ$} &\multirow{1}{*}{$35^\circ$} &\multirow{1}{*}{$24^\circ$}\\
\hline
\end{tabular}}
\end{table*}

\begin{figure}
\DrawFigs{
\centering{
\hspace{-2.8cm}
\begin{tikzpicture}
\draw (0, 0) node[inner sep=0] {\raisebox{0.15cm}{\includegraphics[height=4.5truecm,trim={4.8cm 2.2cm 4.0cm 2.8cm}, clip]{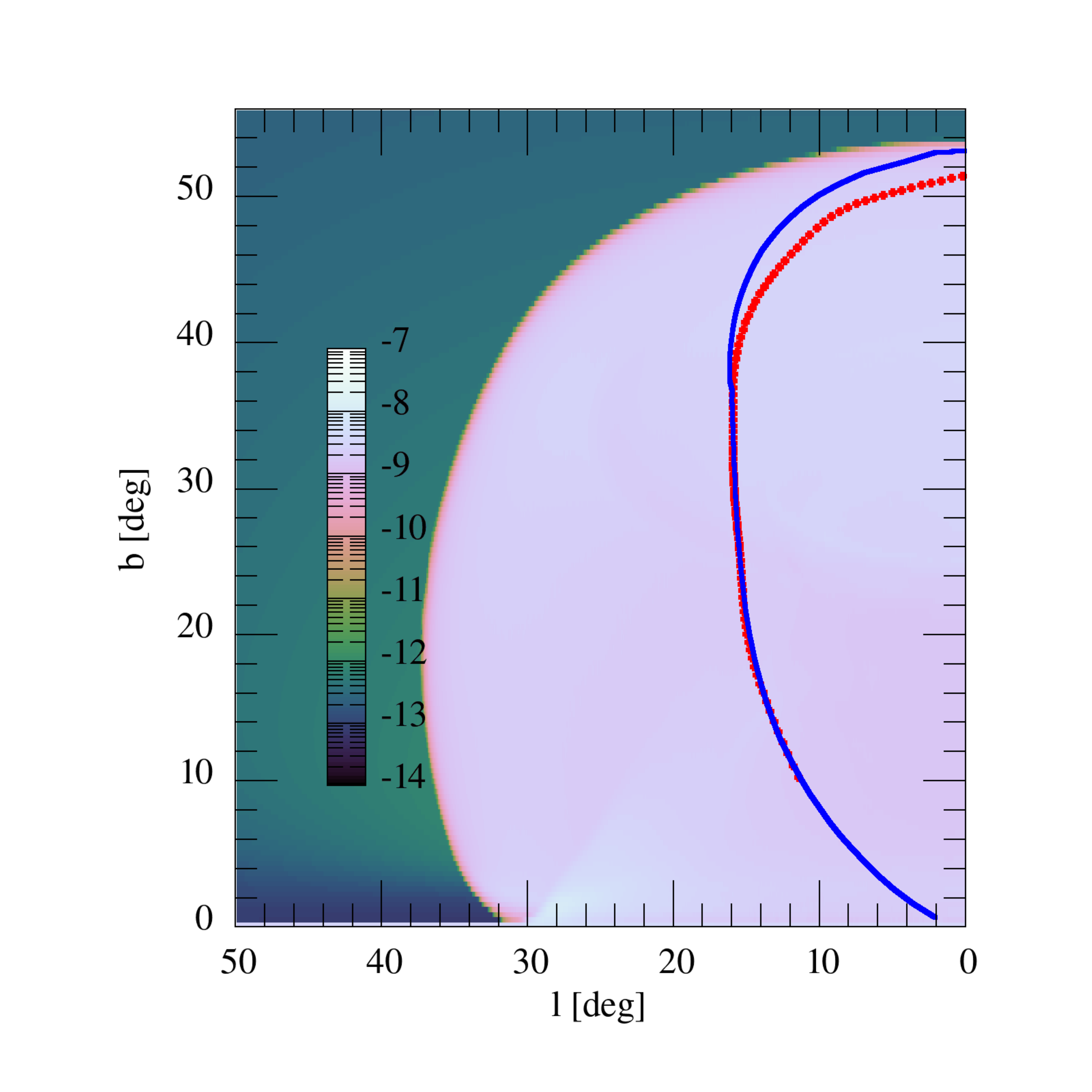}}};
\draw (-0.7, 1.8) node[text=white] {\scriptsize $t=5\Myr$};
\end{tikzpicture}
\hspace{-0.2cm}
\includegraphics[height=4.5truecm,trim={2.1cm 2.5cm 1.5cm 5.2cm}, clip]{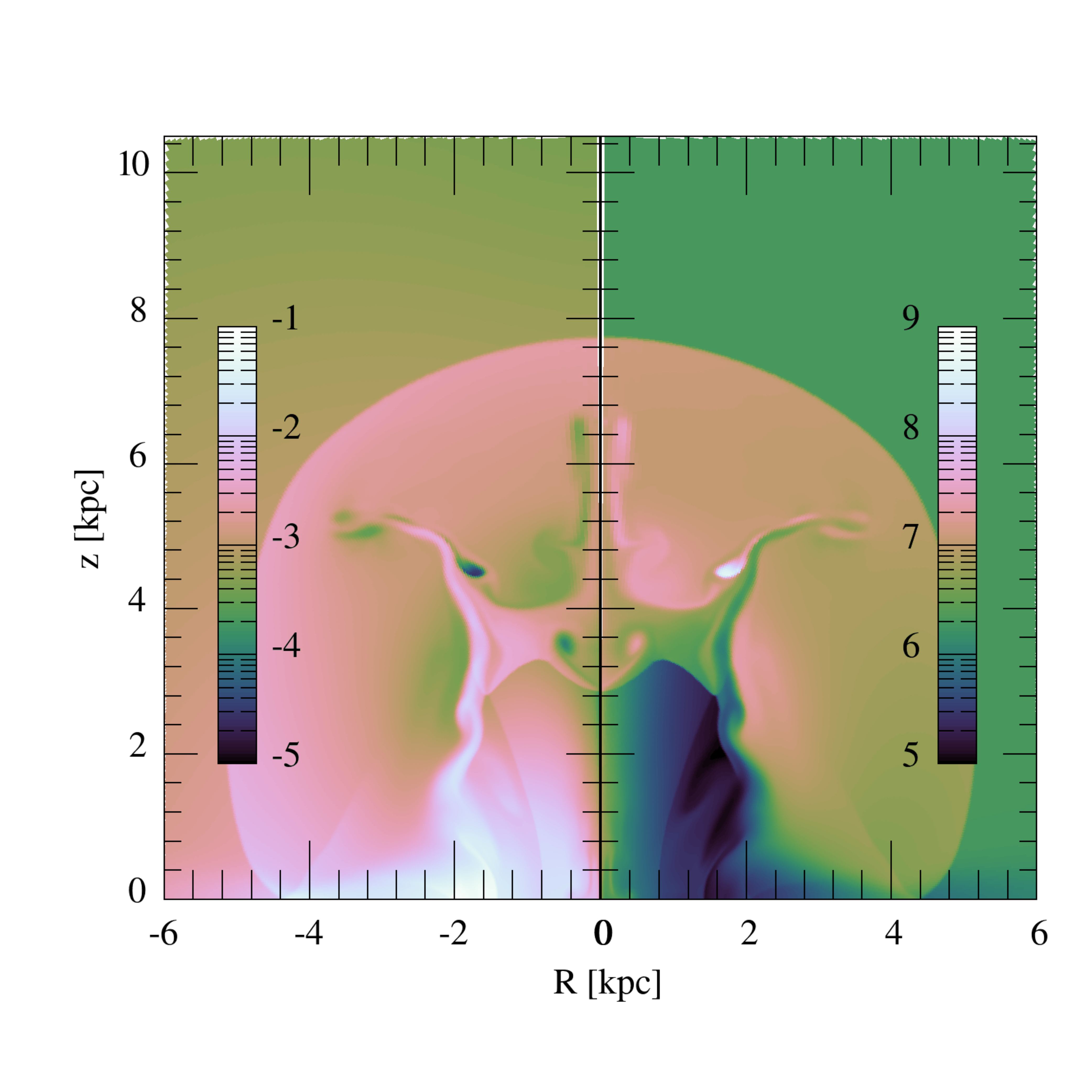}
\hspace{-2.5cm}
}
\\
\centering{
\hspace{-2.5cm}
\begin{tikzpicture}
\draw (0, 0) node[inner sep=0] {\raisebox{0.15cm}{\includegraphics[height=4.5truecm,trim={4.8cm 2.2cm 4.0cm 2.8cm}, clip]{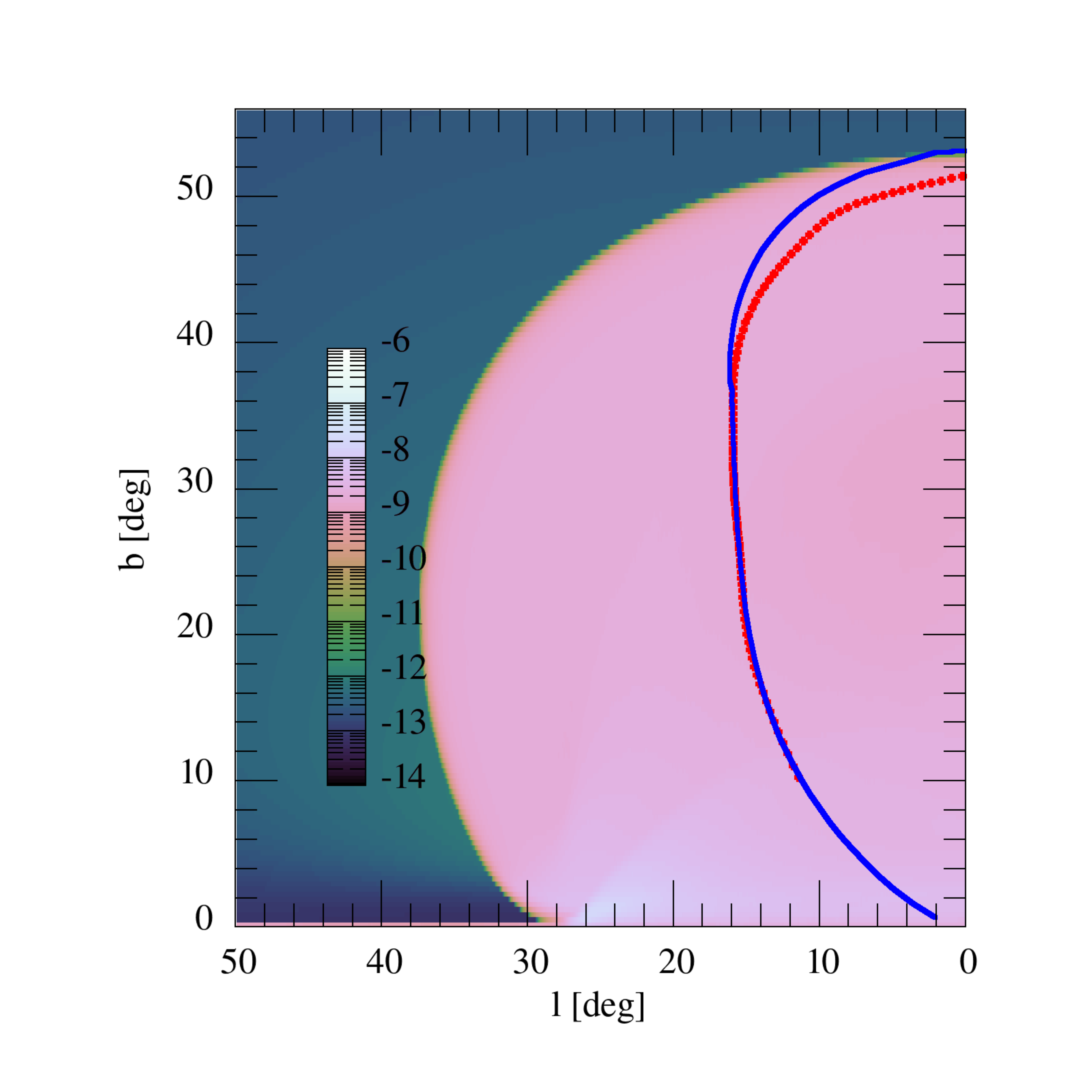}}};
\draw (-0.7, 1.8) node[text=white] {\scriptsize $t=4.3\Myr$};
\end{tikzpicture}
\hspace{-0.2cm}
\includegraphics[height=4.5truecm,trim={2.1cm 2.5cm 1.5cm 5.2cm}, clip]{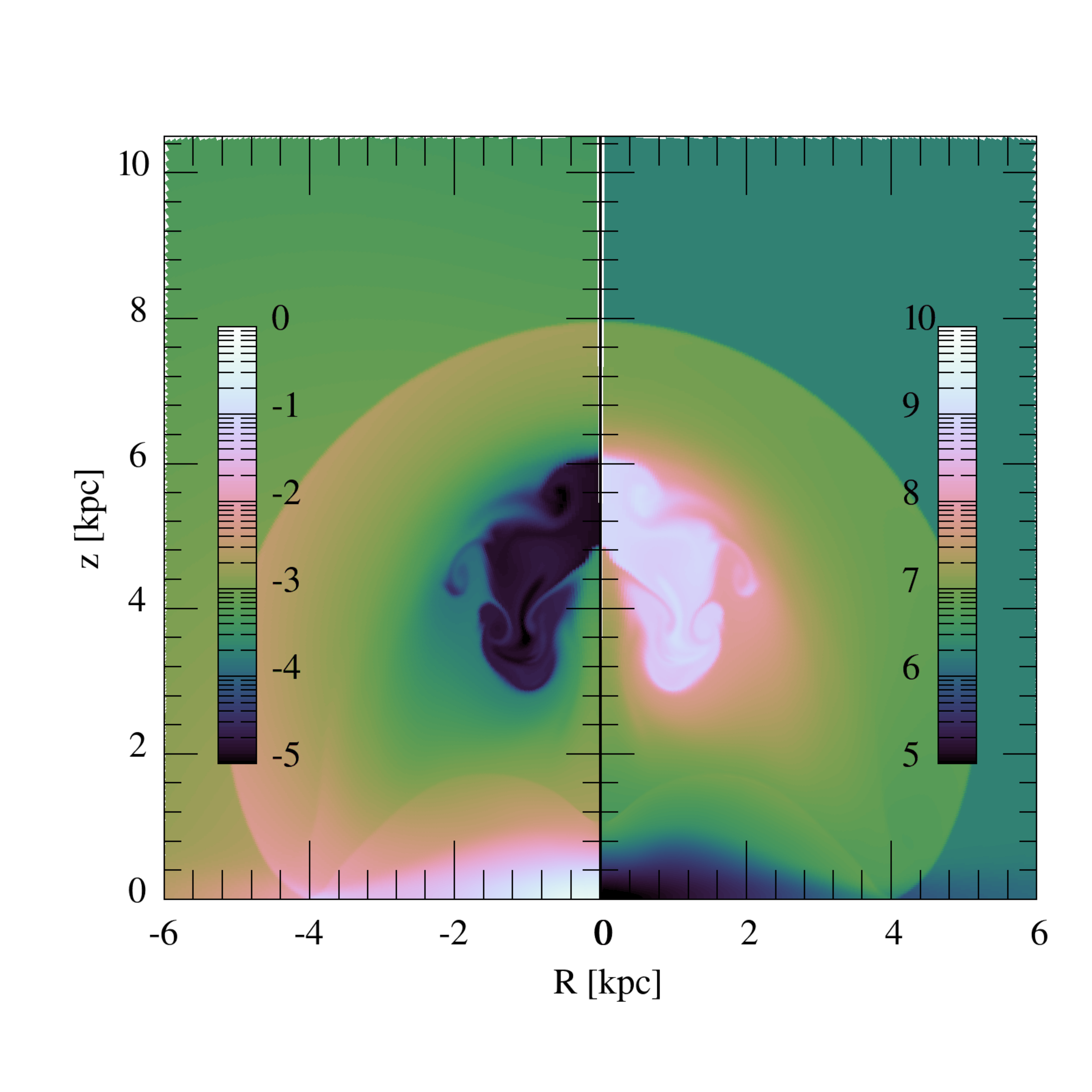}
\hspace{-2.5cm}
}
}
\caption{
Simulated \emph{Fermi}-like bubbles obtained from our nominal models of injection near the CBH
(GC; setup S1 at $t=5\Myr$; top panels) and at intermediate latitudes
(IL; setup S2 at $t=4\Myr$; bottom panels).
Full simulation parameters are listed in \autoref{table:fixedCGM} and \autoref{Tab:NonDirected}, and snapshot times are provided in the upper left labels.
Left panels: projected $2\text{--}10\keV$
surface brightness at the Solar system ($F_X[\text{erg}\se^{-1}\cm^{-2}\sr^{-1}]$ colorbar), with superimposed FB edges based on a gradient filter (of $6^\circ$ width for southeast bubble; from \citetalias{Keshetgurwich17}; red dotted curve) and on projecting a toy model (\citetalias{Keshetgurwich18}; solid blue).
Right panels: spatial distributions of the number density (left half; $n[\text{cm}^{-3}]$ colorbar) and temperature (right; $T[\text{K}]$ colorbar) of the CGM.
The forward shock presents as the large scale, sharp inward jump in the projected $F_X$ and as the external, sharp jumps in the non-projected $n$ and $T$.
\label{fig:base_GC_IL}}
\end{figure}

\autoref{fig:base_GC_IL} shows the \emph{Fermi}-like bubbles obtained for our nominal GC (S1; top panels) and IL (S2; bottom panels) injection simulations.
The figure shows the $2-10$ keV surface brightness maps (left panels) and the spatial distributions of density and temperature (in cylindrical coordinates; right panels).
For comparison with observations, FB edge contours traced by a gradient filter \citepalias[][red dotted curve]{Keshetgurwich17} and obtained from projecting a toy model (\citetalias{Keshetgurwich18}; solid blue) are superimposed on the surface brightness maps.
For this purpose, we adopt the south eastern FB edge, which is better measured than its northern and western counterparts due to the abundance of dust in the northern hemisphere and the elongation of the bubbles to the west.

As the upper panel of \autoref{fig:base_GC_IL} shows, our nominal GC injection model produces bubbles with strong forward shocks that are approximately spherical, and quite inconsistent with the more elongated, thinner morphology of the observed FBs.
Such quasi-spherical bubbles are expected as the shock propagates far from the GC, into an initial CGM density that itself becomes quasi-spherical at $|z|\gtrsim 1\kpc$ distances from the Galactic plane.
For injection near the GC, the initial blastwave is close to spherical and accumulates mass rapidly, so the transition from an initial quasi-ballistic propagation into an approximately self-similar propagation occurs relatively quickly.
Indeed, after $\sim0.5\Myr$, we find an $R\propto t^{0.6}$ and $z\propto t^{0.6}$ evolution, close to the spherical self-similar $r\propto t^{4/7}$ solution expected for the $\rho_h\propto r^{-3/2}$ radial profile implemented by our Galactic model at large radii.
The inner structure of the bubble presents a reverse shock, separated from the forward shock by a deformed contact discontinuity.

As the lower panel of \autoref{fig:base_GC_IL} shows, our nominal IL injection model produces forward shocks similar to those of the GC injection, and again far too spherical to account for the FB observations.
Such injection, representing for example the dissipation of a jet at some intermediate latitude, avoids the initial propagation of the blastwave through the high density regions near the GC, resulting in slightly narrower bubbles.
However, the sideways expansion of the shock is rapid, leading to wide bubbles that are still inconsistent with the FBs.
Here, it takes slightly more time for the bubbles to settle on a power-law temporal evolution, characterised by a more noticeably sublinear and less spherical, $R\propto t^{0.5}$ and $z\propto t^{0.4}$ growth after $\sim 1\Myr$.
The inner structure of the bubbles is noticeably different for IL injection, presenting reverse shocks reflected from the Galactic plane at low latitude and a closed contact discontinuity structure at mid-latitudes.

\begin{figure}
\DrawFigs{
\centering{
\hspace{-2.5cm}
\begin{tikzpicture}
\draw (0, 0) node[inner sep=0] {\raisebox{0.15cm}{\includegraphics[height=4.5truecm,trim={4.8cm 2.2cm 4.0cm 2.8cm}, clip]{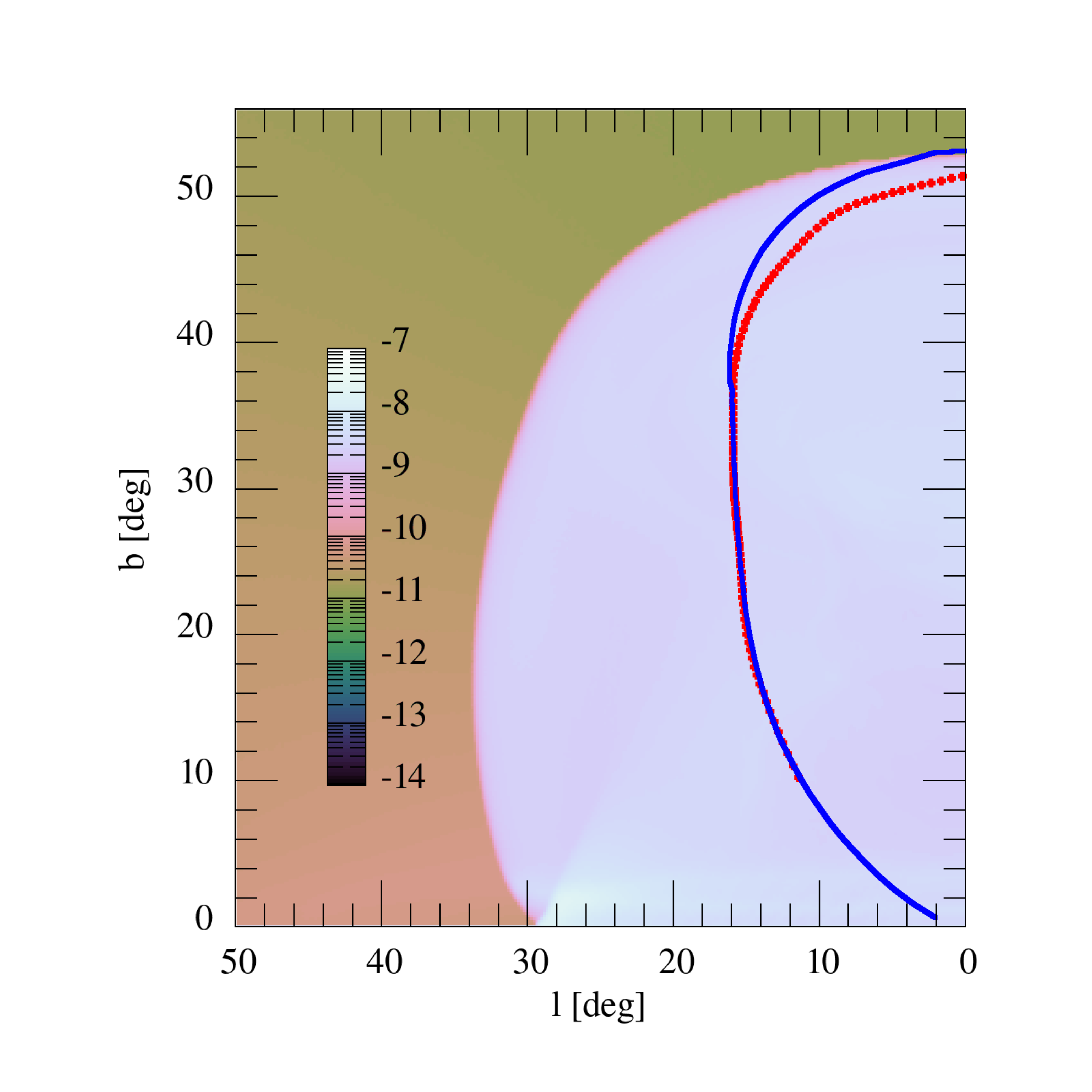}}};
\draw (-0.7, 1.8) node[text=white] {\scriptsize $t=6\Myr$};
\end{tikzpicture}
\hspace{-0.2cm}
\includegraphics[height=4.5truecm,trim={2.1cm 2.5cm 1.5cm 5.2cm}, clip]{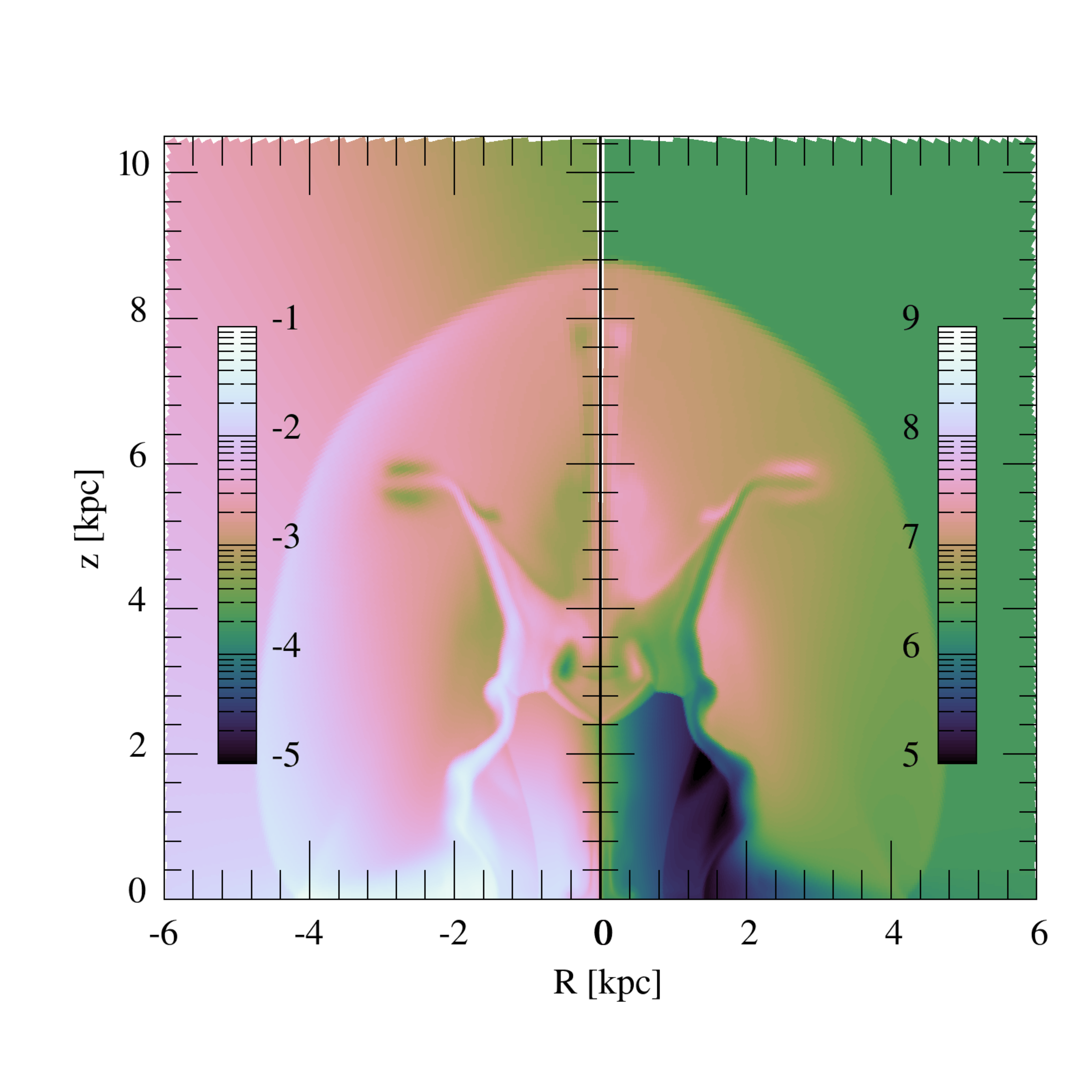}
\hspace{-2.5cm}
}
\\
\vspace{-0.68cm}
\centering{
\hspace{-2.5cm}
\begin{tikzpicture}
\draw (0, 0) node[inner sep=0] {\raisebox{0.15cm}{\includegraphics[height=4.5truecm,trim={4.8cm 2.2cm 4.0cm 2.8cm}, clip]{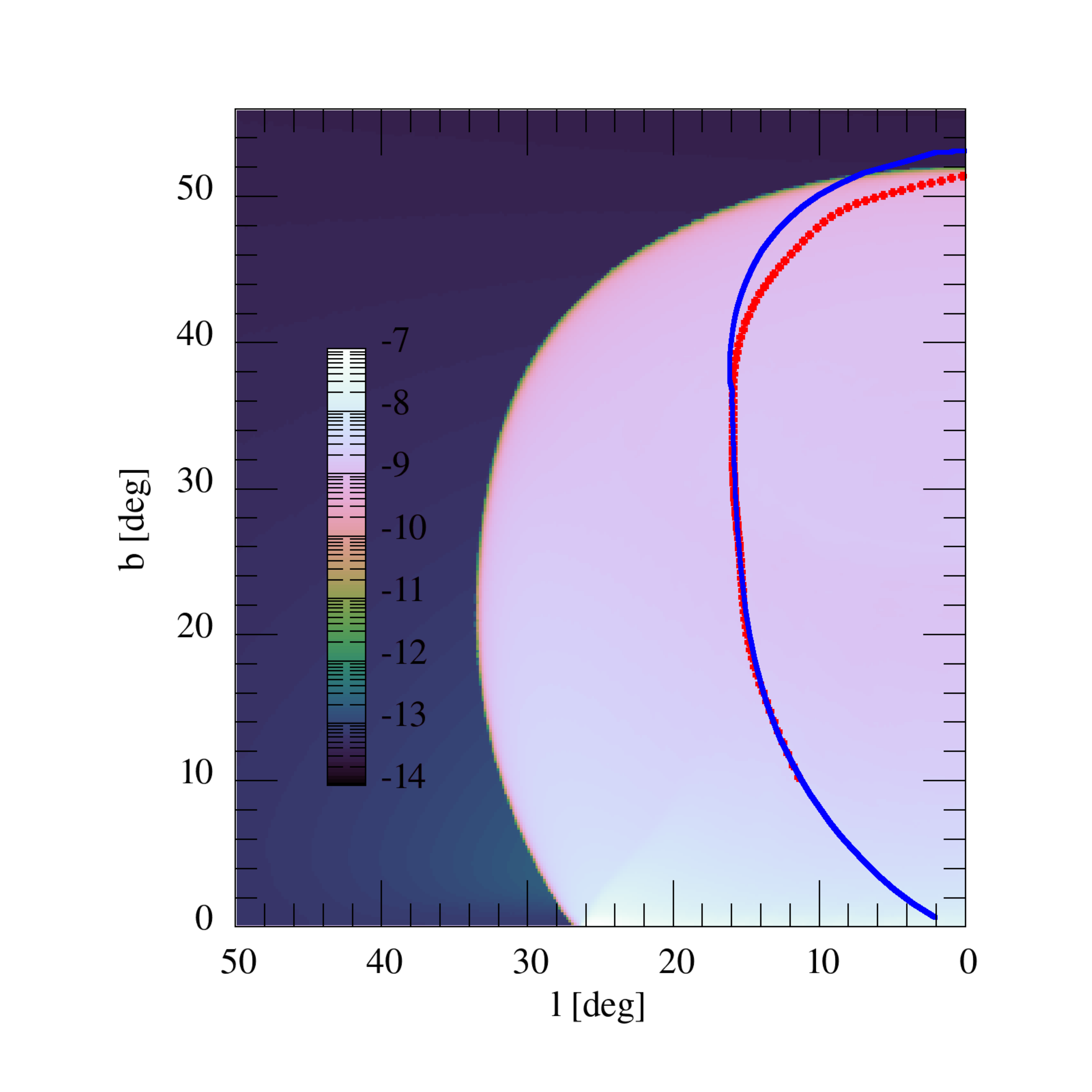}}};
\draw (-0.7, 1.8) node[text=white] {\scriptsize $t=2.8\Myr$};
\end{tikzpicture}
\hspace{-0.2cm}
\includegraphics[height=4.5truecm,trim={2.1cm 2.5cm 1.5cm 5.2cm}, clip]{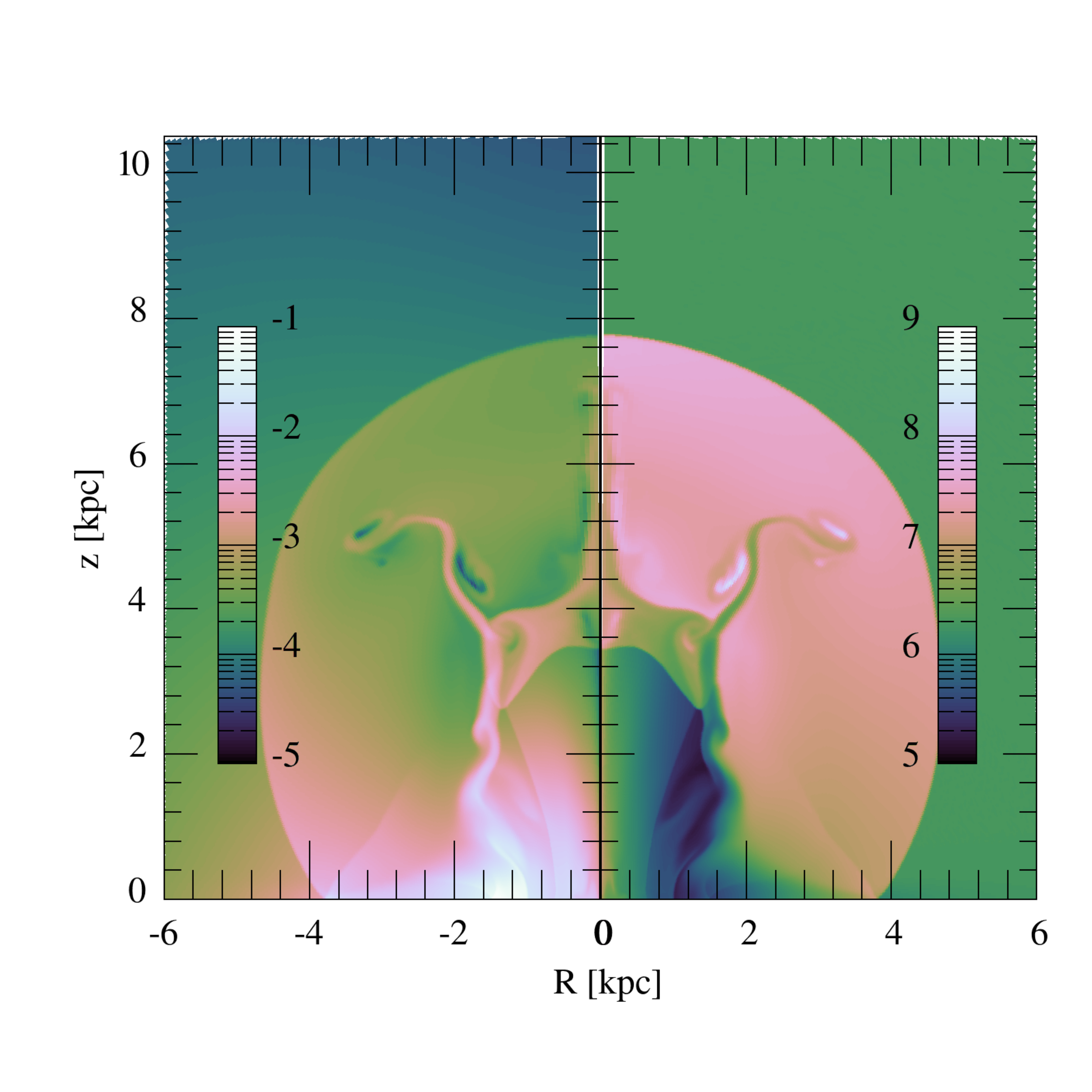}
\hspace{-2.5cm}
}
\\
\vspace{-0.68cm}
\centering{
\hspace{-2.5cm}
\begin{tikzpicture}
\draw (0, 0) node[inner sep=0] {\raisebox{0.15cm}{\includegraphics[height=4.5truecm,trim={4.8cm 2.2cm 4.0cm 2.8cm}, clip]{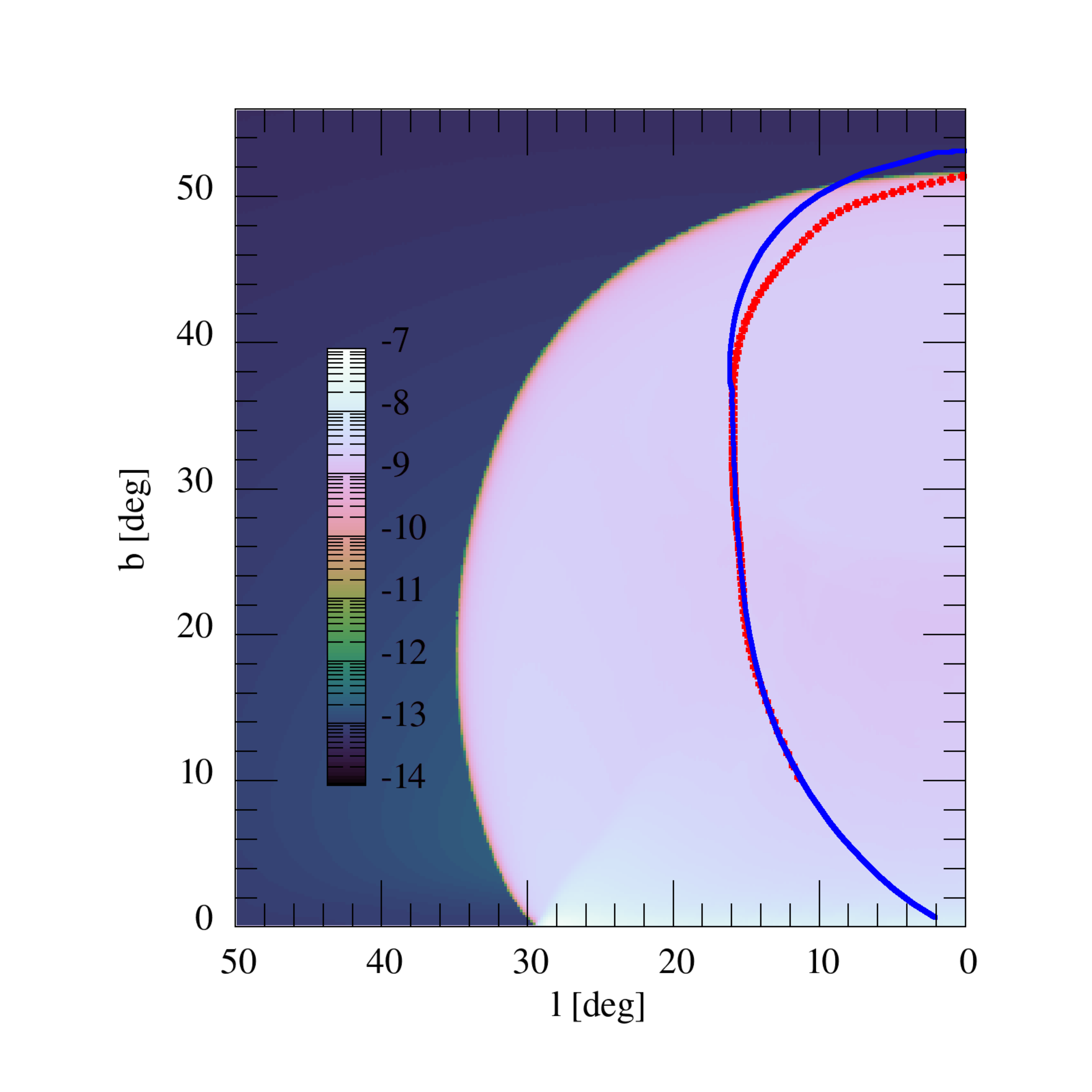}}};
\draw (-0.7, 1.8) node[text=white] {\scriptsize $t=3.5\Myr$};
\end{tikzpicture}
\hspace{-0.2cm}
\includegraphics[height=4.5truecm,trim={2.1cm 2.5cm 1.5cm 5.2cm}, clip]{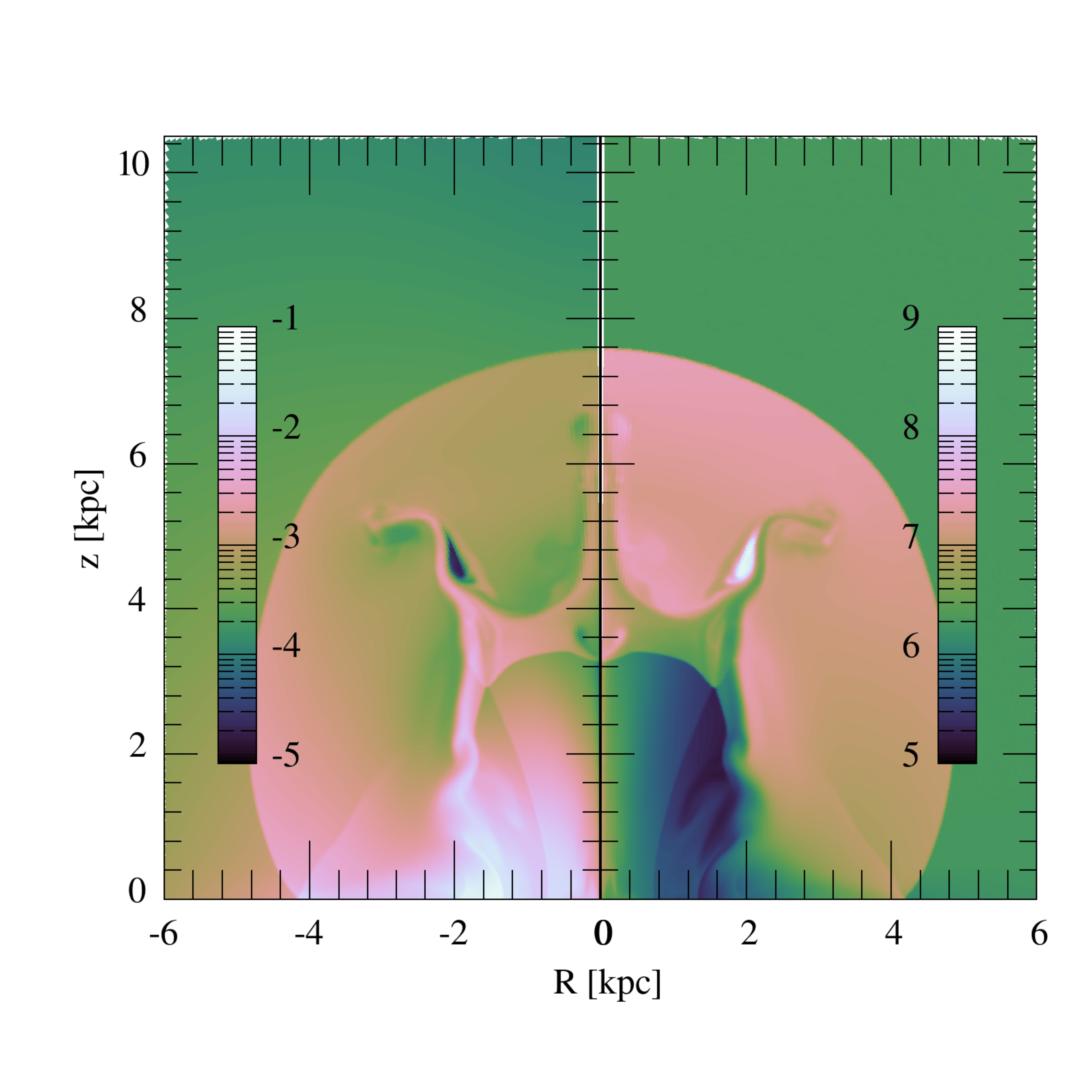}
\hspace{-2.5cm}
}
\\
\vspace{-0.69cm}
\centering{
\hspace{-2.2cm}
\begin{tikzpicture}
\draw (0, 0) node[inner sep=0] {\raisebox{0.15cm}{\includegraphics[height=4.5truecm,trim={4.8cm 2.2cm 4.0cm 2.8cm}, clip]{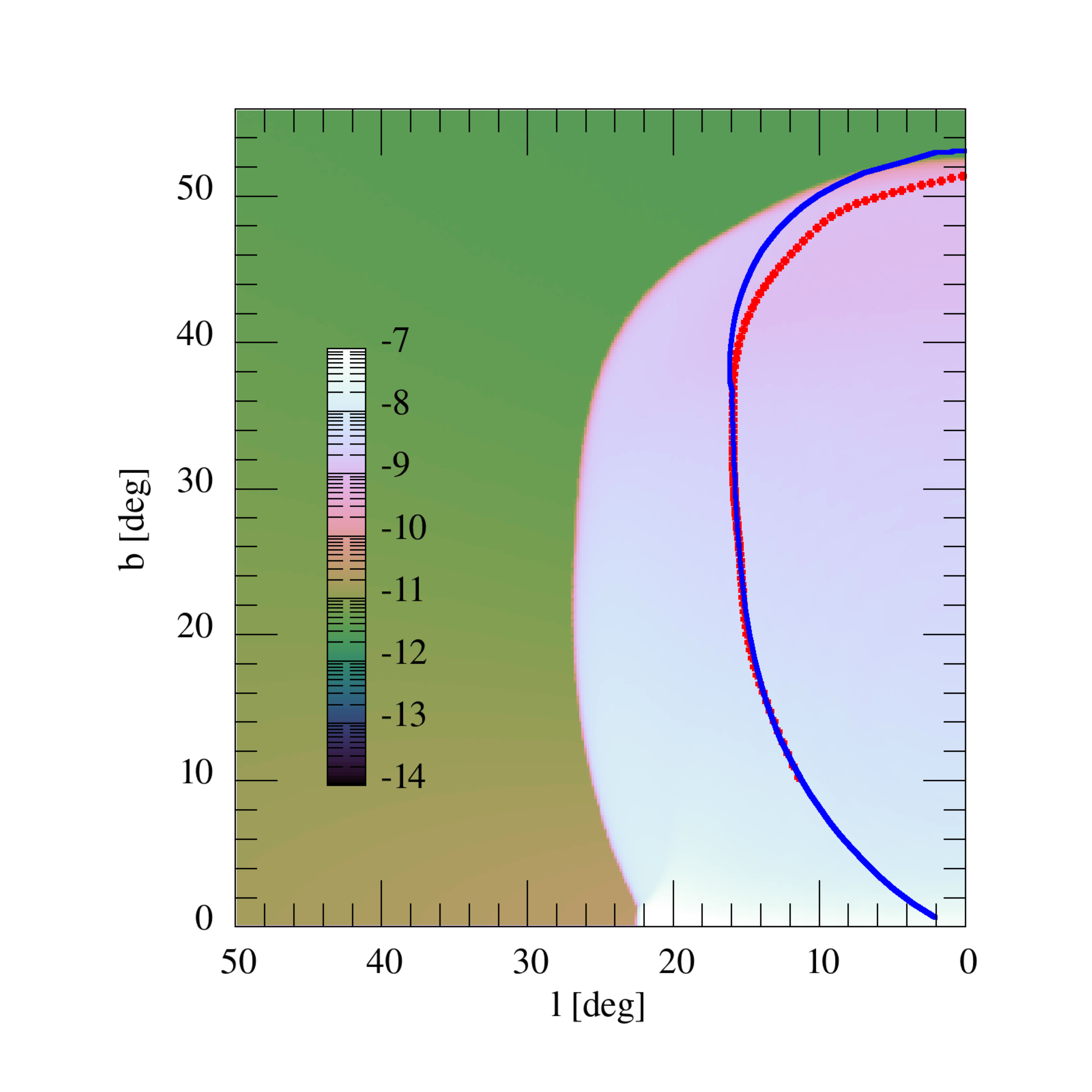}}};
\draw (-0.7, 1.8) node[text=white] {\scriptsize $t=2.8\Myr$};
\end{tikzpicture}
\hspace{-0.2cm}
\includegraphics[height=4.5truecm,trim={2.1cm 2.5cm 1.5cm 5.2cm}, clip]{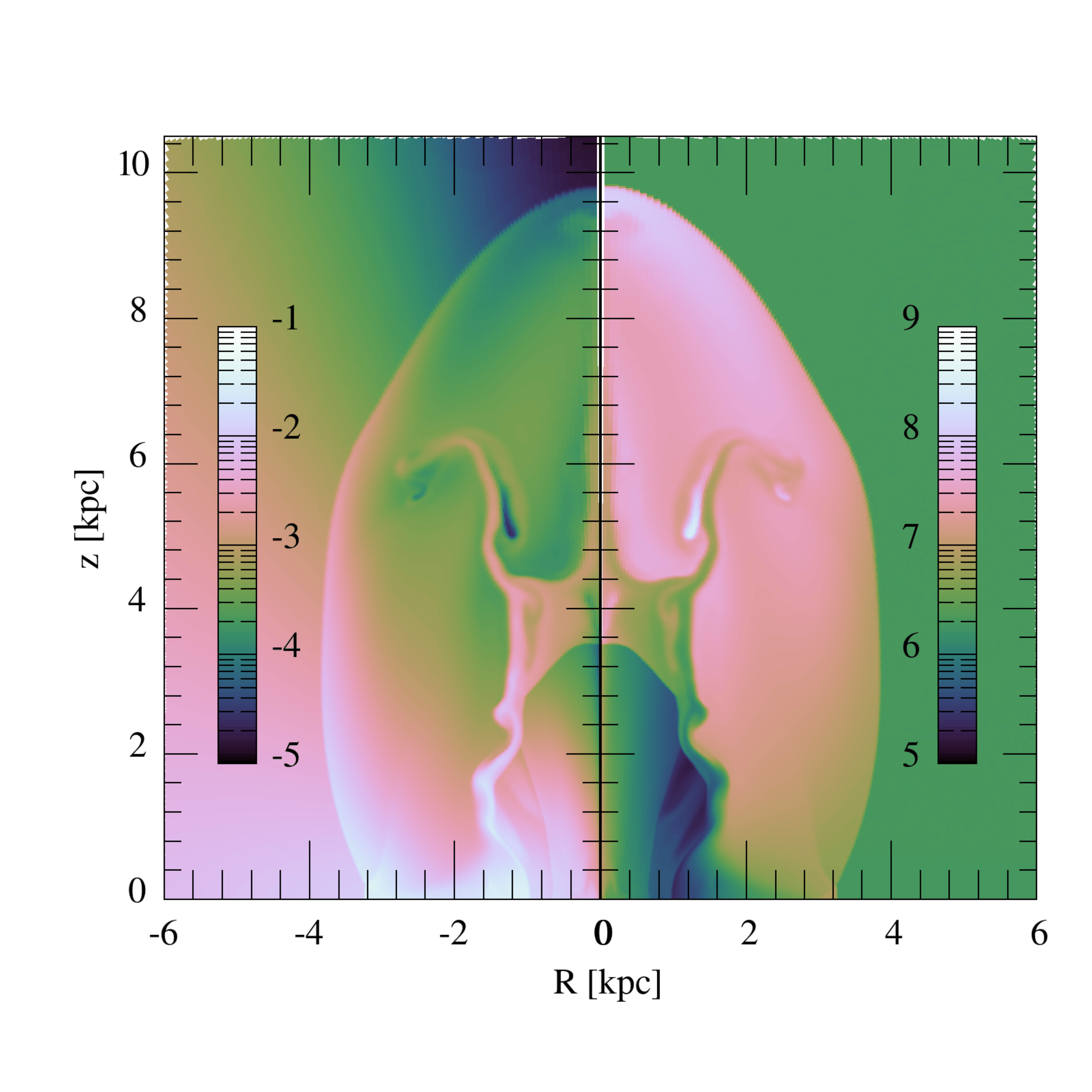}
\hspace{-2.5cm}
}
}
\caption{
Variations on GC injection: each row is similar to the top row of \autoref{fig:base_GC_IL}, and uses the same notations, but with some parameter change giving slightly thinner bubbles: nearly maximal halo
rotation, $f_h=0.9$ (first row), double the {\disk} mass, $M_{d}=1.2\times10^{11}M_\odot$
(second row),
double the bulge mass, $M_{b}=4\times10^{10}M_\odot$ (third row),
and all three modifications combined (bottom row).
Simulation parameters (setups S1a--d, top to bottom) are provided in the middle columns of \autoref{Tab:NonDirected}.
\label{fig:GC_setups}}
\end{figure}
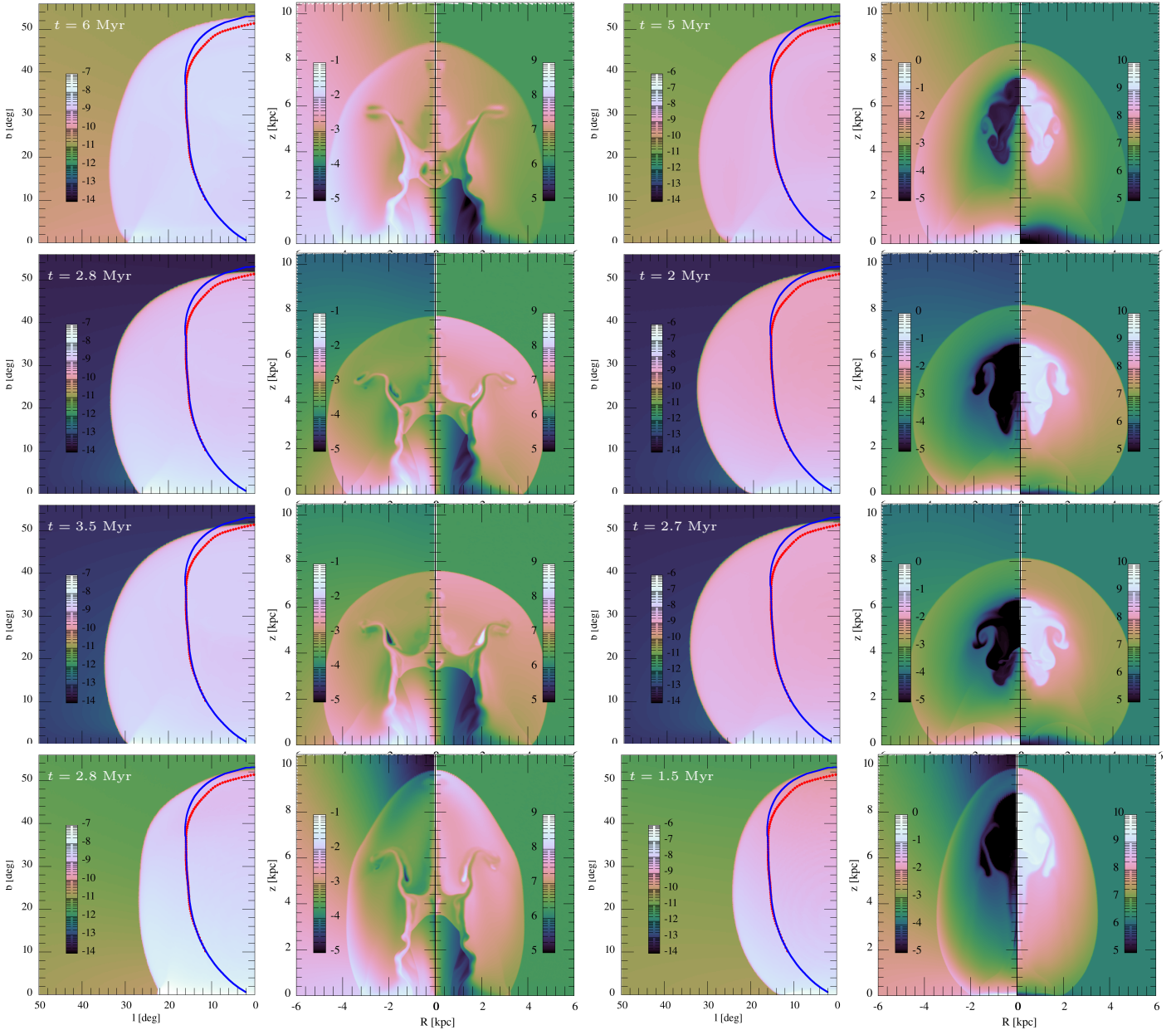

\begin{figure}
\DrawFigs{
\centering{
\hspace{-2.5cm}
\begin{tikzpicture}
\draw (0, 0) node[inner sep=0] {\raisebox{0.15cm}{\includegraphics[height=4.5truecm,trim={4.8cm 2.2cm 4.0cm 2.8cm}, clip]{Figures/solar-p5-HLat-1-2kpc-rp9.pdf}}};
\draw (-0.7, 1.8) node[text=white] {\scriptsize $t=5\Myr$};
\end{tikzpicture}
\hspace{-0.2cm}
\includegraphics[height=4.5truecm,trim={2.1cm 2.5cm 1.5cm 5.2cm}, clip]{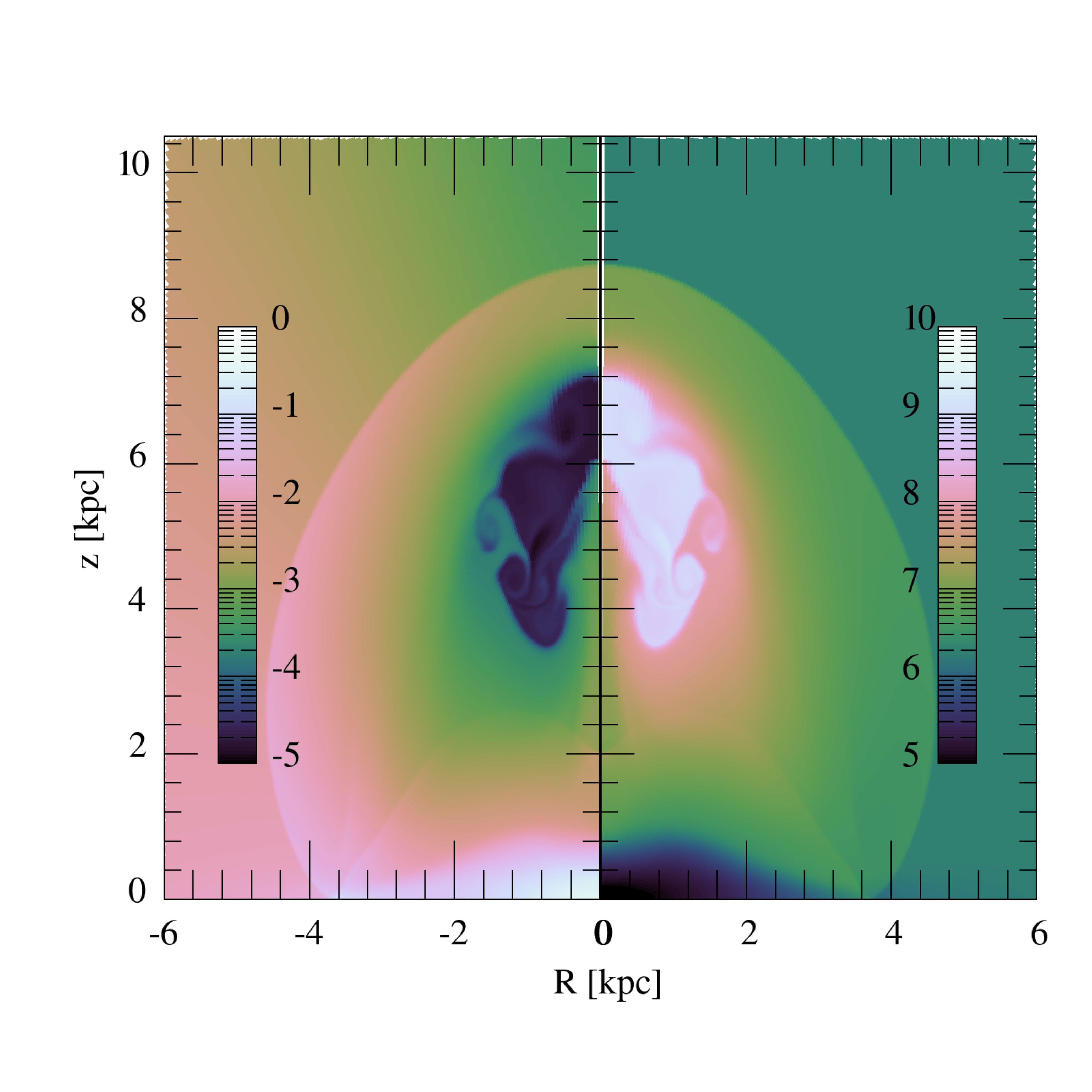}
\hspace{-2.5cm}
}
\\
\vspace{-0.68cm}
\centering{
\hspace{-2.5cm}
\begin{tikzpicture}
\draw (0, 0) node[inner sep=0] {\raisebox{0.15cm}{\includegraphics[height=4.5truecm,trim={4.8cm 2.2cm 4.0cm 2.8cm}, clip]{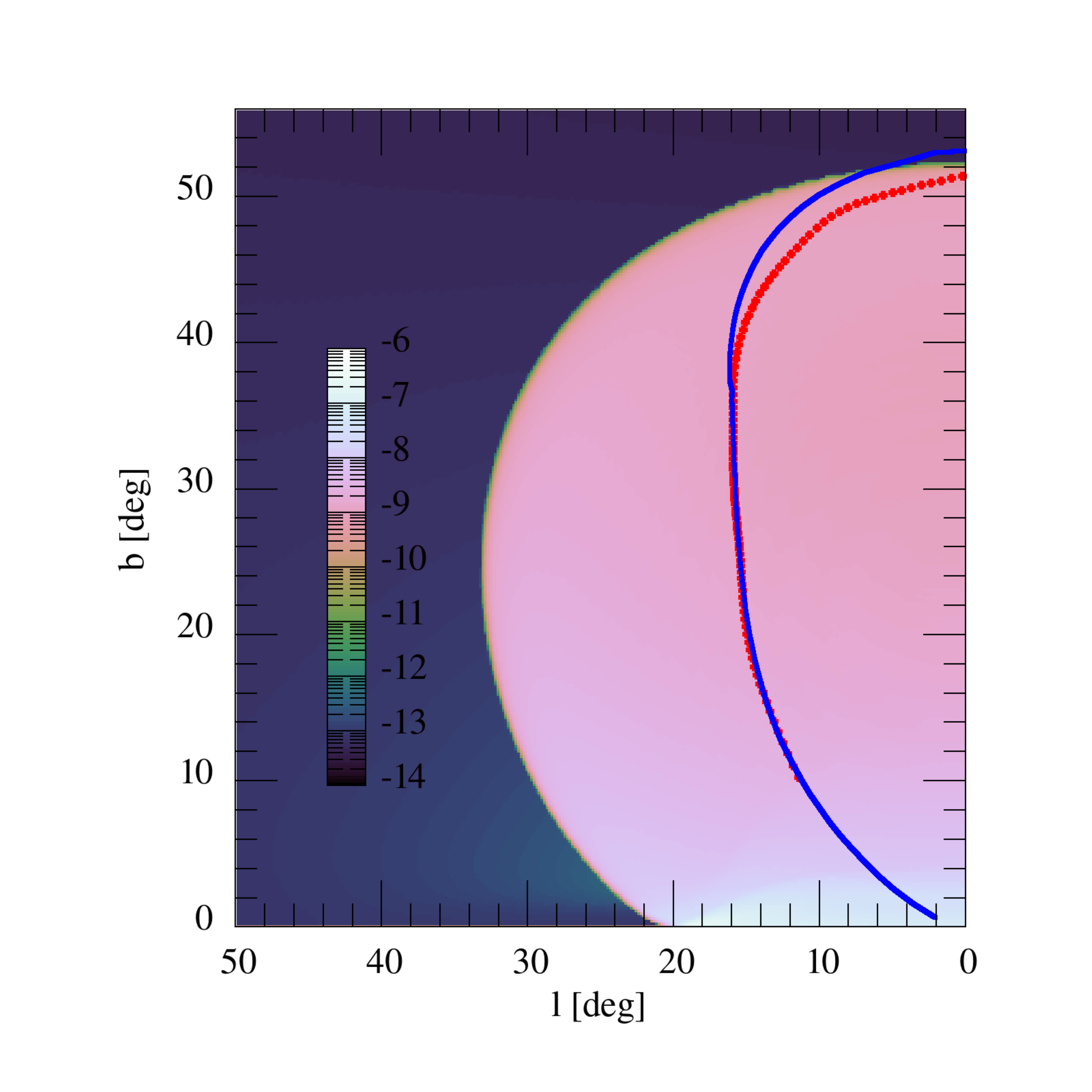}}};
\draw (-0.7, 1.8) node[text=white] {\scriptsize $t=2\Myr$};
\end{tikzpicture}
\hspace{-0.2cm}
\includegraphics[height=4.5truecm,trim={2.1cm 2.5cm 1.5cm 5.2cm}, clip]{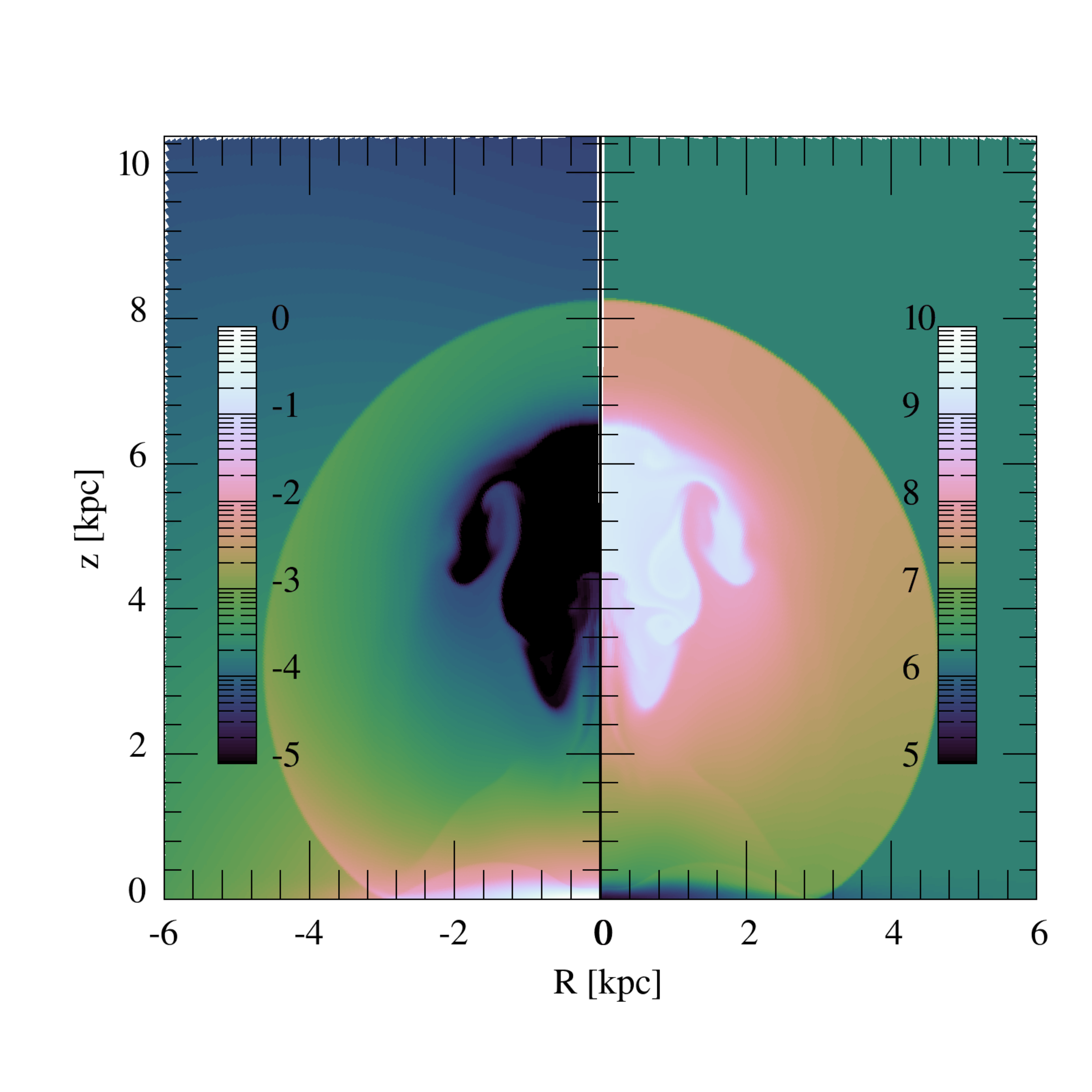}
\hspace{-2.5cm}
}
\\
\vspace{-0.68cm}
\centering{
\hspace{-2.5cm}
\begin{tikzpicture}
\draw (0, 0) node[inner sep=0] {\raisebox{0.15cm}{\includegraphics[height=4.5truecm,trim={4.8cm 2.2cm 4.0cm 2.8cm}, clip]{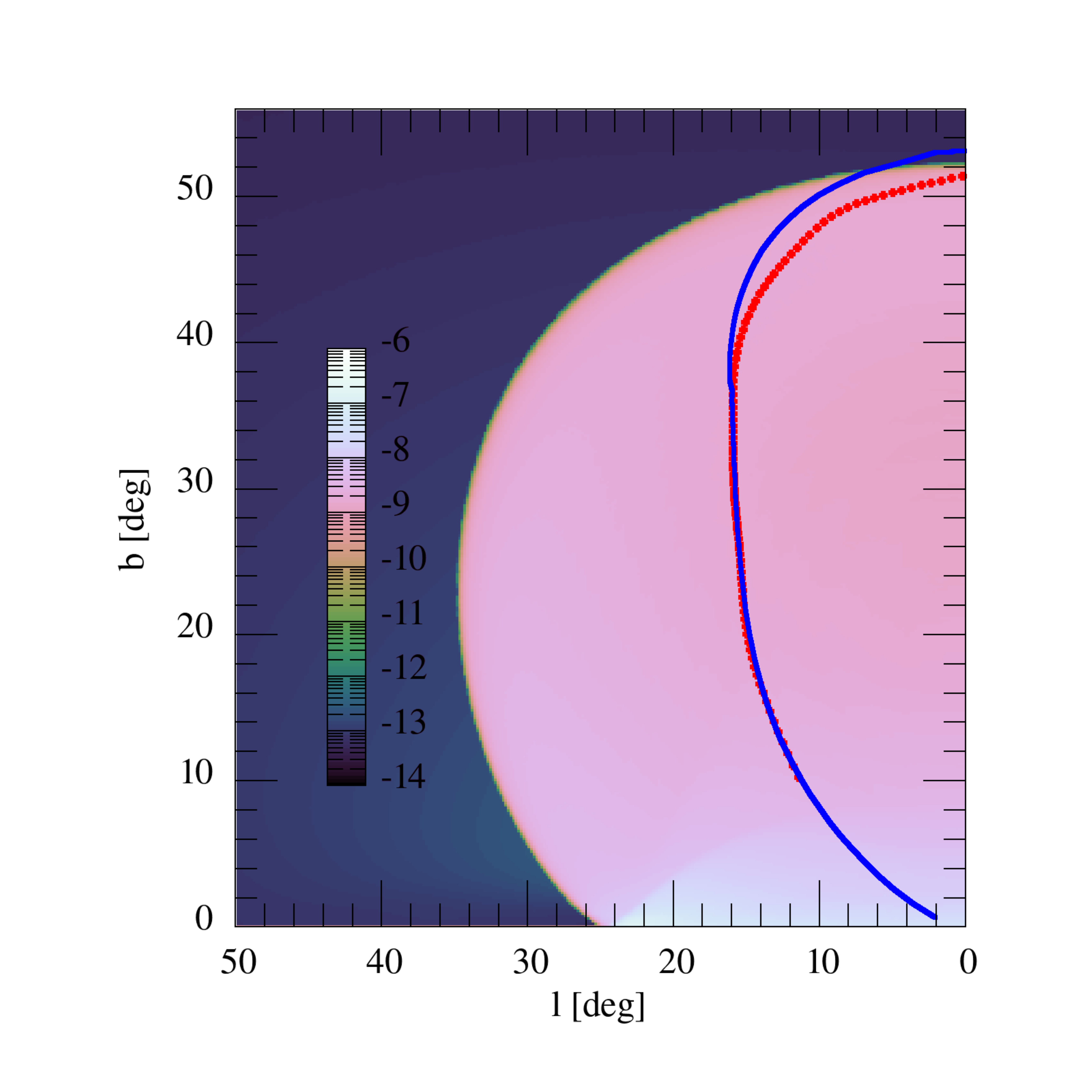}}};
\draw (-0.7, 1.8) node[text=white] {\scriptsize $t=2.7\Myr$};
\end{tikzpicture}
\hspace{-0.2cm}
\includegraphics[height=4.5truecm,trim={2.1cm 2.5cm 1.5cm 5.2cm}, clip]{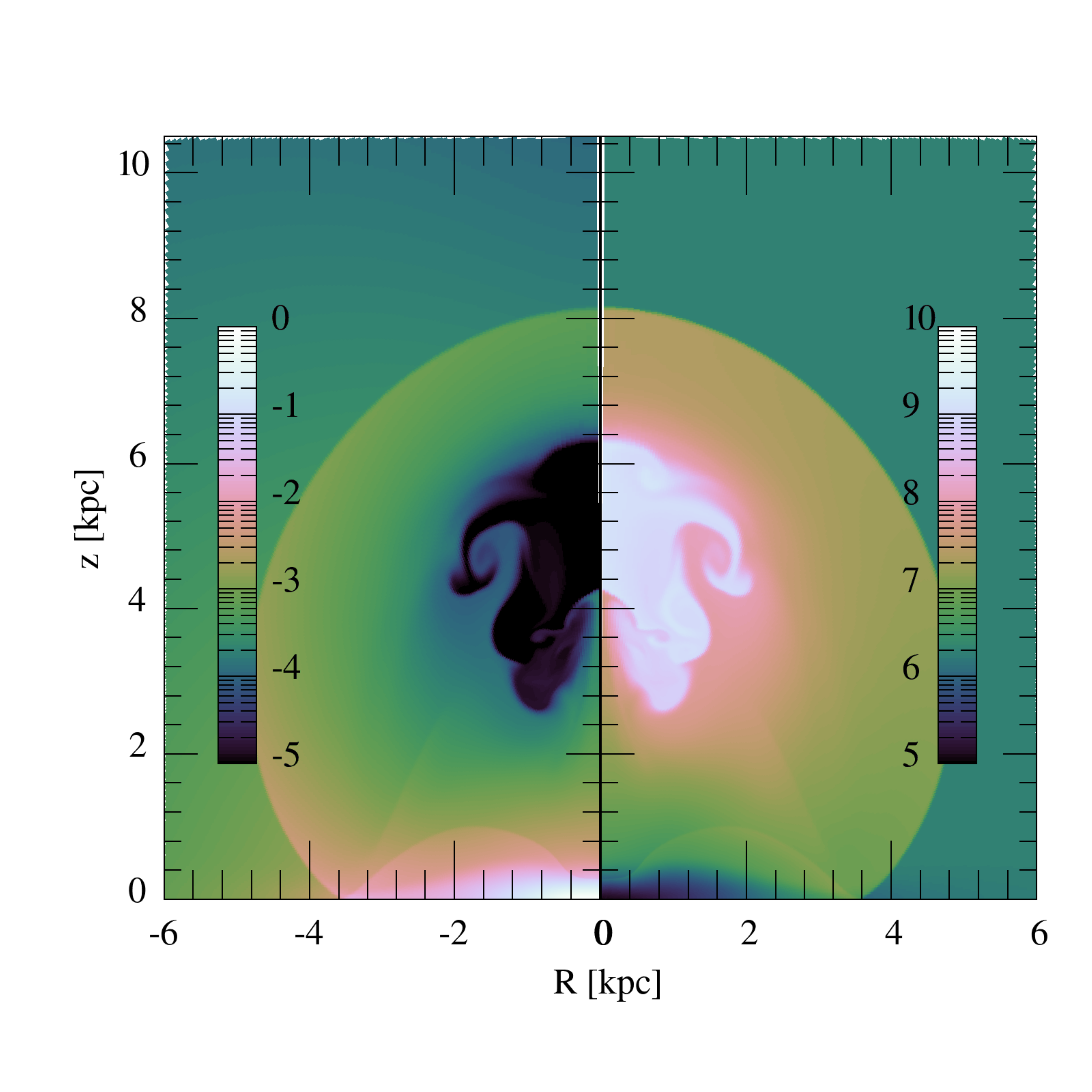}
\hspace{-2.5cm}
}
\\
\vspace{-0.69cm}
\centering{
\hspace{-2.3cm}
\begin{tikzpicture}
\draw (0, 0) node[inner sep=0] {\raisebox{0.15cm}{\includegraphics[height=4.5truecm,trim={4.8cm 2.2cm 4.0cm 2.8cm}, clip]{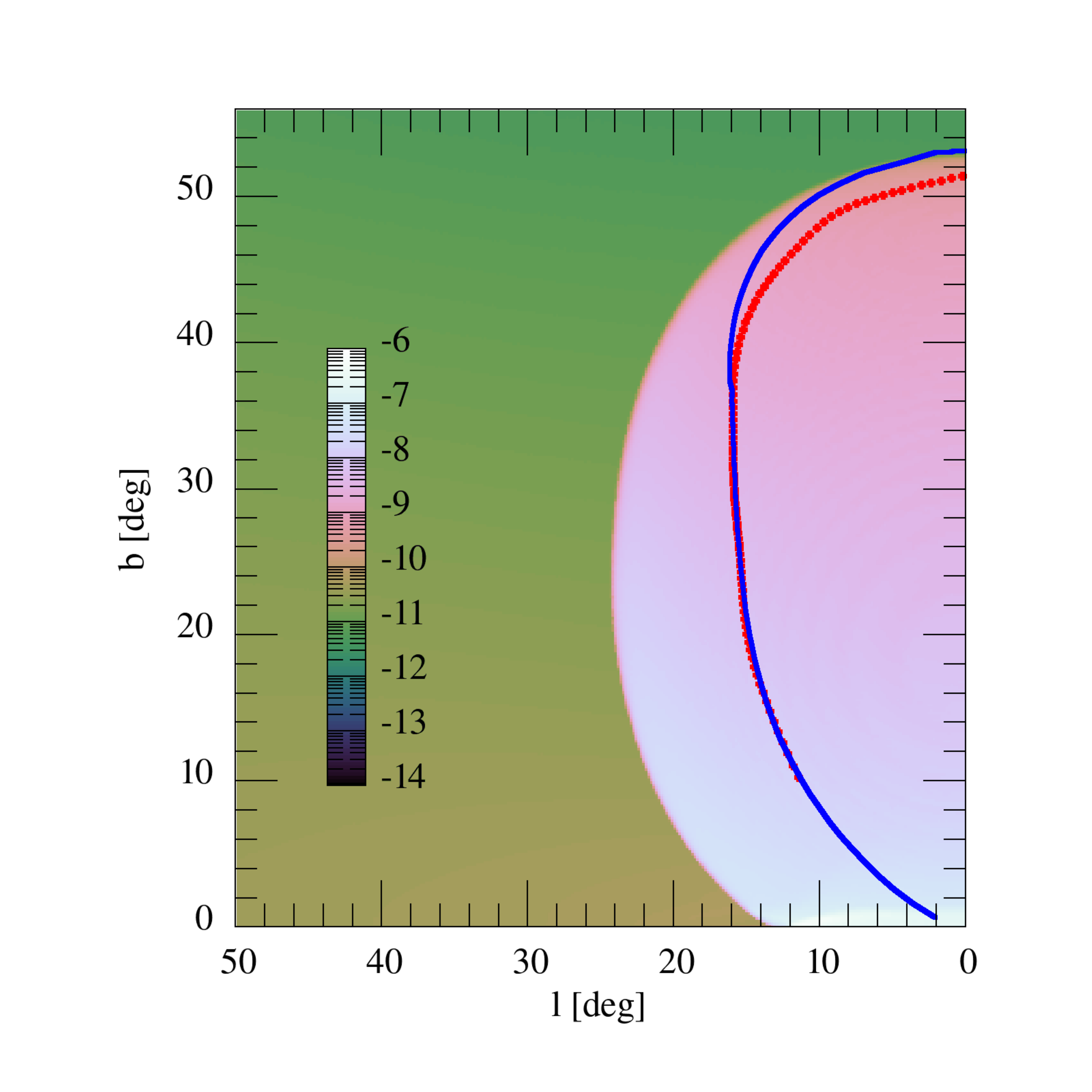}}};
\draw (-0.7, 1.8) node[text=white] {\scriptsize $t=1.5\Myr$};
\end{tikzpicture}
\hspace{-0.2cm}
\includegraphics[height=4.5truecm,trim={2.1cm 2.5cm 1.5cm 5.2cm}, clip]{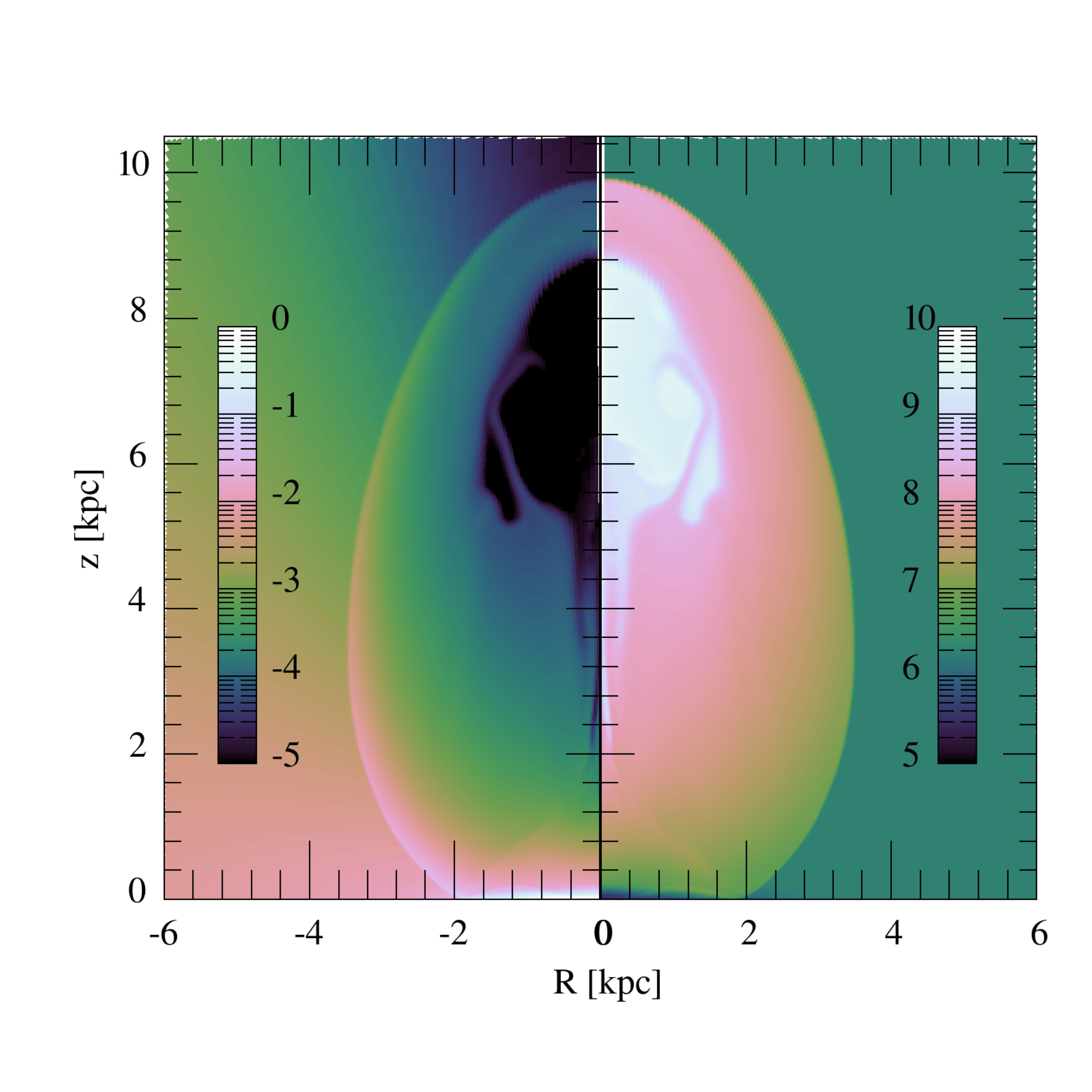}
\hspace{-2.5cm}
}
}
\caption{
Variations on IL injection: each row is similar to the bottom row of \autoref{fig:base_GC_IL}, and uses the same notations, but with some parameter change: $f_h=0.9$ (first row), $M_{d}=1.2\times10^{11}{M}_\odot$ (second row), $M_{b}=4\times10^{10}M_\odot$ (third row),
and the three changes combined (bottom row). Simulation parameters (setups S2a--d, top to bottom) are provided in the right columns of \autoref{Tab:NonDirected}.
\label{fig:IL_setups}}
\end{figure}

After its ballistic stage, the evolution of the bubbles becomes sensitive to the properties of the CGM distribution.
It is therefore natural to ask if there are any plausible modifications to the Galactic model that could alter the simulated bubbles and produce an acceptable match to the observed FBs.
As the FBs are substantially thinner than the bubbles in \autoref{fig:base_GC_IL}, a better fit to the observations requires enhancing the collimation of the simulated bubbles.
Such collimation can be established by considering flatter, more {\disk}-like mass distributions near the origin of the bubbles.
This, in turn, can be facilitated by a faster CGM rotation, a more massive {\disk}, or a more massive bulge.
These three modifications are considered, separately and combined, in \autoref{fig:GC_setups} for GC injection, and in \autoref{fig:IL_setups} for IL injection.

In the first three rows of these two figures, we consider a radical modification of a single parameter:
(i) a nearly maximal halo rotation parameter,
$f_h=0.9$;
(ii) doubling the {\disk} mass, $M_d=1.2\times10^{11}M_\odot$;
and
(iii) doubling the bulge mass, $M_{b}=4\times10^{10}M_\odot$.
The fourth row simultaneously combines all three modifications, to examine their cumulative effect.
The masses of the Galactic stellar components are varied while keeping the dark matter mass of the Galaxy constant; this changes the local gravity near the {\disk}, but does not alter the total mass of the Galaxy (including dark matter) by more than $\sim 10\%$.
The effect of changing the gravity translates to a slightly different gas density profile near the {\disk}.
A similar effect could also be obtained by varying the temperature profile of each component, but our present construction of the CGM profile
does not accommodate a variable temperature.

As the two figures show, these modifications to the CGM model, although extreme, are still insufficient to provide the necessary collimation of the bubbles.
This conclusion holds even when all modifications are combined, for both GC and IL injection, and our results are converged and not sensitive to the addition of viscosity.
In both injection methods, the modifications do somewhat enhance the collimation, in particular when all modifications are combined with IL injection.
However, the resulting bubbles are still too wide and too spherical to account for the observations.
The collimation is associated with taller and younger bubbles, reaching $z\simeq 10\kpc$ and $\tage\sim1.5\Myr$ for the thinnest of the bubbles in the figures.
Extrapolation suggests that a good match to the FB morphology would require unrealistically massive central Galactic components and young bubbles.

We conclude that the rapid injection of energy near the GC, or even at intermediate latitudes above it, cannot explain the morphology of the FBs as forward shocks, for any Galactic model consistent with present observations.
Our results are consistent with previous studies \citet[][hereafter ZG20]{ZhangGuo2020}, which focused on a specific Galactic model and so could not make the above general claim.

\section{Jetted injection}
\label{jettedresults}

As concluded in \autoref{nonjettedresults}, we are unable to reproduce the observed FB geometry using non-directed injection, for any plausible Galactic model.
We are therefore reduced to considering jetted injection scenarios, where a new dimensional scale is introduced into the problem by means of the conserved momentum $p_z=p_j$ deposited in the axial, $z$ direction, or interchangeably by the initial radial velocity $v_{\inj}$ introduced in the small, $0<\theta<\theta_j$ opening cone.

With such jetted injection models, we find that the observed FB geometry can be nicely reproduced in two distinct ways.
A sufficiently energetic, slow initial jet can reproduce acceptable edges for simulated FBs that are presently still in their ballistic stage,
whereas the converse --- an initially fast, low-energy jet --- can reproduce acceptable edges of non-ballistic FBs already in their slowdown phase.

At early times, the injected flow evolves ballistically in both cases, with the head of the bubble (subscript $H$) moving at a near constant velocity $v_H(t)\equiv\dot{z}_H\simeq v_{\inj}$. After the flow accumulates a mass $M$ larger than the initial deposited mass $M_{\inj}$, the flow inevitably starts slowing down, with $v_H(t)\sim t^{-\tv}$ typically following a power-law decay with index $\tv>0$.
It is not yet known if the FBs observed today are in the former, ballistic stage or in the latter, slowdown stage, so we examine both scenarios.
One can further split the ballistic forward evolution of a jetted bubble into two stages according to the sideways expansion being either ballistic or non-ballistic \citep{Irwinetal19}, but we avoid this distinction here.

The mass $M$ swept up by the bubbles is strongly dominated, for plausible Galactic models, by the halo, CGM component, exceeding the swept-up {\disk} mass by about two orders of magnitude.
For a $\rho_h\propto r^{-\ah}$ CGM distribution, the present-day FBs are thus in the ballistic stage if the injected mass $M_j$ satisfies
\begin{equation} \label{eq:BallisticRegime}
\frac{M_j}{2} \equiv \frac{E_{\inj}}{v_{\inj}^2} \gtrsim  \frac{M}{2} \simeq \pi\theta_j^2\int r^2 \rho_h(r)\,dr \simeq \frac{\pi \theta_j^2\rho_h(z_H)z_H^3}{3-\ah} \, ,
\end{equation}
where $z_H\simeq 10\kpc$ is the present-day height of the bubbles, and we approximated $\theta_j$ as small; recall that $E$ and $M$ pertain to the sum of both hemispheres.
Conversely, the FBs have already transitioned into the slowdown phase if $M_j\lesssim M$, which may be written for $\ah=3/2$ as
\begin{equation} \label{eq:NonBallisticRegime}
\frac{E_{\inj}}{3\times 10^{55}\erg} \lesssim \left( \frac{\theta_5\beta_j}{0.01}\right)^{2} z_{10}^3 \rho_4 \simeq (10^3\beta_\perp)^{2} z_{10}^3 \rho_4 \, ,
\end{equation}
where we defined
$z_{10}\equiv z_H/10\kpc$, $\theta_5 \equiv \theta/5^\circ$, $\rho_4\equiv \rho_h(z_H)/4\times 10^{-4}m_p\cm^{-3}$, and the sideways initial velocity $\beta_\perp\equiv \beta_{\inj}\theta_j$.
One can equivalently replace $v_{\inj}$ by the conserved $2E_{\inj}/p_j$.

After developing in \autoref{sec:ToyModels} a simple analytic model for the jetted bubbles in both regimes, we simulate the two scenarios in \autoref{subsec:JettedSims}.
We examine different variants of such bubbles, as described for jetted injection in \autoref{jettedinjection}, and compare the bubbles as they reach $b\simeq 52^\circ$ with the model and with the observed FBs.

\subsection{Stratified model for jetted bubbles}
\label{sec:ToyModels}

Before discussing the simulated bubbles, it is useful to derive some relevant scaling relations using a simple toy model.
Sufficiently far above the Galactic {\disk}, the mass accumulated by the bubble is dominated by halo material, which has an approximately one-dimensional distribution around a narrow jet.
The ambient medium can therefore be approximated as an isothermal, $\rho\propto z^{-\ah}$, planar atmosphere, where in our simulations $\ah\simeq 1.5$.
The fairly narrow FBs, when interpreted as arising from thin initial jets, are thus amenable to a simplified model that approximates the flow behind the head of the bubble as quasi two-dimensional, propagating perpendicular to the $z$ axis.

Such a piecewise planar, or stratified, model is sufficient for capturing the main properties of the simulated flow, as we show for the ballistic bubbles in \autoref{Modelballisticregime}, and for bubbles in their slowdown phase in \autoref{Modelnonballisticregime}.
More sophisticated modeling, for example using the Kompaneets approximation \citep[][and references therein, for a simplified ambient model and $\gamma=4/3$]{Irwinetal19} is less useful here, because we find that (i) the ballistic phase plays an important role in the FBs; (ii) even in the slowdown phase, the injected momentum plays a key role; and (iii) post-shock pressure variations across the bubble surface are not small for slowing-down FBs.

We find that the structure and velocity of the jet are modified in the initial, injection stage, especially in the high $v_j$ regime used to produce bubbles reaching the slowdown phase. We therefore use subscript $\modj$ instead on $j$ to refer to the actual jet parameters immediately after its $0<t<t_j$ launching.

\subsubsection{Ballistic regime}
\label{Modelballisticregime}

\begin{figure*}
\DrawFigs{
\centering{
\includegraphics[height=6.5truecm]{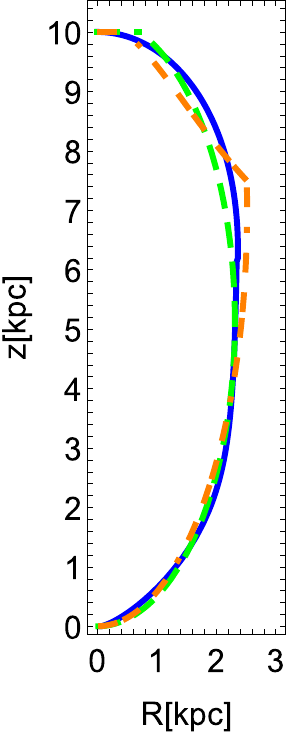}
\includegraphics[height=6.5truecm]{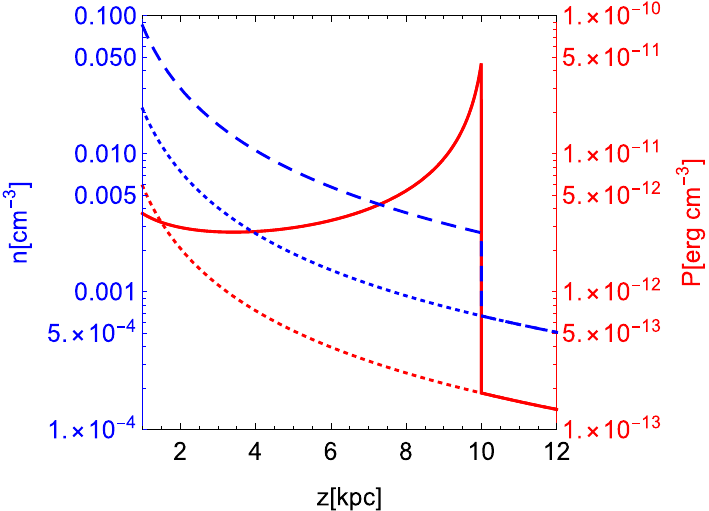}
\includegraphics[height=6.5truecm, clip=true, trim={0cm 0cm 0cm 2cm}]{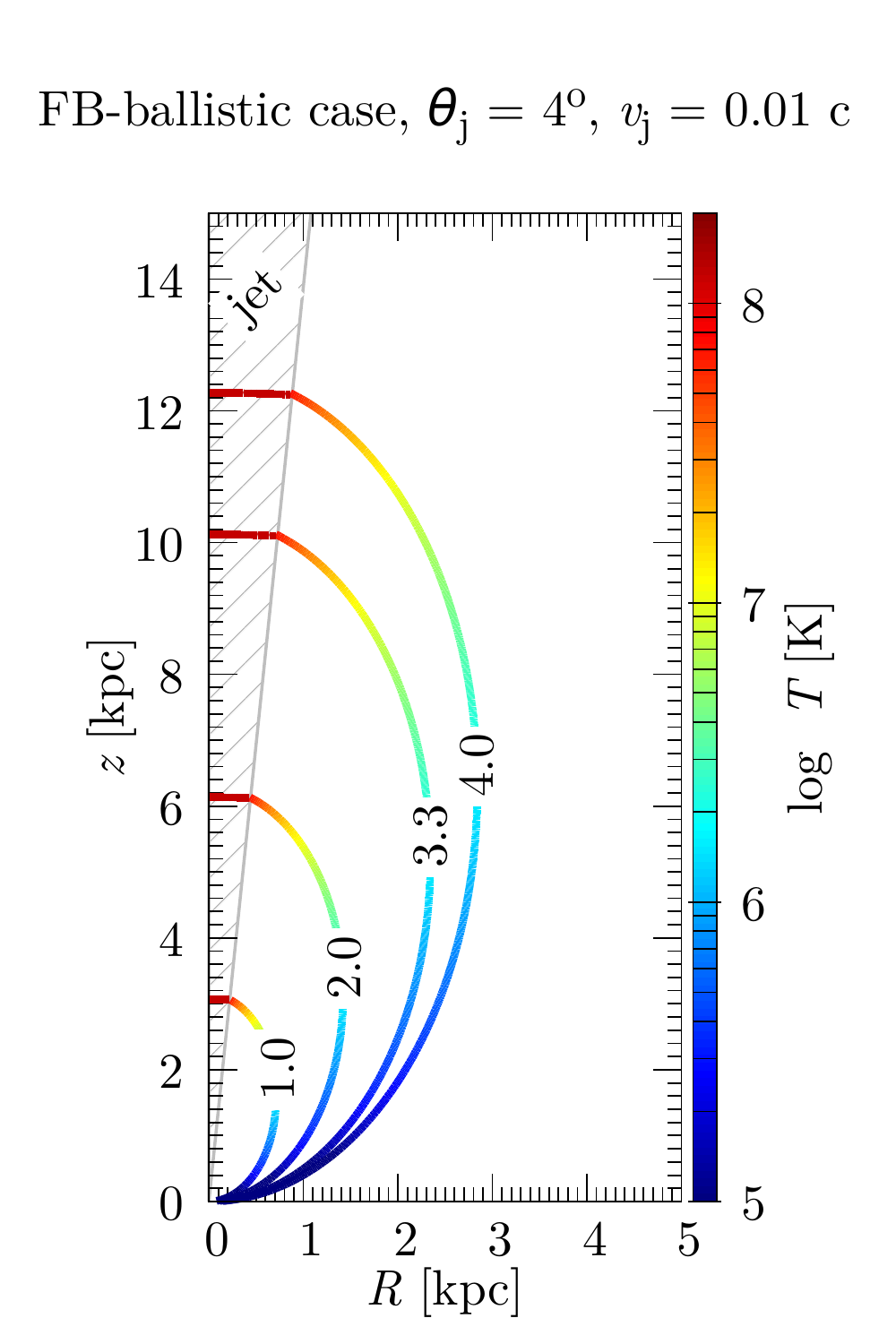}
}
}
\caption{
Left panel: ballistic (with $\theta_{\modj}=4^\circ$; dashed green) and slowdown (with $\theta_{\modj}=2^\circ.5$, $z_{\tr}=7.5\kpc$ and $\Myfq=1/4$; dot-dashed orange) models for the bubble edge at $z_H=10\kpc$, compared with a phenomenological geometry previously used to reproduce the projected observations \citep[][solid blue]{Keshetgurwich18}.
Middle panel: density (dashed blue; left axis) and pressure (solid red; right axis) profiles of the ballistic model with a $\rho_h=4\times10^{-4}m_p (r/10\kpc)^{-1.5}$ CGM and $\beta_{\modj}=0.01$. The profiles before launching the jet are also shown (dotted); subsequent evolution including the reverse shock and contact discontinuity are not.
Right panel: ballistic evolution of the shock front according to Eq.~(\ref{eq:BallisticShape}) at different times as noted on the curves (in Myr). The color in each curve shows the shock temperature from Eq.~(\ref{eq:BallisticTemp}) except within the jet cone where the temperature is given from Eq.~(\ref{eq:BallisticHeadTemp}).
\label{fig:FBModels}}
\end{figure*}

Consider our simple jetted setup deep in the ballistic regime.
Here, the bubble has a height
\begin{equation} \label{eq:BallisticHeight}
z_H(t)\simeq v_{\modj}t \coma
\end{equation}
so an observed FB with $z_H\simeq 10\kpc$ would be of age
\begin{equation} \label{eq:BallisticAge}
\tage \simeq z_H/v_{\modj} \simeq 3.3z_{10}\beta_{-2}^{-1} \Myr \, .
\end{equation}
where $\beta_{-2}\equiv\beta_{\modj}/0.01$.
At its top, the bubble head has a width
\begin{equation}
R\simeq \theta_{\modj} z_H \simeq 0.9 z_{10}\theta_5\kpc \, ,
\end{equation}
where the forward shock is strong, of Mach number
\begin{equation}
\Upsilon\simeq \frac{v_{\modj}}{c_s} \simeq 14 \beta_{-2}\left(\frac{T_h}{2\times 10^6\mbox{ K}}\right)^{-1/2} \, .
\end{equation}
Here, we defined $\theta_5\equiv \theta_{\modj}/5^\circ$.
At lower, $z<z_H$ heights, the bubble initially has a half-width
\begin{equation} \label{eq:R_IC}
R(z,t=z/v_{\modj})\simeq \theta_{\modj} z \, ,
\end{equation}
but later necessarily expands due to the pressure of the shocked CGM.

Let us parameterise the geometry of the bubble as $R_b(z,t)\simeq z\Theta(z,t)$, in the small angle, $\Theta\ll 1$ limit.
A crude way to estimate the shape $R_b(z,t=\tage)$ of the evolved ballistic bubble is to consider each fixed-$z$ slice as a separate layer, independently expanding sideways, \ie in the $R$ direction.
The pressure transferred to this slice as it is first shocked by the head of the bubble is, according to the shock jump conditions, $P(z)\simeq (3/4)\rho_h(z)v_H(z)^2$.
For simplicity, we approximate the subsequent slice evolution as sideways expansion with a characteristic velocity $v_R(z,t)=\MyfR \dot{R}_b$,
where $\MyfR$ is a dimensionless factor of order unity,
chosen as $\MyfR\simeq 1/2$ in light of the nearly linear velocity profile anticipated behind a cylindrical strong shock.
We roughly equate the associated kinetic energy per unit height with the thermal energy initially deposited by the head,
$4\rho_h(z)\pi R_b^2(z,t)v_R^2(z,t)/2 \simeq \pi (z\theta_{\modj})^2 (3/2)P(z)$.

Solving the resulting ODE,
\begin{equation}\label{eq:ShapeODE}
2z\Theta\dot{\Theta}=3v_H(z)\theta_{\modj} \, ,
\end{equation}
with the ballistic behavior $v_H(z)\simeq v_{\modj}$ and the initial condition (\ref{eq:R_IC}), we then obtain
\begin{equation}\label{eq:BallisticShape}
\Theta(z,t)^2 \simeq 3\left( \frac{v_{\modj} t}{z}-1 \right)\theta_{\modj} +\theta_{\modj}^2 \, ,
\end{equation}
in good agreement with the FB shape inferred from observations.
For small $\theta_{\modj}$, and for $z$ not too close to $z_H$, we may approximate the bubble shape by
\begin{equation}\label{eq:BallisticR}
R_b(z,t)^2\simeq 3\left[z_H(t)-z\right]\theta_{\modj}z \, .
\end{equation}
Note that for a fixed $z\gg R$, at late times $R(z)\sim t^{1/2}$ agrees with the expected self-similar cylindrical expansion.

For small $\theta_{\modj}$, the maximal half-width
\begin{equation}\label{eq:BallisticRmax}
R_{\max}(t)\simeq \frac{v_{\modj}t}{2}\sqrt{3\theta_{\modj}} \, ,
\end{equation}
is obtained at the bubble half-height, $z\simeq z_H/2$.
Hence, for the evolved bubble
\begin{equation} \label{eq:BallisticRmaxTage}
R_{\max}(t=\tage)\simeq \frac{z_H}{2}\sqrt{3\theta_{\modj}}\simeq 2.6\,\theta_5^{1/2}z_{10}\kpc \, .
\end{equation}
Equation (\ref{eq:BallisticRmax}) also implies that the latitudinal and longitudinal widths of the bubble at any given time are
\begin{align} \label{eq:BallisticWidths}
 \Delta b & \simeq  \frac{z_H}{R_\odot} \simeq 20.1^\circ \:\beta_{-2}\: t_{\rm Myr} \, ; \nonumber \\
\Delta l &\sim \frac{R_{\rm max(t)}}{R_\odot} \simeq 5.3^\circ\: \beta_{-2}\: t_{\rm Myr}\: \theta_5^{1/2}\,.
\end{align}

Non-ballistic corrections, becoming gradually larger as the swept up mass $M$ approaches the injected mass $M_{\inj}$ near the slowdown phase, can render the bubbles somewhat older and wider than estimated here.
The geometry of the bubble edge
for $z_H=10\kpc$ is demonstrated in \autoref{fig:FBModels} (left panel).
This one-zone approximation for the sideways expansion leaves out substructure associated with the contact discontinuity and the reverse shock.

The shock is fastest at the top of the bubble, where the downstream temperature is
\begin{equation} \label{eq:BallisticHeadTemp}
T_i\simeq \frac{3\mu_m m_p}{16k_B} v_{\modj}^2 \simeq 1.2 \times 10^8 \beta_{-2}^2\mbox{ K}  \, .
\end{equation}
Note that the electron temperature may be considerably lower than the ion temperature close to the shock, as the ion--electron equilibration
time-scale
can be shorter than the dynamical time-scale ($z_H/\tage$) of the shock \citepalias[\eg][]{Keshetgurwich18}.
The sideways velocity $v_R=f_R z\dot{\Theta}$ of the shock at a given $z$ and $t$ is given by
\begin{equation}
v_R \simeq \frac{3 v_k \theta_k}{4\Theta(z)} \simeq
\frac{3}{4}v_k\bigg/\sqrt{1+\frac{3}{\theta_k}\left[\frac{z_H(t)}{z}-1\right]}
\,,
\end{equation}
indicating that the temperature behind a strong sideways shock,
\begin{equation}
    T_{R} \simeq \frac{3}{16} \frac{\mu_m m_p}{k_B} \left( \frac{3 v_k \theta_k}{4\Theta } \right)^2 \,,
    \label{eq:BallisticTemp}
\end{equation}
is constant at a given $\Theta$, but gradually declines in time at a fixed $z$.
The evolution of the shape and temperature is shown in the right panel of Fig \ref{fig:FBModels}.
Note that, the above temperature is only derived from the sideways expansion and is expected to fail near $\Theta \sim \theta_k$ since the shock velocity in this region will be dominated by vertical motion of the jet.

\autoref{fig:FBModels} (middle panel) illustrates the vertical profiles $n(z)\simeq 4(\mu_m m_p)^{-1}\rho_h(z)\theta_{\modj}^2/\Theta(z)^2$ of particle number density and $P(z)\simeq (3/4)\rho_h(z)v_{\modj}^2\theta_{\modj}^2/\Theta(z)^2$ of pressure, for the one-zone model, assuming a homogeneous isothermal expansion near the axial.

The above arguments suggest that the FBs may indicate a ballistic flow arising from an abrupt injection of energy and momentum, provided that the ratio $E_{\inj}/v_{\inj}^2\propto p_{\inj}^2/E_{\inj}$ is sufficiently high to satisfy (\ref{eq:BallisticRegime}), that the opening angle of the jet is of order $4^{\circ}$, and that the FB age is given by (\ref{eq:BallisticAge}).
This possibility is studied numerically in \autoref{ballisticregime}.

\subsubsection{Slowdown regime}
\label{Modelnonballisticregime}

Next, consider the same jetted setup, but now deep in the slowdown regime.
Here, the nearly planar shock of area $\sim \pi \theta_{\modj}^2 z_H^2$ (the exact value is immaterial for the present argument) at the head of the jet propagates at an increasingly slower velocity $v_H(t)\propto t^{-\tv}$ into the approximately $\rho_h\propto z^{-\ah}$ CGM.

We approximate the momentum of the shocked mass, $p_z\simeq \Myfp\pi\theta_{\modj}^2\rho_h(z_H)z_H^3v_M/(3-\ah)$, as some fraction $\Myfp$ of the product of the swept-up mass $M$ and the velocity $v_M\simeq (3/4)v_H$ inferred from the shock jump conditions just behind the shock, where all velocities are measured in the Galactic frame. Fresh mass accreted through the shock raises the momentum of the bubble at a rate $\dot{p}_z^+=\pi(\theta_{\modj} z_H)^2 \rho_h(z_H)v_H v_M$, so momentum conservation implies that $dp_z/dt+\dot{p}_z^+=0$.

The resulting ODE,
\begin{equation} \label{eq:HeadSlowDown}
\frac{\partial\ln[\rho_h(z_H)z_H^3 v_H]}{\partial \ln z_H}=-\MyfqInv \, ,
\end{equation}
where $\Myfq\equiv \Myfp/(3-\ah)$, leads to
$v_H\propto z_H^{-(\MyfqInv+3-\alpha)}$,
and hence to the late-time power-law indices
\begin{equation} \label{eq:zHvsT}
0<\tz \equiv \frac{d\ln z_H}{d\ln t}\simeq \frac{\Myfq}{1+(4-\ah)\Myfq} \, ,
\end{equation}
and
\begin{equation}
0<\tau_v \equiv -\frac{d\ln v_H}{d\ln t} = 1-\tz = \frac{1+(3-\alpha)\Myfq}{1+(4-\alpha)\Myfq} \fin
\end{equation}
For $\ah=3/2$, typical estimates $1/5\lesssim\Myfq\lesssim1/3$ yield a small $2/15\lesssim\tz\lesssim 2/11$ and a near-unity $9/11\lesssim\tv\lesssim 13/15$, consistent with the simulations shown below.

\begin{figure}
\DrawFigs{
\centering{
\includegraphics[width=8truecm]{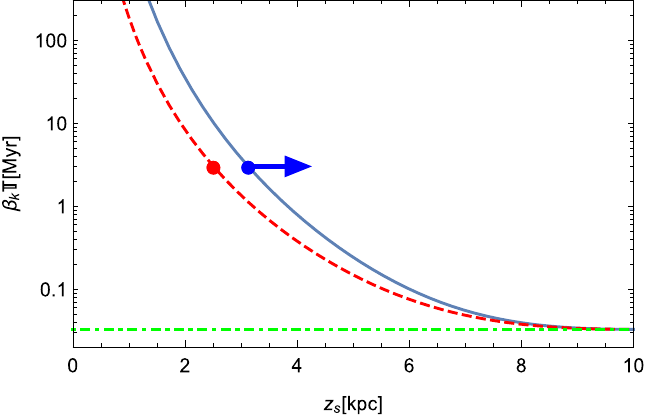}
}
}
\caption{
Bubble age $\tage$ as a function of the height $z_{\tr}$ of the transition from ballistic to slowdown phases, weighted by the normalized velocity $\beta_{\modj}\equiv z_{\tr}/ct_{\tr}<1$ of the jet before the transition, shown for $\Myfq=1/4$ (solid blue) and $\Myfq=1/3$ (dashed red) with a bubble height $z_H=10\kpc$.
A given bubble age thus imposes lower limits on $z_{\tr}$ ({\disks} and arrow) and on $\beta_{\modj}$ (dot-dashed), as illustrated for $\tage=3\Myr$.
\label{fig:FBTlim}}
\end{figure}

More precisely, twice integrating Eq.~(\ref{eq:HeadSlowDown}) yields
\begin{equation} \label{eq:NonBalZH}
z_H\left(t>t_{\tr}\right)\simeq \left[ 1+\left(\frac{t}{t_{\tr}}-1\right)\tzInv\right]^{\tz}
z_{\tr} \, ,
\end{equation}
where $z_{\tr}$ and $t_{\tr}$ are the height and time of the transition from ballistic to slowdown phases, derived toward the end of this subsection.
The implied age of the bubble,
\begin{equation}\label{eq:NonBallisticAge}
\tage\simeq \left[\tv+\left(\frac{z_H}{z_{\tr}}\right)^{1/\tz}\tz\right]\: t_{\tr} \coma
\end{equation}
is the sum of a linear, $t_{\tr}\propto z_{\tr}$ ballistic period and a non-linear, $\propto (z_H/z_{\tr})^{1/\tz}$ slowdown period.
As $\tz$ is a small number, a low transition height $z_{\tr}$ would imply that $\tage$ is dominated by the slowdown phase.
A given bubble age thus imposes a lower limit on $z_{\tr}$,
\begin{equation} \label{eq:NonBallisticAgeLimit}
\frac{z_{\tr}}{z_H}
> \left( \frac{z_H \tz}{c\beta_{\modj} \tage} \right)^{\tz/\tv}
> \left( \frac{z_H \tz}{c\tage} \right)^{\tz/\tv} \coma
\end{equation}
and slowdown at $z_{\tr}<z_H$ would require $\beta_{\modj}\tage >z_H/c$,
as illustrated in \autoref{fig:FBTlim}.
For example, for a $z_H=10\kpc$, $\tage=3\Myr$ bubble and $\Myfq=1/4$, the FBs can be in the slowdown phase only if $z_{\tr}\gtrsim 3.1\beta_{\modj}^{-2/11}\kpc>3.1\kpc$ (blue arrow in the figure); requiring slowdown implies that the average normalized velocity till $t_{\tr}$ is $\beta_{\modj}>z_H/c\tage\simeq 0.01$.

To estimate the shape of the bubble in the slowdown regime, assume for simplicity that the sideways expansion remains ballistic for some duration after $t_{\tr}$.
We solve the same sideways expansion ODE (\ref{eq:ShapeODE}) as in the ballistic case, but now incorporating the slowing down vertical velocity
\begin{equation}
v_H(z>z_{\tr})\simeq (z/z_{\tr})^{-(\MyfqInv+3-\alpha)}v_{\modj} \coma
\end{equation}
which corresponds to a Mach number
\begin{equation} \label{eq:slowdownMach}
\Upsilon\simeq \frac{v_{\modj}}{c_s} \simeq 14 (z/z_{\tr})^{-(\MyfqInv+3-\alpha)}\beta_{-2}\left(\frac{T_h}{2\times 10^6\mbox{ K}}\right)^{-1/2} \,
\end{equation}
at the top of the bubble.
The initial sideways-expansion condition (\ref{eq:R_IC}) of the ballistic regime now pertains only to $t<t_{\tr}$, and is supplemented by the slowdown-phase condition
\begin{equation}
R_b[z=z_H(t>t_{\tr}), t>t_{\tr}]\simeq\theta_{\modj} z \fin
\end{equation}
For $z<z_{\tr}$, the bubble geometry is still formally given by (\ref{eq:BallisticShape}),
whereas above the transition height the bubble narrows sharply with increasing $z$,
\begin{equation} \label{eq:ThetaNB}
\Theta(z>z_{\tr},t)^2
\simeq  3 \left\{\left[\frac{z_H(t)}{z}\right]^{1/\tz}-1\right\} \tz \theta_{\modj} + \theta_{\modj}^2 \, .
\end{equation}

The combined solution (\ref{eq:BallisticShape}) and (\ref{eq:ThetaNB}) is illustrated in \autoref{fig:FBModels} for $\Myfq=1/4$. For small $\theta_{\modj}$ and for $z$ not too close to $z_H$, we may approximate
\begin{equation}
R_b(z>z_{\tr},t)^2
\simeq  3\theta_{\modj}\tz z_H^{1/\tz} z^{-\tv/\tz} \, .
\end{equation}
Anticipating the widest part of the bubble at $z\simeq z_{\tr}$, Eq.~(\ref{eq:BallisticShape}) indicates that
\begin{equation} \label{eq:NonBallisticRmaxOft}
R_{\max}(t) \simeq \left( 3\theta_{\modj} v_{\modj} z_{\tr} t \right)^{\frac{1}{2}} \simeq 2.0 \left(\beta_{-2}\theta_5 t_1\frac{z_{\tr}}{5\kpc} \right)^{\frac{1}{2}}\kpc
\end{equation}
in the $t\equiv t_1\Myr\gg t_{\tr}$ regime.
As the bubble necessarily narrows from $z_{\tr}$ to $z_H$, the observed shape of the FBs indicates that $z_{\tr}\gtrsim z_H/2$.

In addition to the stratified flow approximation, the above solution to the shape of the bubble in the slowdown phase is also based on the simplifying assumptions of a sudden transition from ballistic to slowdown behaviors, and a fixed head opening-angle $\Theta[z_H(t),t]=\theta_{\modj}$ at all times. In practice, the transition is smooth, and the tip of the head may gradually expand.
Consequently, while the non-ballistic profile in \autoref{fig:FBModels} provides a reasonable approximation to the simulated slowing-down bubbles, the sharp break at $z_{\tr}$ is smoothed out in our simulations.
Therefore, instead of estimating the maximal width of the evolved bubble at $z=z_{\tr}$,
\begin{equation}\label{eq:NonBallisticRmax}
R_{\max}(t=\tage) \simeq (3\tz \theta_{\modj})^{1/2}\left(\frac{z_H}{z_{\tr}}\right)^{1/2\tz} z_s\coma
\end{equation}
we may for simplicity use $z\simeq z_H/2$, such that
\begin{equation}\label{eq:NonBallisticRmaxB}
R_{\max}(t=\tage) \simeq \left( \frac{3}{2}\theta_{\modj}v_{\modj}z_H \tage \right)^{1/2}  \fin
\end{equation}
As $v_{\modj}\tage>z_H$ in the slowdown regime, nominal parameters yield $\theta_\modj\lesssim 2^{\circ}$.

A jet generating FBs in the slowdown regime is faster and lower in energy and in mass than its ballistic counterpart, and thus more sensitive to the numerical setup.
Assuming that the setup loads the jet with an additional, spurious mass $M_0$ and slows it down to a velocity $v_{\modj}<v_{\inj}$, we equate the total injected mass with an uncertainty factor $\Myfm$ times the swept-up mass to estimate the slowdown at time $t_{\tr}\equiv z_{\tr}/v_{\modj}$ and height
\begin{equation} \label{eq:zTransition}
z_{\tr} \simeq  \left( \frac{M_0+2E_{\inj}/v_{\inj}^2}{2\pi\theta_{\modj}^2 C_\rho \Myfm} \right)^{\frac{1}{3-\ah}} \simeq 10 \frac{\left(E_{55}/4\Myfq C_4\right)^{2/3}}{\left(\beta_{-2}\theta_5\right)^{4/3}} \kpc \coma
\end{equation}
where $C_\rho\equiv z^\ah\rho_h(z)\equiv C_4(10\kpc)^\ah\rho(z=10\kpc)/(4\times 10^{-4}m_p\cm^{-3})$ is a constant,
and $E_{55}\equiv E_{\inj}/10^{55}\erg$.
The last expression in Eq.~(\ref{eq:zTransition}) pertains to $\ah=3/2$, approximates $\Myfm\simeq \Myfq$, and assumes a negligible $M_0$.
Equation (\ref{eq:zTransition}) is more accurate than the conditions (\ref{eq:BallisticRegime}) and (\ref{eq:NonBallisticRegime}) for determining the phase of the bubble.

\subsubsection{Ballistic or slowing-down FBs?}
\label{subsec:BallisticOrSlowDown}

In spite of the lowered dimensionality of the stratified toy models outlined in \autoref{Modelballisticregime} and \autoref{Modelnonballisticregime}, in which each layer of constant $z$ evolves independently, these models capture much of the bubble structure and scaling we find in the simulations described below in \autoref{ballisticregime} and \autoref{nonballisticregime}, respectively.

In the case of FBs in the slowdown phase, the numerical launching of the faster, less energetic jet is more delicate, and the outcome is more sensitive to details.
For simplicity, above we neglected the finite, $t_{\inj}\simeq 0.4\Myr$ injection stage of the jet, during which interactions with ambient gas over a substantial, $z_{\inj} \lesssim t_{\inj} v_{\inj} \simeq 1.2\: (t_{\inj}/0.04\Myr)(\beta_{\inj}/0.1)\kpc$ height can modify the jet parameters.
In particular, for a sufficiently high $v_{\inj}$, the mass $M_0$ accumulated during this stage can exceed the injected mass $M_{\inj}$, the jet can be slowed down by a factor $v_{\modj}/v_{\inj}$ of a few, and it can become broadened or pinched by a noticeable factor $\theta_{\modj}/\theta_{\inj}$.

For a fast injected jet, the strong initial shear may disrupt the jet, reshape it, focus it, or broaden it via partial thermalisation of the jet-base through viscosity or KHI.
Consider in particular the viscous radial acceleration of gas around the jet, boosting the initial $v_R=0$ to $v_R(t)\simeq (t v_{\inj} \mu)/(\rho \Delta R \Delta z)$, where $\Delta R\simeq (t\mu/\rho)^{1/2}$ is the width of the shear layer and $\Delta z$ is the typical scale for flow variations in the $z$ direction. Then the effective opening angle becomes
\begin{equation} \label{eq:thetaEff}
\theta_{\eff}\simeq v_R(t)/v_{\inj} \simeq \left( t \mu/\rho \right)^{1/2}/\Delta z \propto \mu^{1/2} \coma
\end{equation}
with some dependence on $t_{\inj}$ arising from $t$ and possibly also from $\rho$ and $\Delta z$, depending on details.

We conclude that the FBs are either presently in the ballistic stage, or were at the ballistic stage as recently as when the bubbles were at a height $z_{\tr}\gtrsim 5\kpc$, based on four different observations: (i) the bubbles are not very old, so Eq.~(\ref{eq:NonBallisticAgeLimit}) constrains $z_{\tr}$; (ii) the Mach number is high, so Eq.~(\ref{eq:slowdownMach}) similarly constrains $z_{\tr}$; (iii) the bubbles are fairly cylindrical, as discussed below Eq.~(\ref{eq:NonBallisticRmaxOft}); and (iv) the initial opening angle cannot be too small, so $z_{\tr}$ is constrained by Eq.~(\ref{eq:NonBallisticRmax}),
\begin{equation}
\frac{z_{\tr}}{z_H} \simeq \left( \frac{R_{\max}/z_H}{\sqrt{3\theta_{\modj}\tz} } \right)^{\frac{2\tz}{2\tz-1}}
 \simeq 0.9 \left(\frac{4R_{\max}}{z_H}\right)^{-\frac{4}{9}} \theta_5^{2/9} \coma
\end{equation}
where the last estimate pertains to $\Myfq=1/4$.
Due to this weak dependence upon $\theta_{\modj}$, a small, say $z_{\tr}<1\kpc$ would require an early jet opening angle smaller than an arcsecond.
Adopting $z_{\tr}\gtrsim 5\kpc$, Eq.~(\ref{eq:zTransition}) implies that
\begin{equation}
\frac{E_{55}}{(\beta_{-2}\theta_5)^2} \gtrsim 0.4(4\Myfq)C_4 \fin
\end{equation}

In both ballistic and slowdown scenarios, a high Mach number $\Upsilon$ at the top of the bubble requires $\beta_{-2}\gtrsim 0.4\Upsilon_5$, where $\Upsilon_5\equiv \Upsilon/5$, and the thickness of the bubbles requires $\theta_{\modj}\lesssim 4^{\circ}$. These results become approximate equalities in the ballistic case, whereas an extended slow-down phase requires a much faster and narrower jetted injection.
The FBs are in the ballistic stage if they are energetic or slow, $E_{55}\gtrsim 3(\beta_{-2}\theta_5)^2$, in which case $E_{55}\gtrsim 2\beta_{-2}^2\gtrsim 0.2\Upsilon_5^2$ and $\tage\simeq 3.3\beta_{-2}^{-1}\Myr\lesssim 8\Upsilon_5^{-1}\Myr$.
The FBs are in the slowdown phase if $E_{55}\lesssim 3(\beta_{-2}\theta_5)^2$, in which case $0.05\Upsilon_5^2\theta_5^2\lesssim E_{55}\lesssim 2\beta_{-2}^2$ and $\tage\simeq 1.4(\beta_{-2}\theta_5)^{-1}\Myr\lesssim 4(\Upsilon_5\theta_5)^{-1}\Myr$.
Such slowing-down FBs can still be energetic, provided that $\beta_{-2}$ is large, but their age would then be of order a Myr only if $\theta_5$ is small.

\subsection{Jetted simulations}
\label{subsec:JettedSims}

\begin{figure}
\DrawFigs{
\centering{
\hspace{-0.7cm}
\begin{tikzpicture}
\draw (0, 0) node[inner sep=0] {\raisebox{0.15cm}{\includegraphics[height=4.3truecm,trim={4.8cm 2.2cm 4.0cm 2.8cm}, clip]{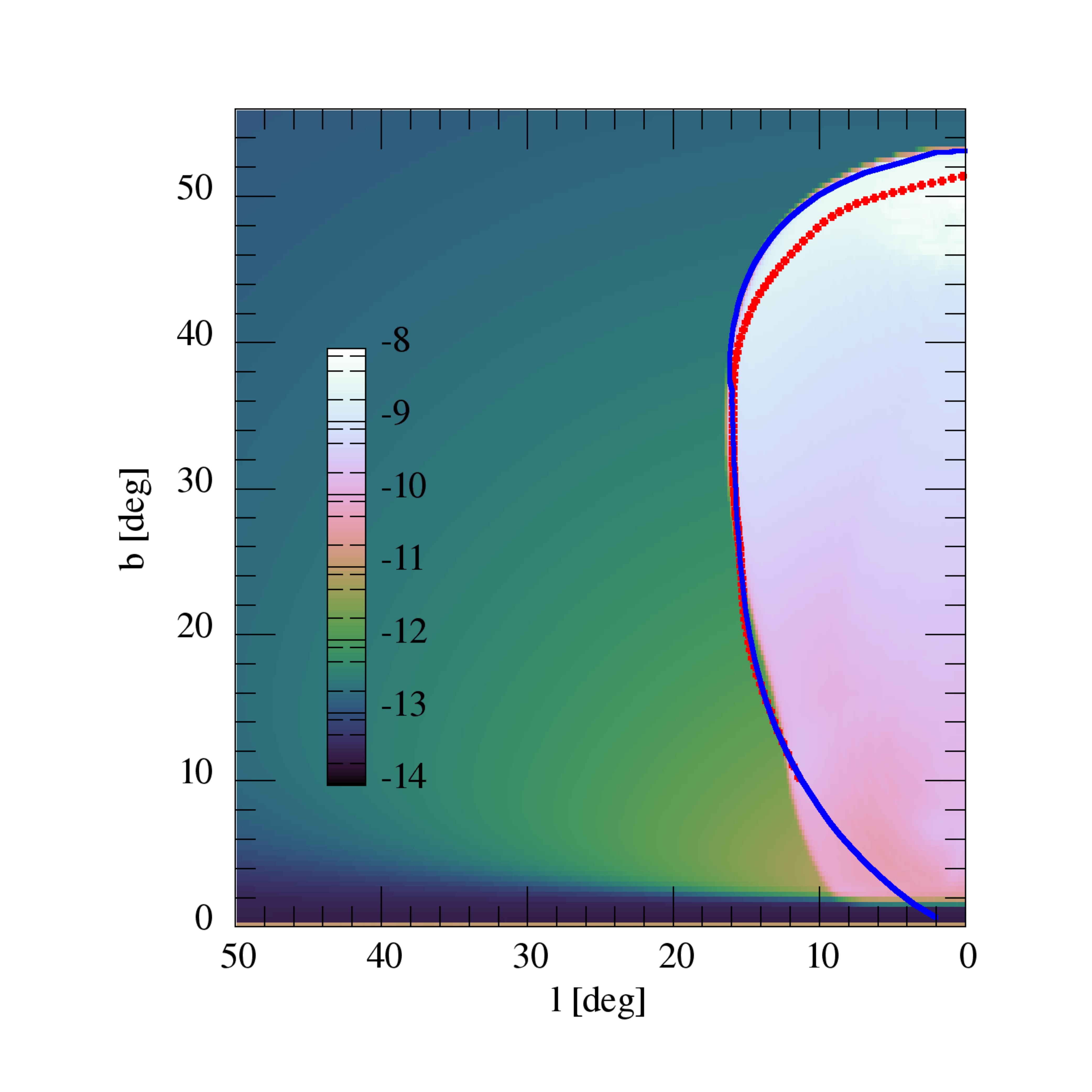}}};
\draw (-0.7, 1.8) node[text=white] {\scriptsize $t=3.3\Myr$};
\end{tikzpicture}
\hspace{-0.1cm}
\includegraphics[height=4.3truecm,trim={2.1cm 2.5cm 1.5cm 5.2cm}, clip]{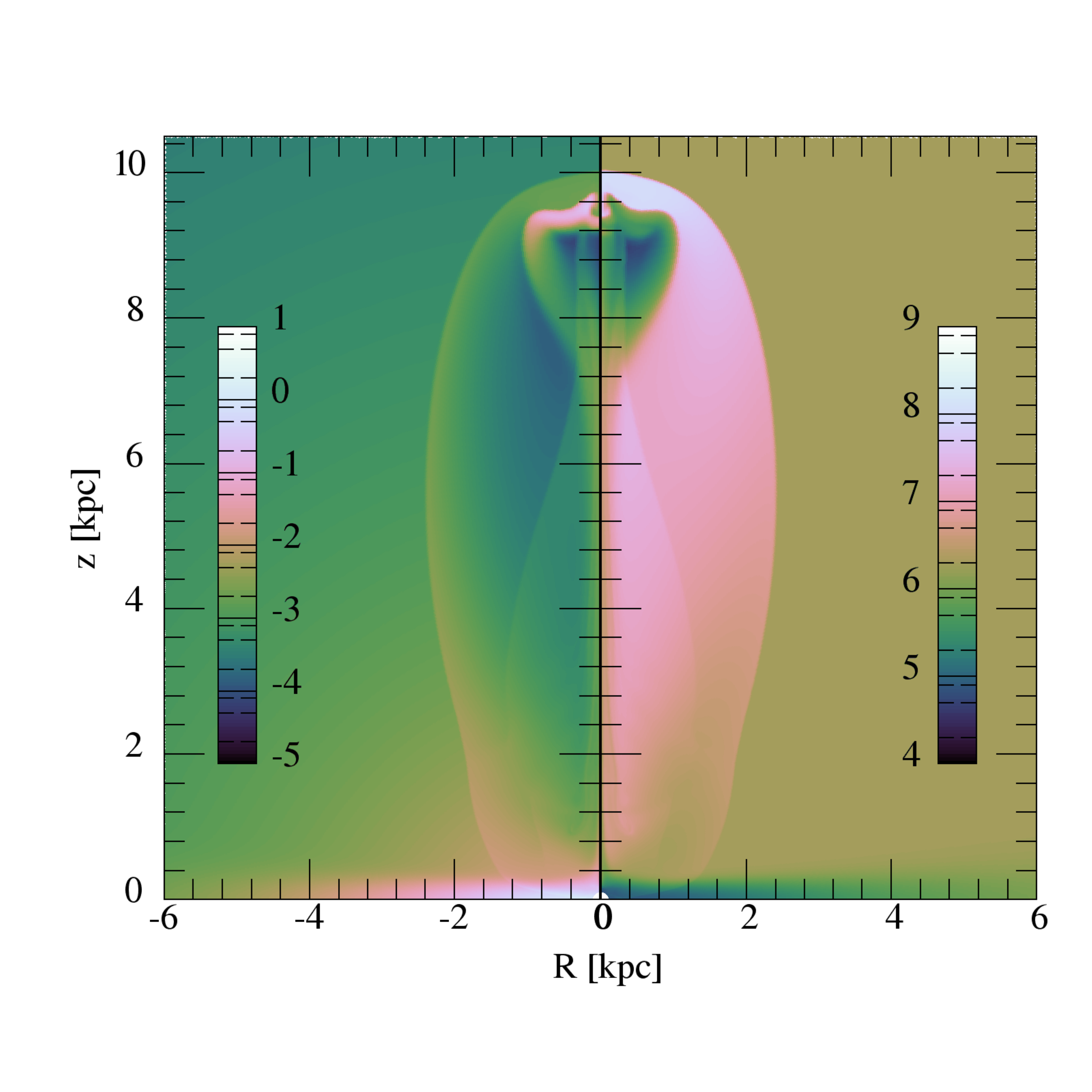} \\
\hspace{-0.8cm}
\begin{tikzpicture}
\draw (0, 0) node[inner sep=0] {\raisebox{0.15cm}{\includegraphics[height=4.5truecm,trim={4.8cm 2.2cm 4.0cm 2.8cm}, clip]{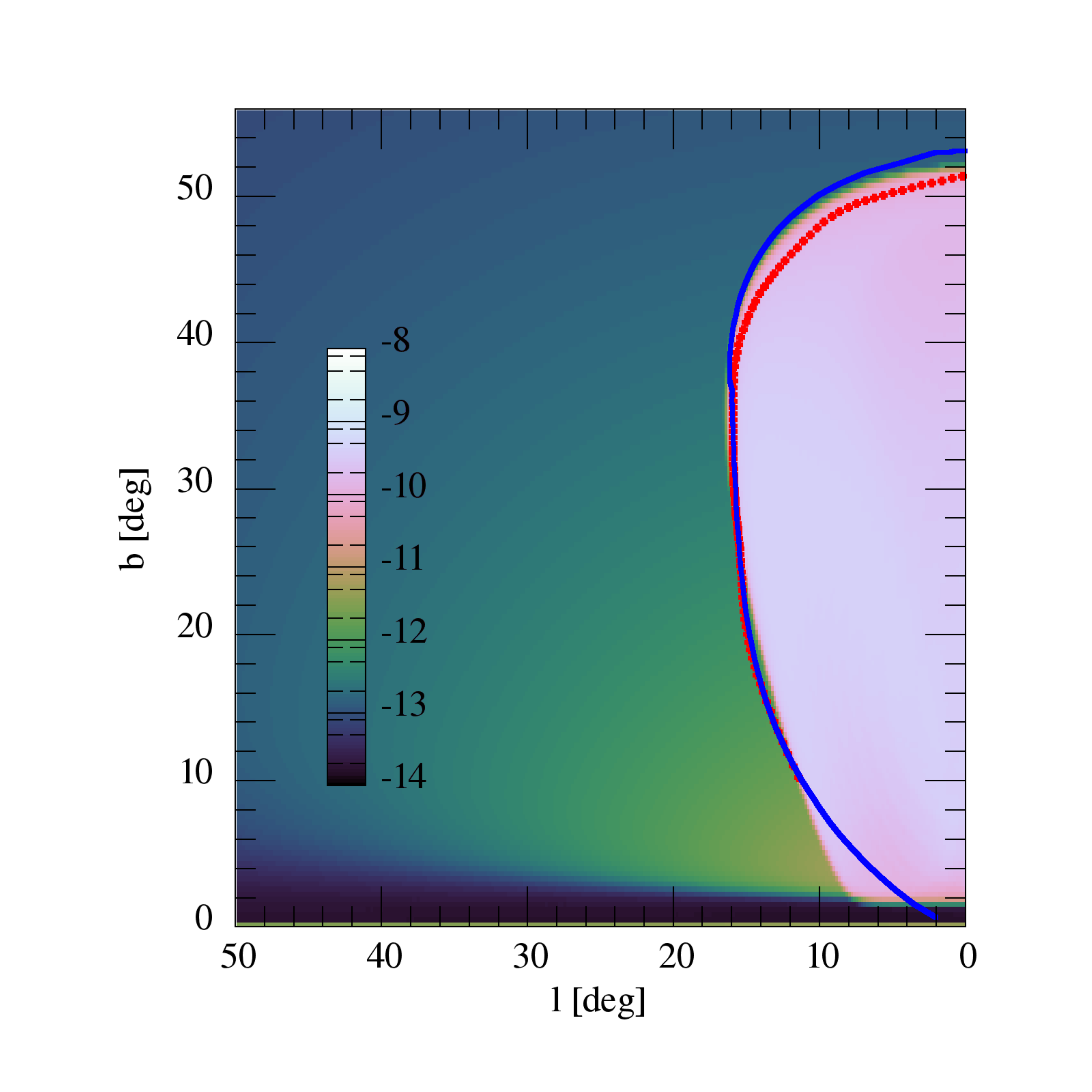}}};
\draw (-0.7, 1.8) node[text=white] {\scriptsize $t=2.4\Myr$};
\end{tikzpicture}
\hspace{-0.3cm}
\includegraphics[height=4.5truecm,trim={2.1cm 2.5cm 1.5cm 5.2cm}, clip]{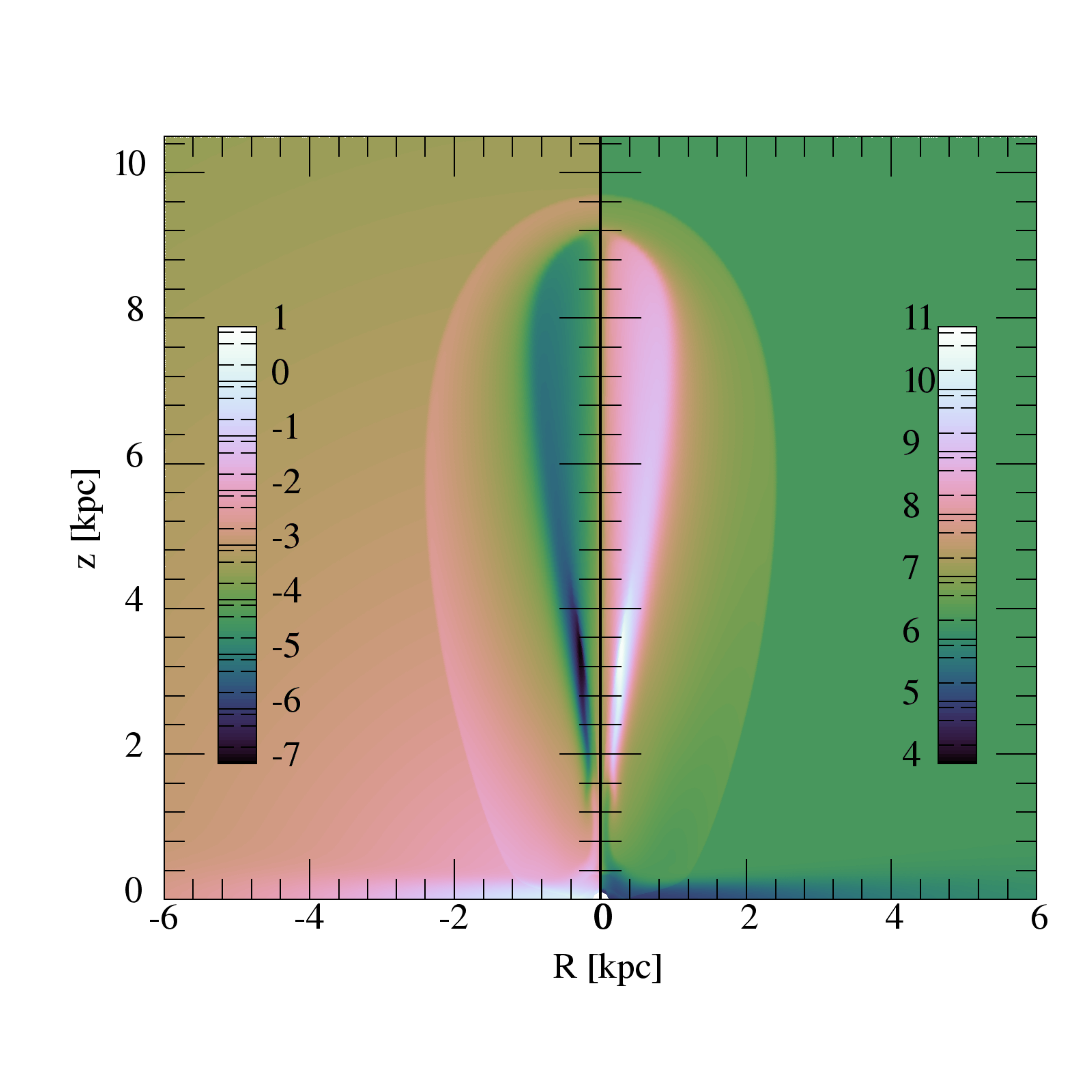}
}
}
\caption{
Simulated FBs generated by jetted injection, in their ballistic (top; setup J1)
and slowdown (bottom; setup J2)
phases.
The simulation parameters are provided in \autoref{Tab:Jetted}. Notations are the same as in \autoref{fig:base_GC_IL}.
The slowdown bubble is somewhat younger as its ballistic stage was very short due to the much faster injection.
\label{fig:NominalJetted}
}
\end{figure}

We carry out a suite of simulations that generate \emph{Fermi}-like bubbles in both ballistic and slowdown phases.
The parameters and resulting bubble properties for a sample of such simulations are listed in \autoref{Tab:Jetted}.
\autoref{fig:NominalJetted} shows our nominal ballistic (setup J1; top row) and slowdown (J2; bottom panel) simulated bubbles.
Unlike non-directed injection, jetted injection gives rise to forward shocks that do agree well with the observed FB edges, as inferred from the stratified models of \S\ref{sec:ToyModels} and confirmed by our simulations.
As the figure demonstrates, this agreement holds in both ballistic and slowdown regimes, albeit with different parameters, as seen in the table.

The figure indicates that the inner structure of the ballistic and slowing-down bubbles differ substantially, the former showing an inner cylindrical shock surrounded at its top by an irregular contact discontinuity surface, and the latter showing a bubble-like contact discontinuity trailing the shock.
The two types of bubbles differ qualitatively also in their evolution and parametric dependence, as outlined in \S\ref{sec:ToyModels} and explored numerically below.

\subsubsection{Ballistic bubbles}
\label{ballisticregime}

\begin{figure*}
\DrawFigs{
\centering{
\includegraphics[height=4.7truecm,trim={2.1cm 2.5cm 1.5cm 5.2cm}, clip]{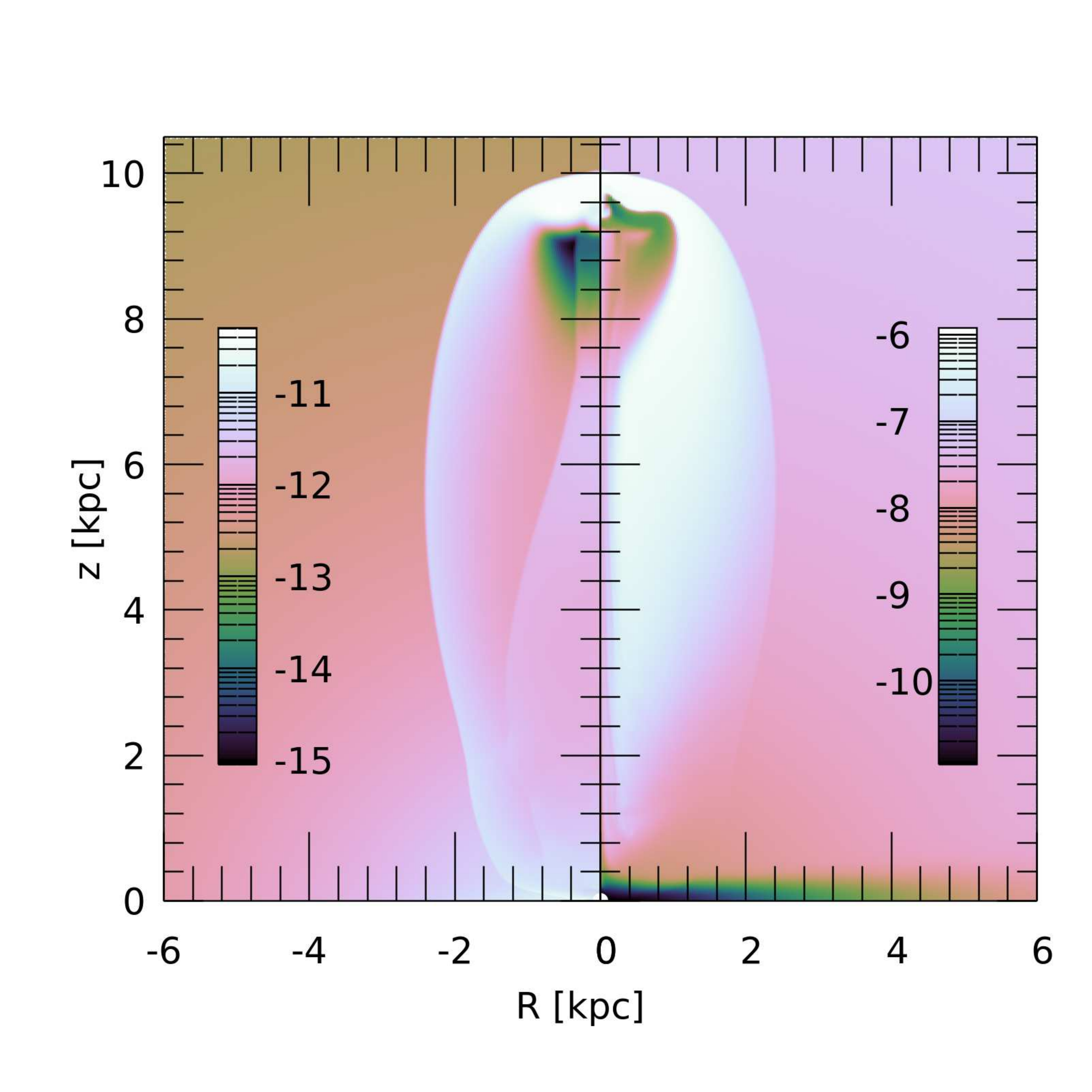}
\includegraphics[height=4.6truecm]{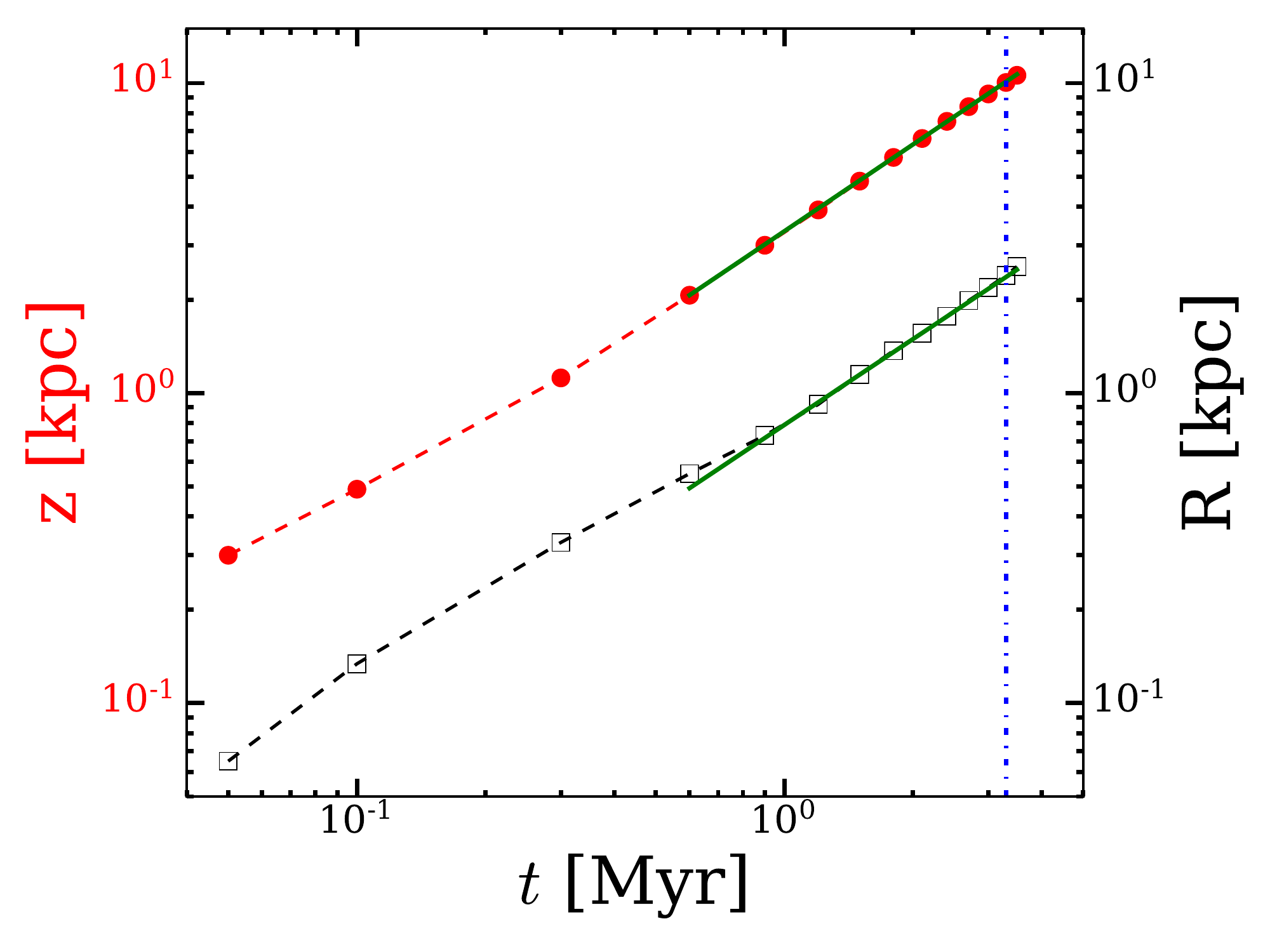}
\includegraphics[height=4.6truecm]{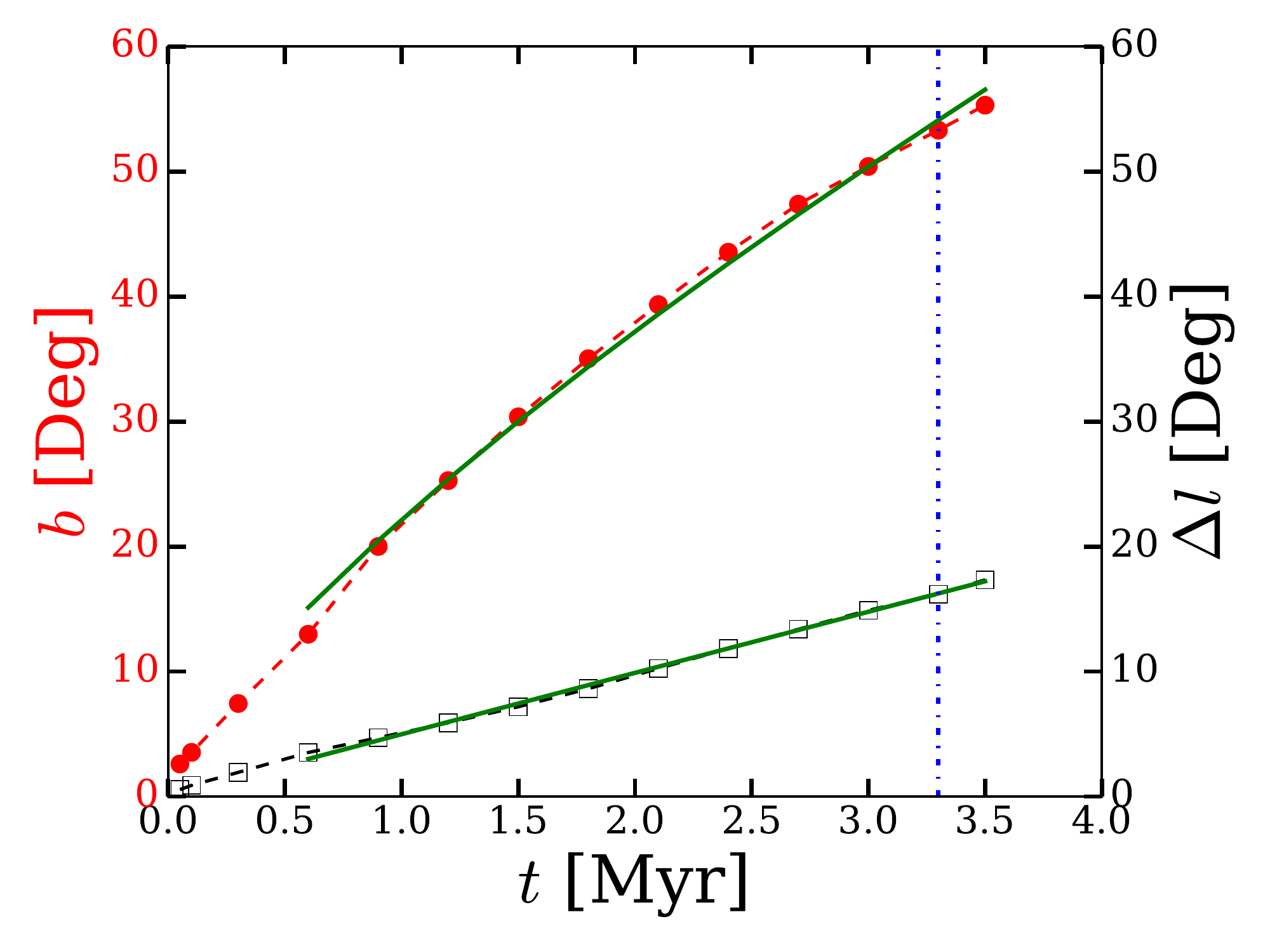}
}
}
\caption{
Thermal structure (left panel) and spatial (middle) and projected (right) evolution of our fiducial ballistic bubble (setup J1).
Left panel: logarithmic pressure (in erg$\cm^{-3}$; left half of left panel) and entropy (shown as the adiabat $Pn^{-\gamma}$ in erg$\cm^{2}$ units; right half)
distributions highlight the inner shock and contact discontinuity.
Middle and right panels: the temporal evolution of the height (red {\disks}
with left axis) and half-width (black squares; right axis) of the bubble is shown, along with power-law fits (solid green for $t>0.6\Myr$; see text) and the cutoff time (dot-dashed blue) at the designated latitude.
\label{fig:ballisticCombo}}
\end{figure*}

In our fiducial ballistic simulation, denoted J1, an energy $3\times 10^{56}\erg$ is injected with
$\beta_{\inj}=0.01$ inside an opening angle $\theta_{\inj}=5^\circ$ for a duration $t_{\inj} = 0.04\Myr$, resulting in a \emph{Fermi}-like bubble reaching the designated latitude at $\tage\simeq 3.3\Myr$, as depicted in \autoref{fig:NominalJetted} (top row).
The contact discontinuity (seen towards the top of the bubble) and inner shock are also evident in the pressure and entropy distributions presented in \autoref{fig:ballisticCombo} (left panel).
These distributions indicate a non-homogeneous inner structure, with a very low pressure region
behind the bubble head.
However, the pressure just behind the shock and throughout the bubble is fairly uniform, varying by only a factor of a few throughout the surface of the bubble.

The low pressure region can be identified as the unperturbed jet material that has been pinched by the high pressure region behind the ballistically expanding shock. In a simple picture, this region should be conical in shape, with a head half-width $R_{lp} \simeq z_H\: \theta_j$ and a length $\Delta z_{lp}\simeq v_j R_{lp}/c_s$ given by the $R_{lp}$ sound crossing time, where $c_s \simeq (3\gamma/16)^{1/2}v_j$ is the sound speed behind the shock. For $v_j = 0.01\:c$ and $\theta_j = 5^\circ$, $\Delta z_{lp}\simeq 1.5$ kpc is consistent with the simulated region in Figure \ref{fig:NominalJetted}.

The figure also shows the temporal evolution of the bubble, in space (middle panel) and in projection (right).
After the initial $\sim 0.6\Myr$, the height and maximal half-width of the bubble are well fit (green curves) by $z_H \sim 3.33t_1^{0.928\pm0.004}$~kpc
and $R_{\max} \sim 0.79t_1^{0.92\pm0.02}$~kpc, where $t_1\equiv t/\mbox{Myr}$.
This evolution is consistent with the expected linear behavior (\ref{eq:BallisticHeight}) and (\ref{eq:BallisticRmax}), with $v_{\modj}\simeq v_{\inj}$ and $\theta_{\modj}\simeq \theta_{\inj}$ almost unaffected by the initial transients, and with only a $\sim8\%$ slowdown during $0<t<\tage$.
The projected coordinates are well fit at this stage by $b\simeq 22^\circ.2\:t_1^{0.75\pm0.02}$ and $\Delta l\sim 5^\circ\: t_1^{0.99\pm0.02}$, in agreement with the model (see Eq \ref{eq:BallisticWidths}).

\begin{table*}
\scriptsize
\centering
\caption{\label{Tab:Jetted}
Parameters varied in the select jetted injection simulations presented in \autoref{fig:NominalJetted} (top panel) and \autoref{fig:ballistic_setups}. Many additional simulations (not shown) with varying parameters are used to approximate parameter dependencies, as illustrated in Figures \ref{fig:ballistic_plots} and \ref{fig:nonballistic_plots}.
}
\scalebox{1.2}{
\begin{tabular}{|l l|p{0.4cm} p{0.4cm}p{0.4cm}p{0.4cm}p{0.4cm} |p{0.4cm}p{0.4cm}p{0.4cm}p{0.4cm}p{0.4cm}|}
\hline
\textbf{Parameter} & \textbf{Definition}   & \multicolumn{5}{c|}{\textbf{Ballistic bubbles}}  & \multicolumn{5}{c|}{\textbf{Slowing-down bubbles}}  \\
&  & J1    & J1a  & J1b & J1c   &J1d     & J2    & J2a   & J2b   & J2c  &J2d    \\
\hline
$M_d$ [$10^{10}M_\odot$] & {\Disk} mass & \multicolumn{5}{c|}{\multirow{1}{*}{6.0}}  & {\multirow{1}{*}{6.0}} & {\multirow{1}{*}{6.0}} & {\multirow{1}{*}{6.0}} & {\multirow{1}{*}{6.0}} & {\multirow{1}{*}{12.0}}\\
$M_b$ [$10^{10}M_\odot$] & Bulge mass & \multicolumn{5}{c|}{\multirow{1}{*}{2.0}}  & \multicolumn{5}{c|}{\multirow{1}{*}{2.0}}\\
$f_h$  &  CGM rotation    & \multicolumn{5}{c|}{\multirow{1}{*}{0}}  & \multicolumn{5}{c|}{\multirow{1}{*}{0}}\\
\hline
$E_{\inj}$[$10^{56}$erg] & Injected energy & \multirow{1}{*}{3.0}  & \multirow{1}{*}{2.0}  & \multirow{1}{*}{3.0} & \multirow{1}{*}{3.0} & \multirow{1}{*}{3.0}
& \multirow{1}{*}{0.25} & \multirow{1}{*}{0.3}  & \multirow{1}{*}{0.3} & \multirow{1}{*}{0.3}& \multirow{1}{*}{0.25}\\
$\beta_{\inj}=v_{\inj}/c$ & Injected velocity & \multirow{1}{*}{0.01}  & \multirow{1}{*}{0.01}  & \multirow{1}{*}{0.015} & \multirow{1}{*}{0.01} & \multirow{1}{*}{0.01}
& \multirow{1}{*}{0.1} & \multirow{1}{*}{0.1}  & \multirow{1}{*}{0.12} & \multirow{1}{*}{0.1} & \multirow{1}{*}{0.1}\\
$t_{\inj}$ [Myr] & Injection duration   & \multirow{1}{*}{0.04} & \multirow{1}{*}{0.04} & \multirow{1}{*}{0.04} & \multirow{1}{*}{0.04} & \multirow{1}{*}{0.01}
& \multirow{1}{*}{0.04} & \multirow{1}{*}{0.04}  & \multirow{1}{*}{0.04} & \multirow{1}{*}{0.06} & \multirow{1}{*}{0.04}\\
$\theta_j$ & Injection opening angle  &
\multirow{1}{*}{$5^\circ$}  & \multirow{1}{*}{$5^\circ$}  & \multirow{1}{*}{$5^\circ$} & \multirow{1}{*}{$3^\circ$} & \multirow{1}{*}{$5^\circ$}
& \multirow{1}{*}{$4^\circ$} & \multirow{1}{*}{$4^\circ$} & \multirow{1}{*}{$4^\circ$}& \multirow{1}{*}{$4^\circ$} & \multirow{1}{*}{$4^\circ$} \\
$r_0$ [kpc] & Inner boundary &
\multicolumn{5}{c|}{\multirow{1}{*}{0.1}} & \multicolumn{5}{c|}{\multirow{1}{*}{0.1}} \\
$\mu_{\max}$ [g$\se^{-1}\cm^{-1}$] & Viscosity (when included) limit &
--- & --- & --- & --- & --- &      10 & 10 &10 &10 &10 \\
\hline
$\tage$ [Myr] & Bubble age (approximate) & \multirow{1}{*}{3.3} &\multirow{1}{*}{3.3} &\multirow{1}{*}{2.2} &\multirow{1}{*}{3.3}  &\multirow{1}{*}{3.2}
&\multirow{1}{*}{2.4} &\multirow{1}{*}{1.5} &\multirow{1}{*}{2.5} &\multirow{1}{*}{2.0}   &\multirow{1}{*}{1.0}\\
$\Delta l$  & Bubble half width (approximate) &\multirow{1}{*}{$16^\circ$}   &\multirow{1}{*}{$16^\circ$} &\multirow{1}{*}{$16^\circ$} &\multirow{1}{*}{$13^\circ$} &\multirow{1}{*}{$16^\circ$}
&\multirow{1}{*}{$16^\circ$} &\multirow{1}{*}{$12^\circ$}  &\multirow{1}{*}{$17^\circ$} &\multirow{1}{*}{$15^\circ$} &\multirow{1}{*}{$14^\circ$}\\
\hline
\end{tabular}}
\end{table*}

The
ballistic bubbles and their internal structures are not much affected by viscosity, as the associated jet velocities ($\beta_j \sim 0.01$) are relatively low. We have confirmed this behavior with several viscous simulations, of Spitzer viscosity capped at $\mu_{\max} = 10$~g cm$^{-1}$ s$^{-1}$ (as discussed in sec \ref{sec:Method}) and even with an excessive, fixed $\mu = 40$~g cm$^{-1}$ s$^{-1}$ throughout the simulation box.

\begin{figure}
\DrawFigs{
\vspace{-0.3cm}
\centering{
\hspace{-0.7cm}
\begin{tikzpicture}
\draw (0, 0) node[inner sep=0] {\raisebox{0.15cm}{\includegraphics[height=4.5truecm,trim={4.8cm 2.2cm 4.0cm 2.8cm}, clip]{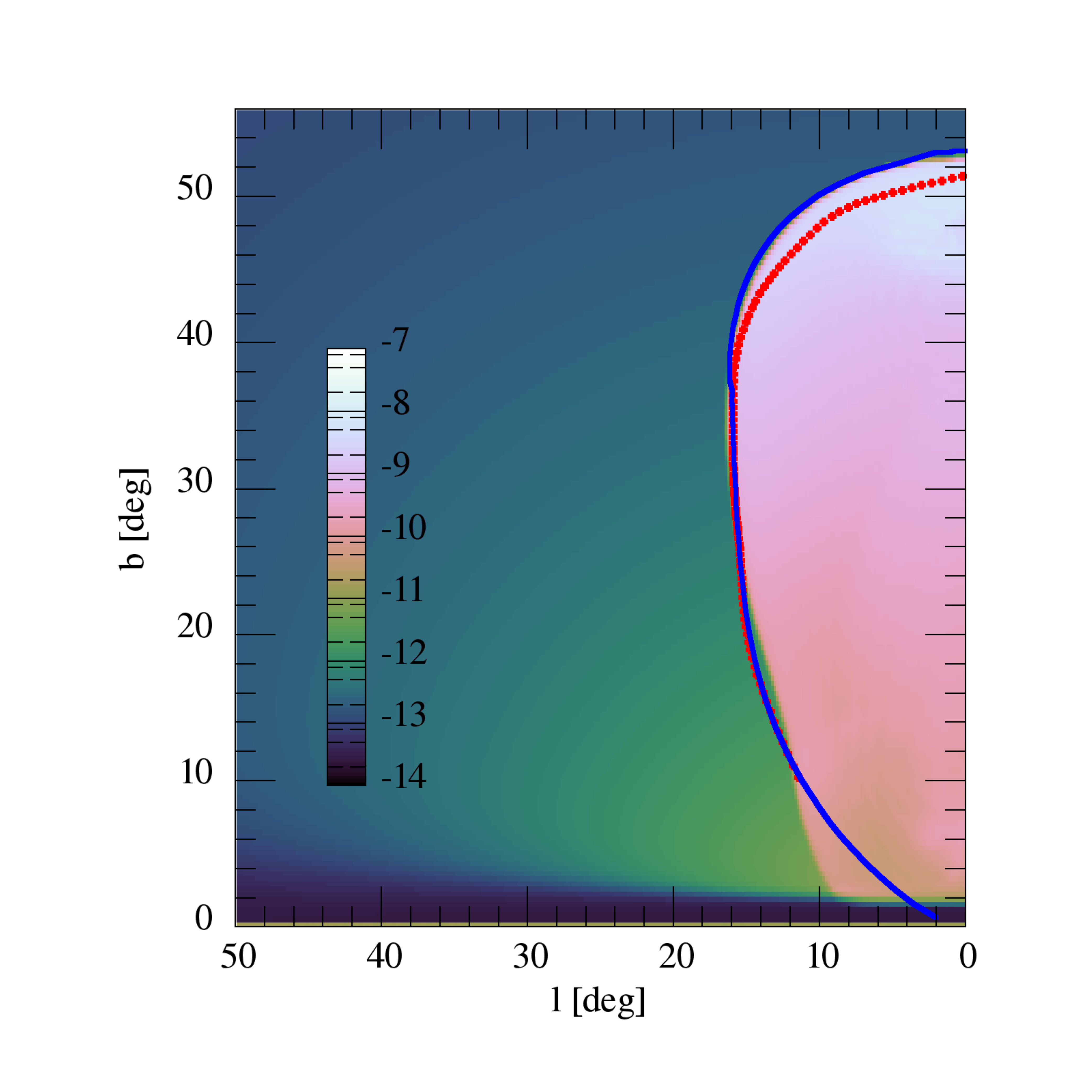}}};
\draw (-0.7, 1.8) node[text=white] {\scriptsize $t=3.3\Myr$};
\end{tikzpicture}
\hspace{-0.2cm}
\includegraphics[height=4.5truecm,trim={2.1cm 2.5cm 1.5cm 5.2cm}, clip]{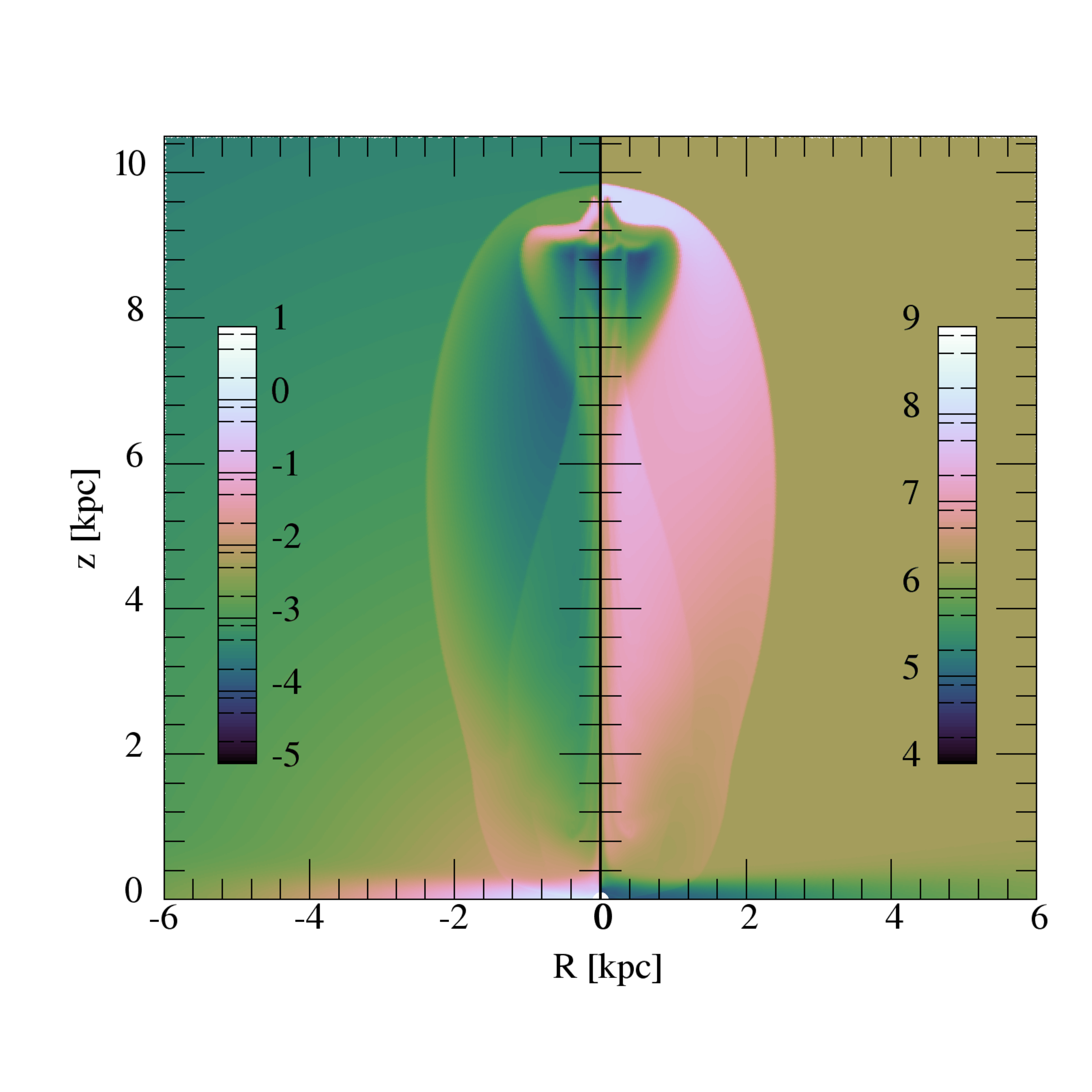}
\hspace{-0.5cm}
}
\\
\vspace{-0.68cm}
\centering{
\hspace{-0.7cm}
\begin{tikzpicture}
\draw (0, 0) node[inner sep=0] {\raisebox{0.15cm}{\includegraphics[height=4.5truecm,trim={4.8cm 2.2cm 4.0cm 2.8cm}, clip]{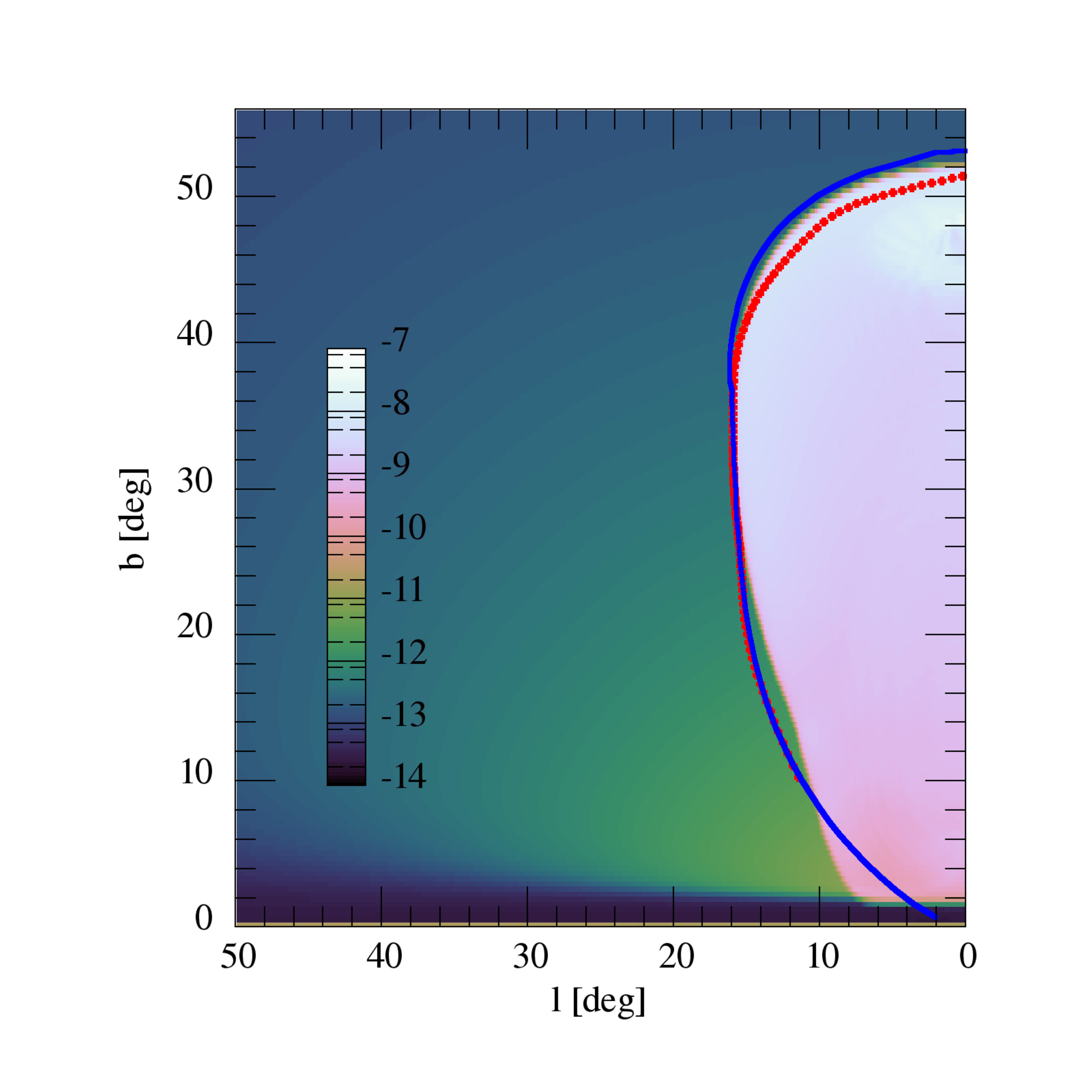}}};
\draw (-0.7, 1.8) node[text=white] {\scriptsize $t=2.2\Myr$};
\end{tikzpicture}
\hspace{-0.2cm}
\includegraphics[height=4.5truecm,trim={2.1cm 2.5cm 1.5cm 5.2cm}, clip]{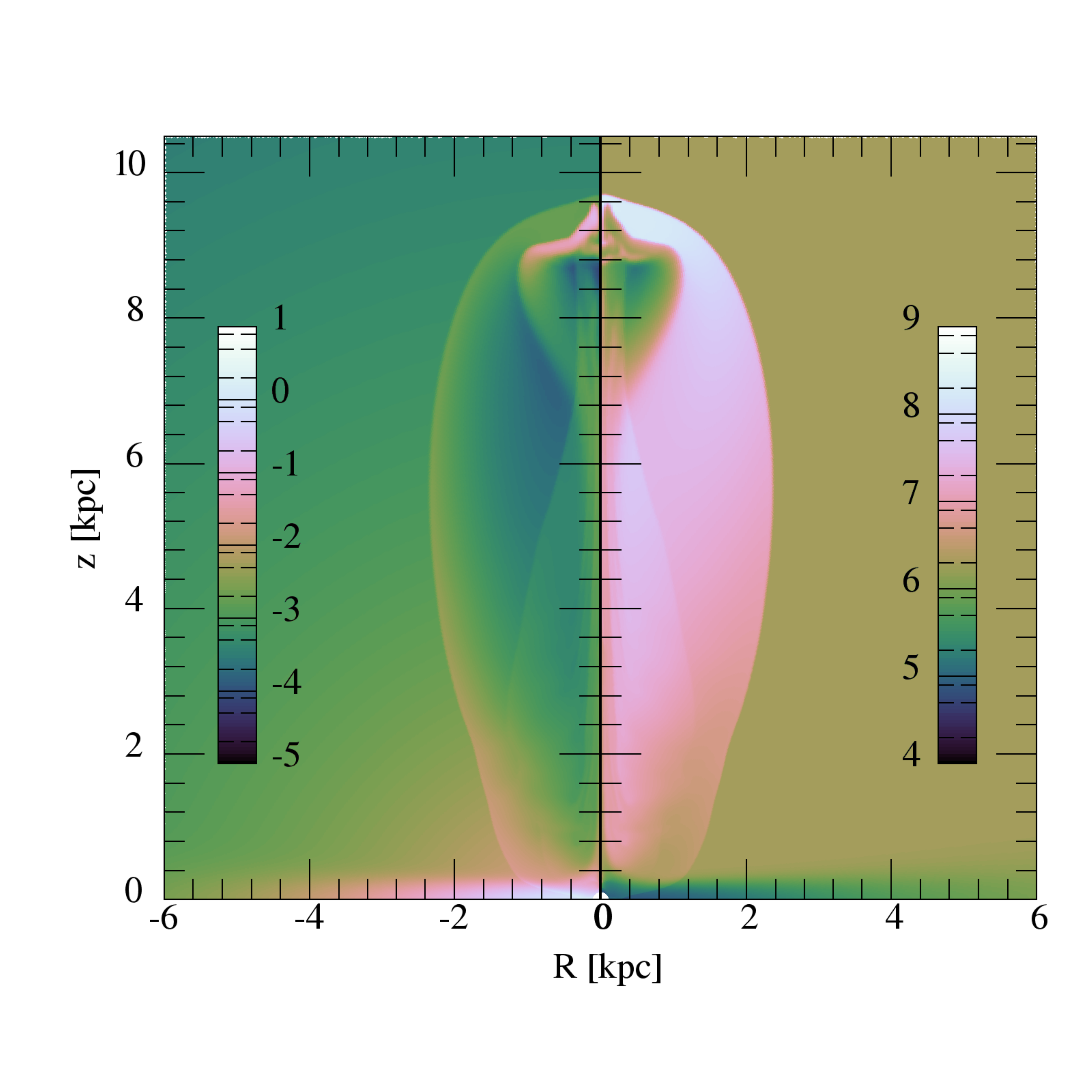}
\hspace{-0.5cm}
}
\\
\vspace{-0.75cm}
\centering{
\hspace{-0.7cm}
\begin{tikzpicture}
\draw (0, 0) node[inner sep=0] {\raisebox{0.15cm}{\includegraphics[height=4.4truecm,trim={4.8cm 2.2cm 4.0cm 2.8cm}, clip]{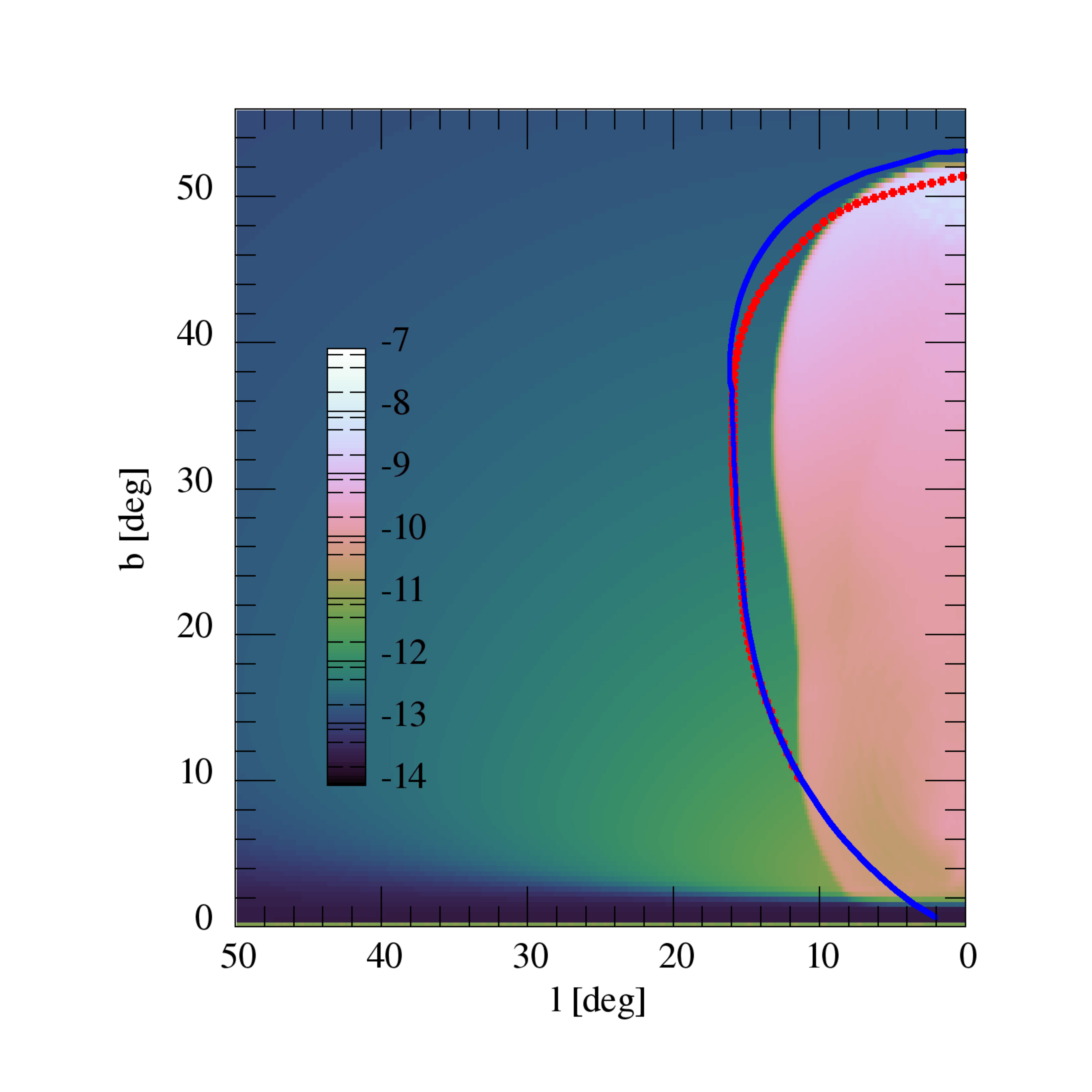}}};
\draw (-0.7, 1.8) node[text=white] {\scriptsize $t=3.3\Myr$};
\end{tikzpicture}
\hspace{-0.15cm}
\includegraphics[height=4.3truecm,trim={2.1cm 2.5cm 1.5cm 5.2cm}, clip]{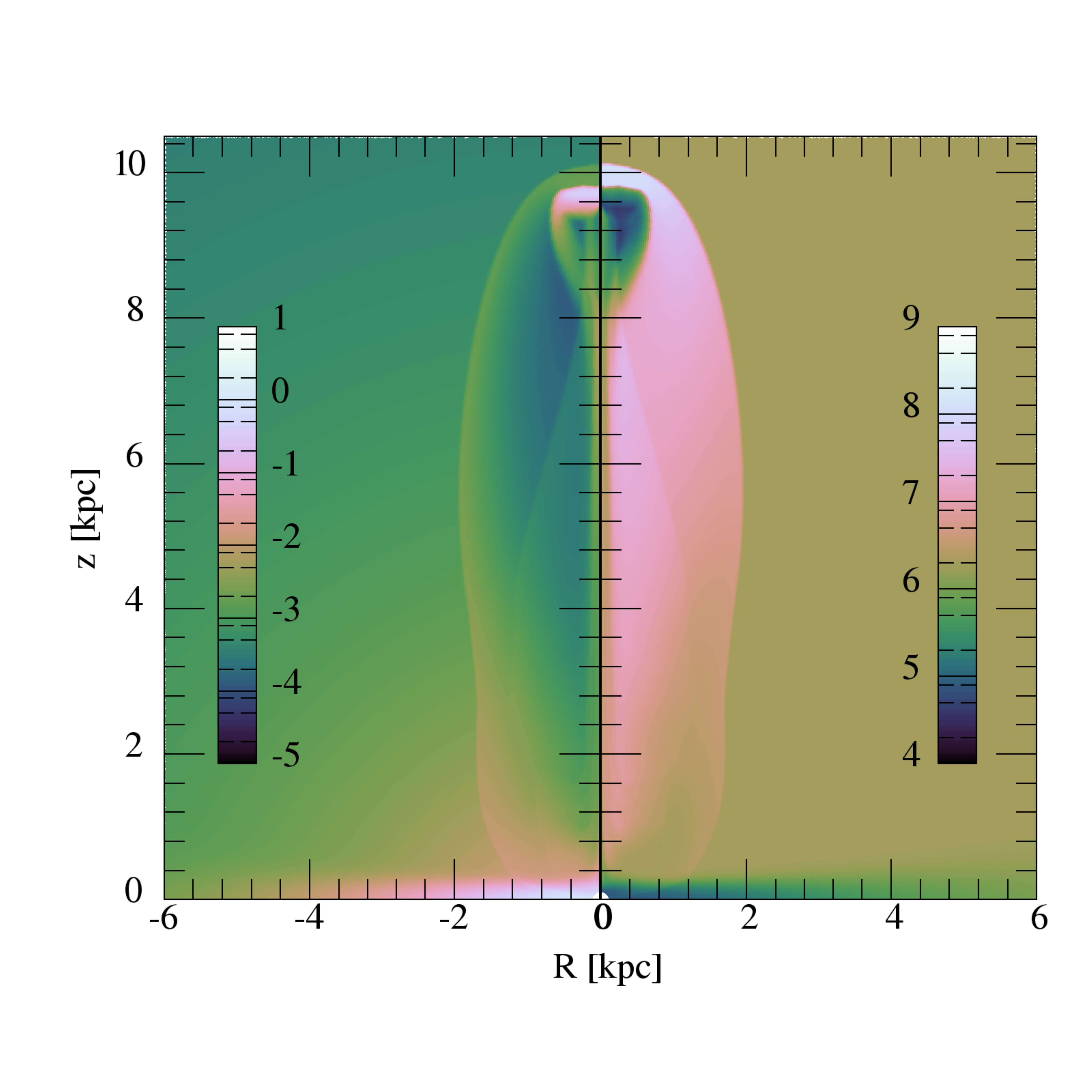}
\hspace{-0.5cm}
}
\\
\vspace{-0.69cm}
\centering{
\hspace{-0.55cm}
\begin{tikzpicture}
\draw (0, 0) node[inner sep=0] {\raisebox{0.15cm}{\includegraphics[height=4.5truecm,trim={4.8cm 2.2cm 4.0cm 2.8cm}, clip]{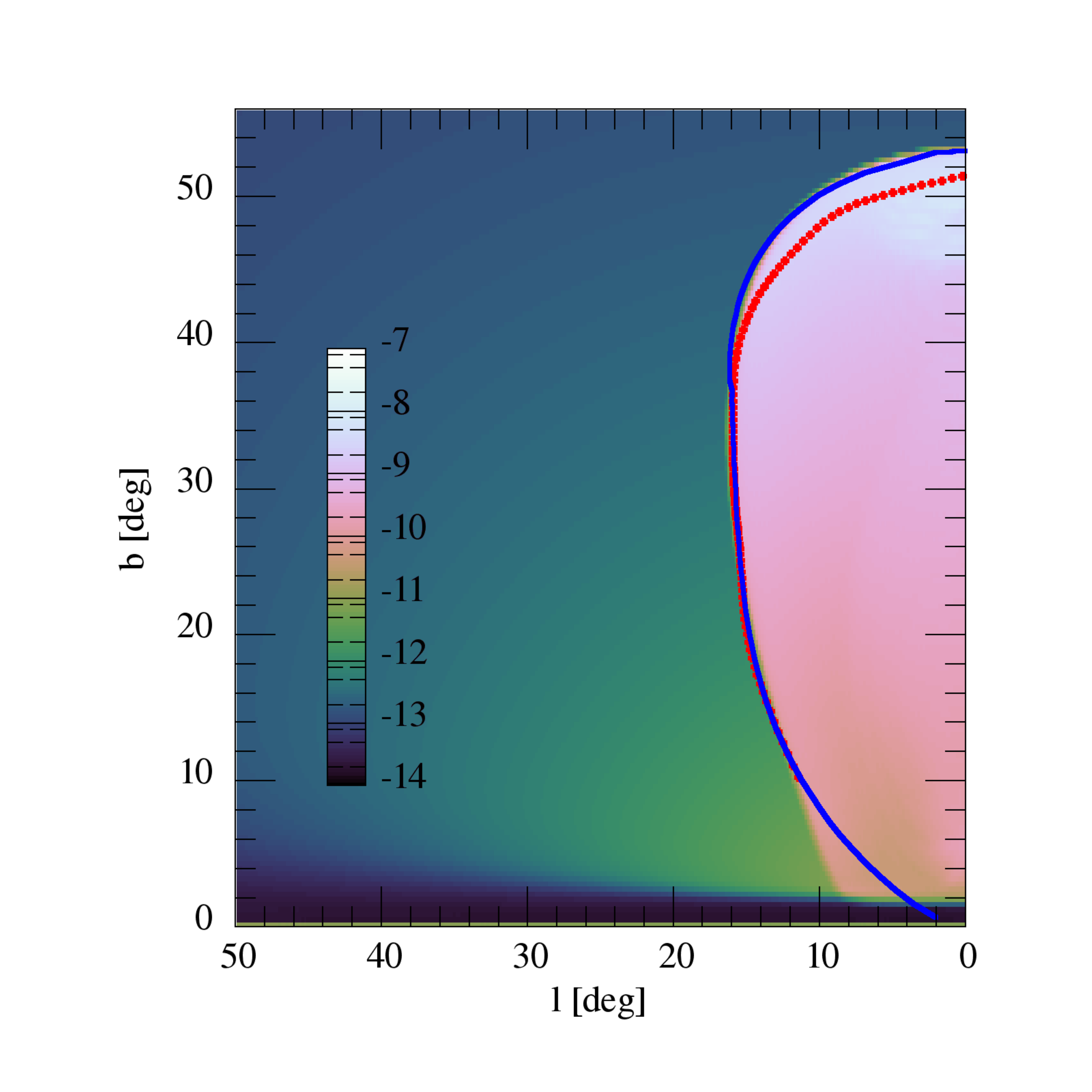}}};
\draw (-0.7, 1.8) node[text=white] {\scriptsize $t=3.3\Myr$};
\end{tikzpicture}
\hspace{-0.3cm}
\includegraphics[height=4.5truecm,trim={2.1cm 2.5cm 1.5cm 5.2cm}, clip]{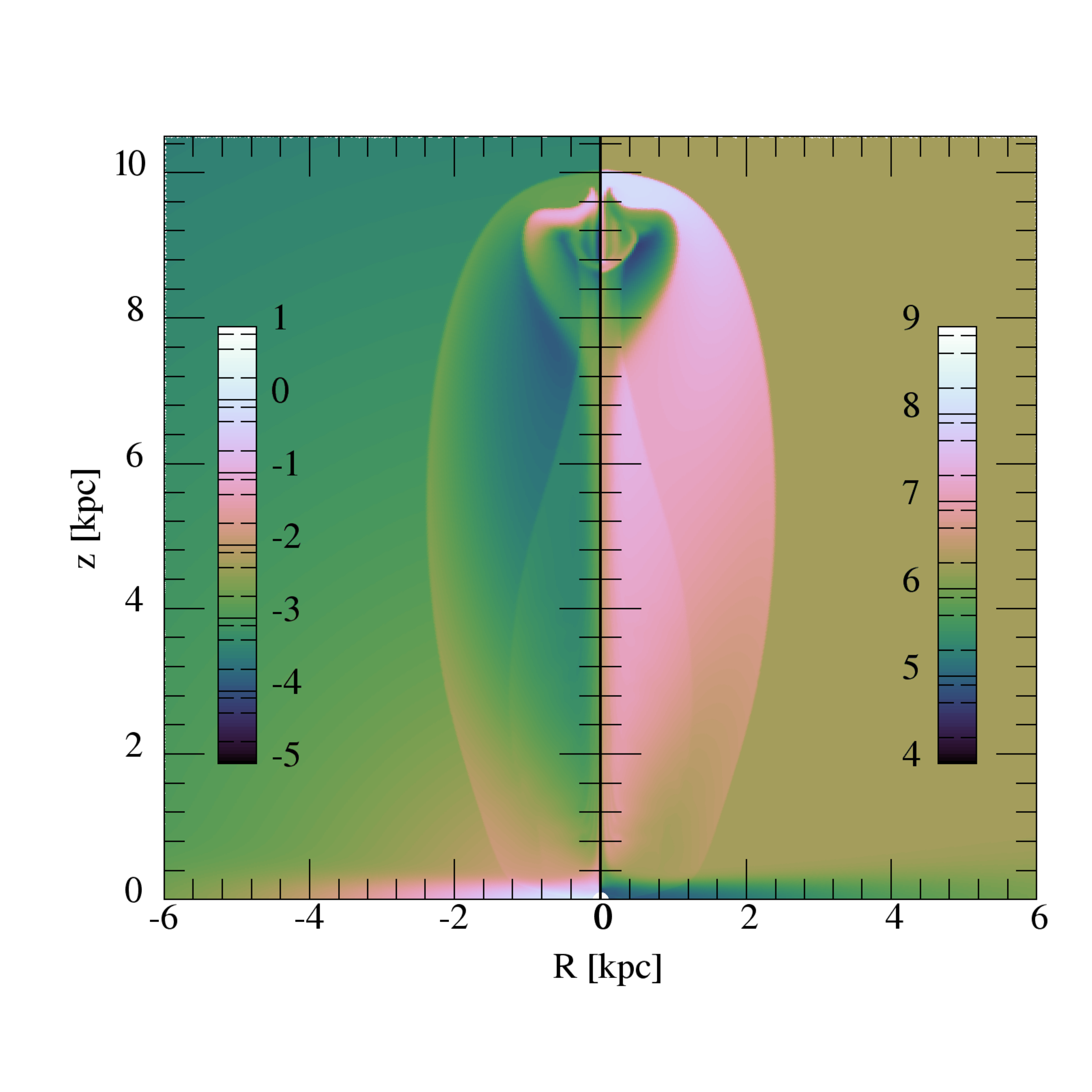}
\hspace{-0.5cm}
}
}
\caption{Select variations of the ballistic bubble setup, with parameters identical to J1 except $E_{\inj}$=2$\times10^{56}\erg$ (J1a; first row), $\beta_{\inj}=0.015$ (J1b; second row), $\theta_{\inj}=3^\circ$ (J1c; third row), and $t_j=0.01$Myr (J1d; fourth row).
Notations are the same as in \autoref{fig:base_GC_IL}.
\label{fig:ballistic_setups}
}
\end{figure}

We carry out a suite of simulations that generate \emph{Fermi}-like bubbles in their ballistic stage.
A representative sample of such simulations, including our fiducial setup J1 and its variants J1a through J1d, are summarised in the middle columns of \autoref{Tab:Jetted}, and are presented in \autoref{fig:ballistic_setups} at the times (displayed in the upper left corner of each row) when each bubble reaches the designated latitude.
The table provides the central measured properties of each simulated bubble: its age $\tage$ and maximal half-width $\Delta{l}$.

\begin{figure}
\centering{
\DrawFigs{
\raisebox{-0.cm}{\includegraphics[height=4.1truecm]{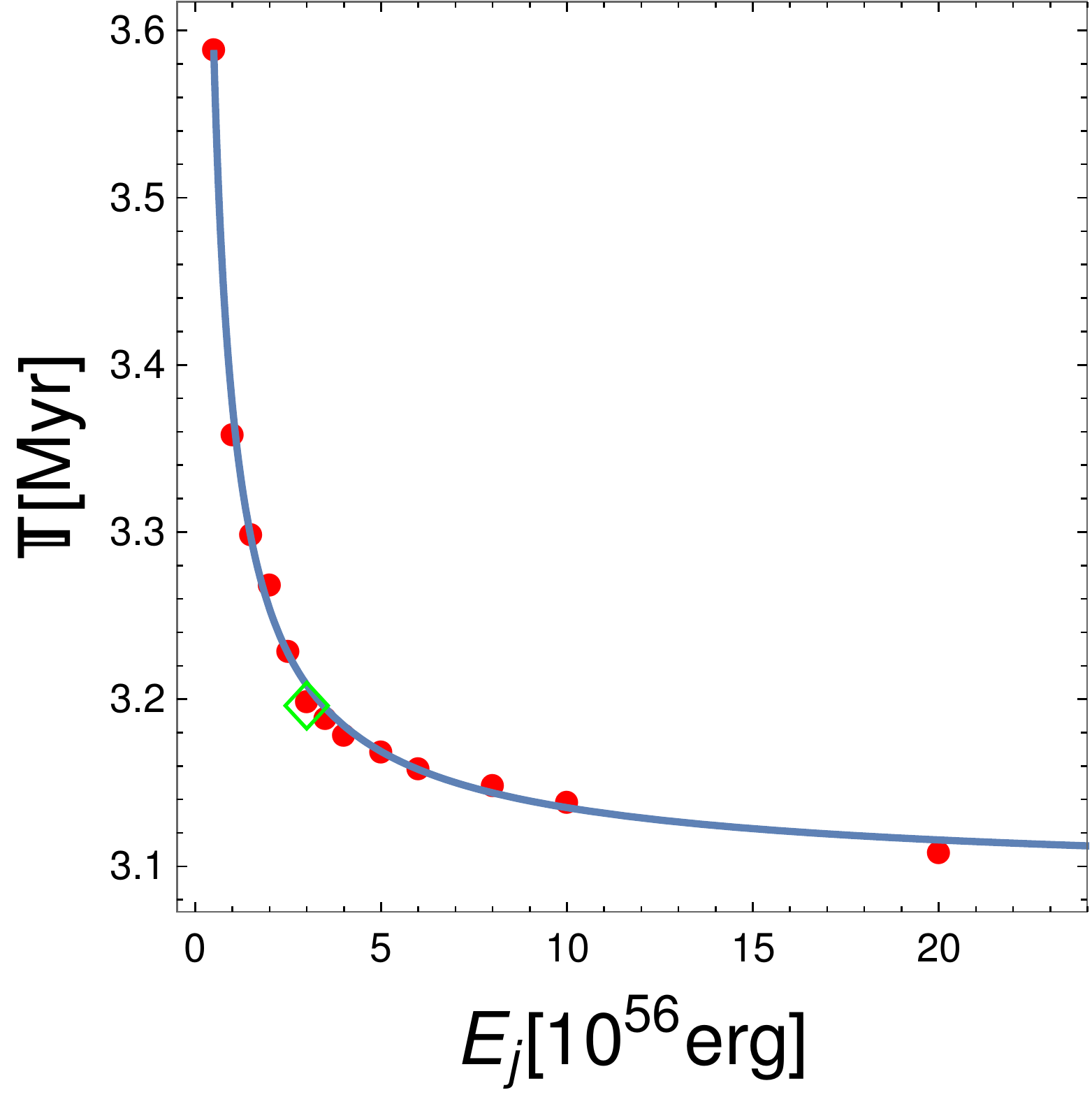}}
\raisebox{-0.cm}{\includegraphics[height=4.1truecm]{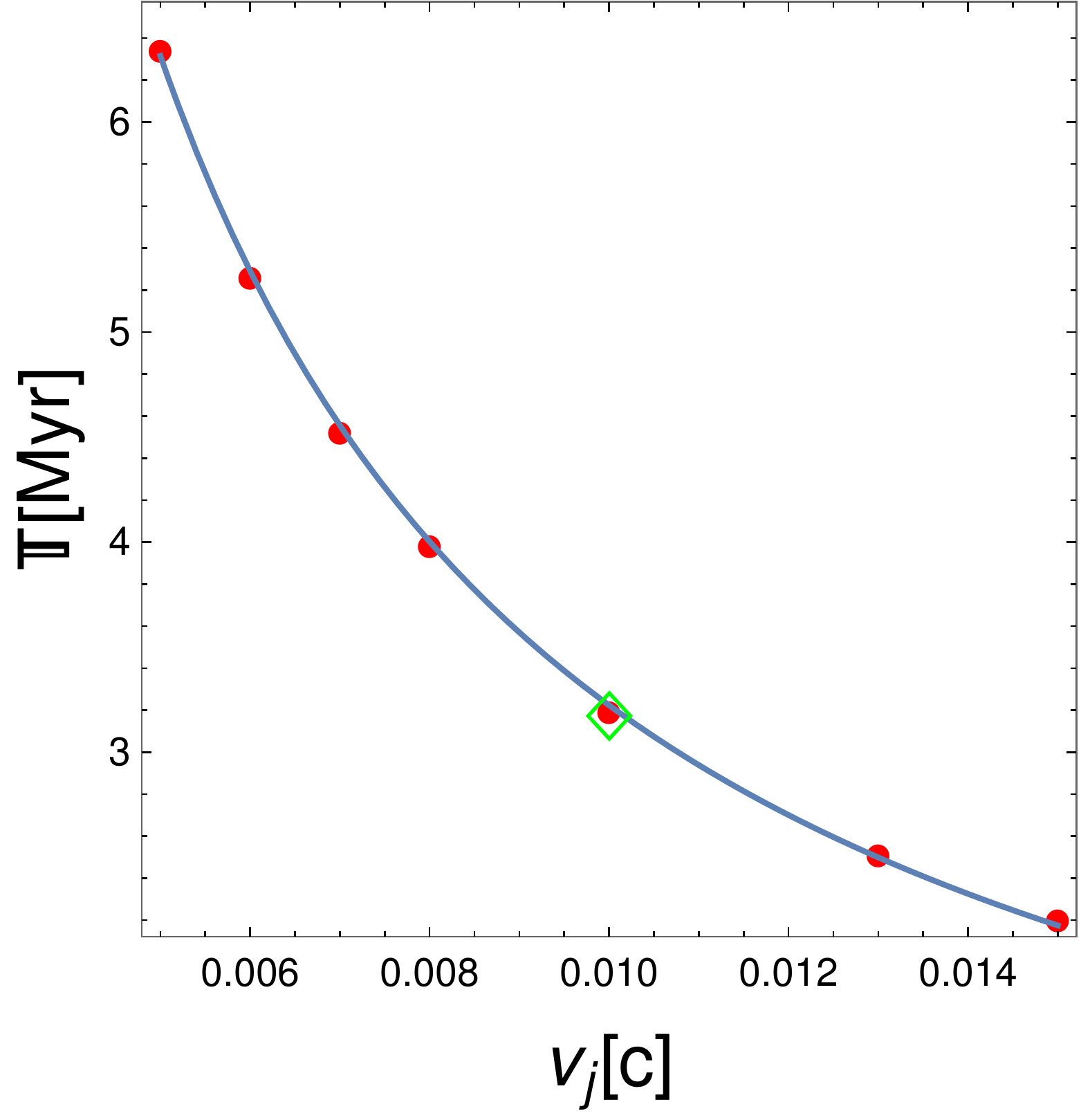}}
}
\\
\centering{
\raisebox{-0.cm}{\includegraphics[height=4.1truecm]{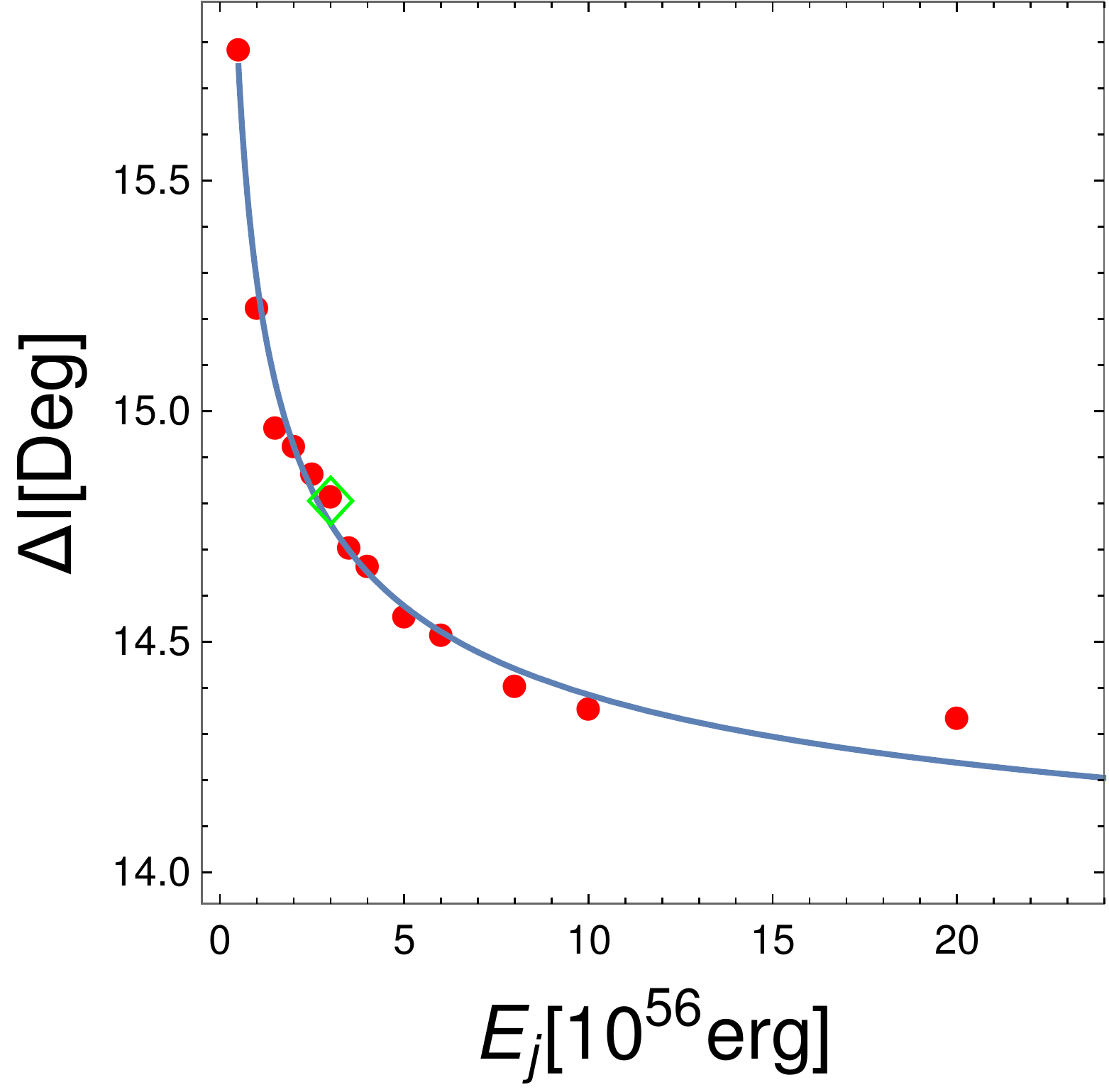}}
\raisebox{-0.cm}{\includegraphics[height=4.1truecm]{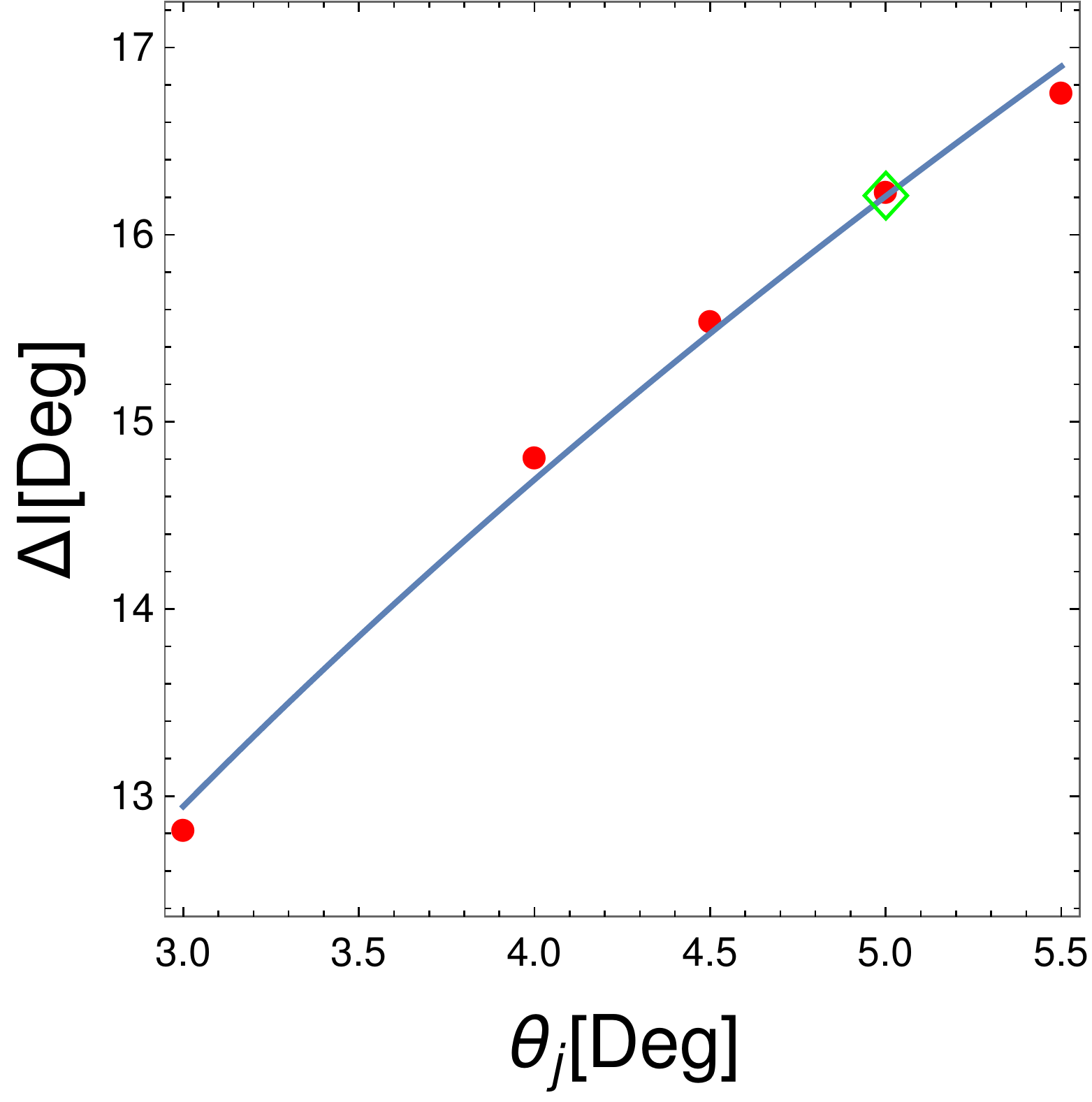}}
}
}
\caption{Best fit values (red {\disks}) of $\tage$ (top row) and $\Delta l$ (bottom) for setup J1 (greed diamond) modified by single-parameter variations in $E_{\inj}$ (left panels), v$_{\inj}$ (top right) or $\theta_{\inj}$ (bottom right).
Best fit profiles (see text) are also shown (solid blue curve).
\label{fig:ballistic_plots}}
\end{figure}

By separately varying each of the setup parameters, one at a time, we derive the functional dependencies of the bubble age
\begin{align} \label{eq:BallisticFitTage}
\tage \simeq & 3.2\left(\frac{E_{\inj}}{3\times10^{56}\erg}\right)^{-0.030\pm0.004} \beta_{-2}^{-0.96\pm0.01} \theta_5^{-0.003\pm0.031} \nonumber \\
    & \times t_{0.04}^{0.007\pm0.004} \Myr \coma
\end{align}
its maximal half-width
\begin{align} \label{eq:BallisticFitRmax}
R_{\max} \simeq & 2.2\left(\frac{E_{\inj}}{3\times10^{56}\erg}\right)^{-0.032\pm0.005} \beta_{-2}^{-0.11\pm0.02} \theta_5^{0.40\pm0.03} \nonumber \\
    & \times t_{0.04}^{-0.017\pm0.002} \kpc \coma
\end{align}
and maximal longitude
\begin{align}
\Delta l \simeq & 16^{\circ}.3\left(\frac{E_{\inj}}{3\times10^{56}\erg} \right)^{-0.024\pm0.004} \beta_{-2}^{-0.07\pm0.02} \theta_5^{0.41\pm0.03} \nonumber \\
 & \times t_{0.04}^{-0.014\pm0.005}  \coma
\end{align}
valid in the close vicinity of the fiducial J1 setup.
Such individual parametric scans are illustrated in \autoref{fig:ballistic_plots}.
Here, we defined $t_{0.04}\equiv t_{\inj}/0.04\Myr$.
The fits (\ref{eq:BallisticFitTage}) and (\ref{eq:BallisticFitRmax}) are in good agreement with the model Eqs.~(\ref{eq:BallisticAge}) and (\ref{eq:BallisticRmaxTage}), showing that the anticipated age $\tage\propto \beta_{\inj}^{-1}$ and half-width $R_{\max}\propto \theta_{\inj}^{1/2}$ of the ballistic bubble are nearly independent of other parameters.

As expected, these scaling relations break down when the setup parameters are modified substantially away from the J1 setup.
In particular, if (only) the energy is lowered considerably, non-ballistic corrections emerge, and we obtain
\begin{equation}
\tage \sim \left[3.1+(E_{\inj}/2.1\times10^{55}\erg)^{-0.80\pm0.03}\right]\Myr
\end{equation}
and
\begin{equation}
\Delta l \sim \left[13^{\circ}.8 + 1^{\circ}.0(E_{\inj}/3.0\times10^{56}\erg)^{-0.39\pm0.01}\right] \coma
\end{equation}
as shown in \autoref{fig:ballistic_plots}.
These results indicate that $\lesssim7\%$ non-ballistic corrections are present in the J1 bubble.


The properties of the simulated ballistic bubbles are approximately independent of
the injection parameters $t_\inj$ and $r_0$ (in the nominal ranges
$0.25<t_{0.04}<2$ and $0.05<r_0<0.25$).
The results are also fairly independent of the Galactic model parametrs.
In particular, even extreme changes to the {\disk} mass and halo rotation parameters have a rather negligible effect on the shape and age of the bubble.

\subsubsection{Slowing-down bubbles}
\label{nonballisticregime}

In our fiducial slowing-down FB simulation, denoted J2, an energy of $2.5\times 10^{55}\erg$ is injected with velocity $\beta_{\inj}=0.1$ inside an opening angle $\theta_{\inj}=4^\circ$ for a duration of $t_{\inj} = 0.04\Myr$. This results in
a \emph{Fermi}-like bubble reaching the designated latitude at $\tage\simeq 2.4\Myr$, as depicted in \autoref{fig:NominalJetted} (bottom row).
For such a lower energy, higher velocity jet, the amount of thermalisation near the jet-base is not numerically resolved, due to unresolved KHI. To deal with such artefacts, we introduce viscosity
adopting Spitzer viscosity with a $\mu_{\max}=10\mbox{ g} \se^{-1}\cm^{-1}$ upper limit, as discussed in section \ref{sec:Method}.

\begin{figure*}
\DrawFigs{
\centering{
\includegraphics[height=4.7truecm,trim={2.1cm 2.5cm 1.5cm 5.2cm}, clip]{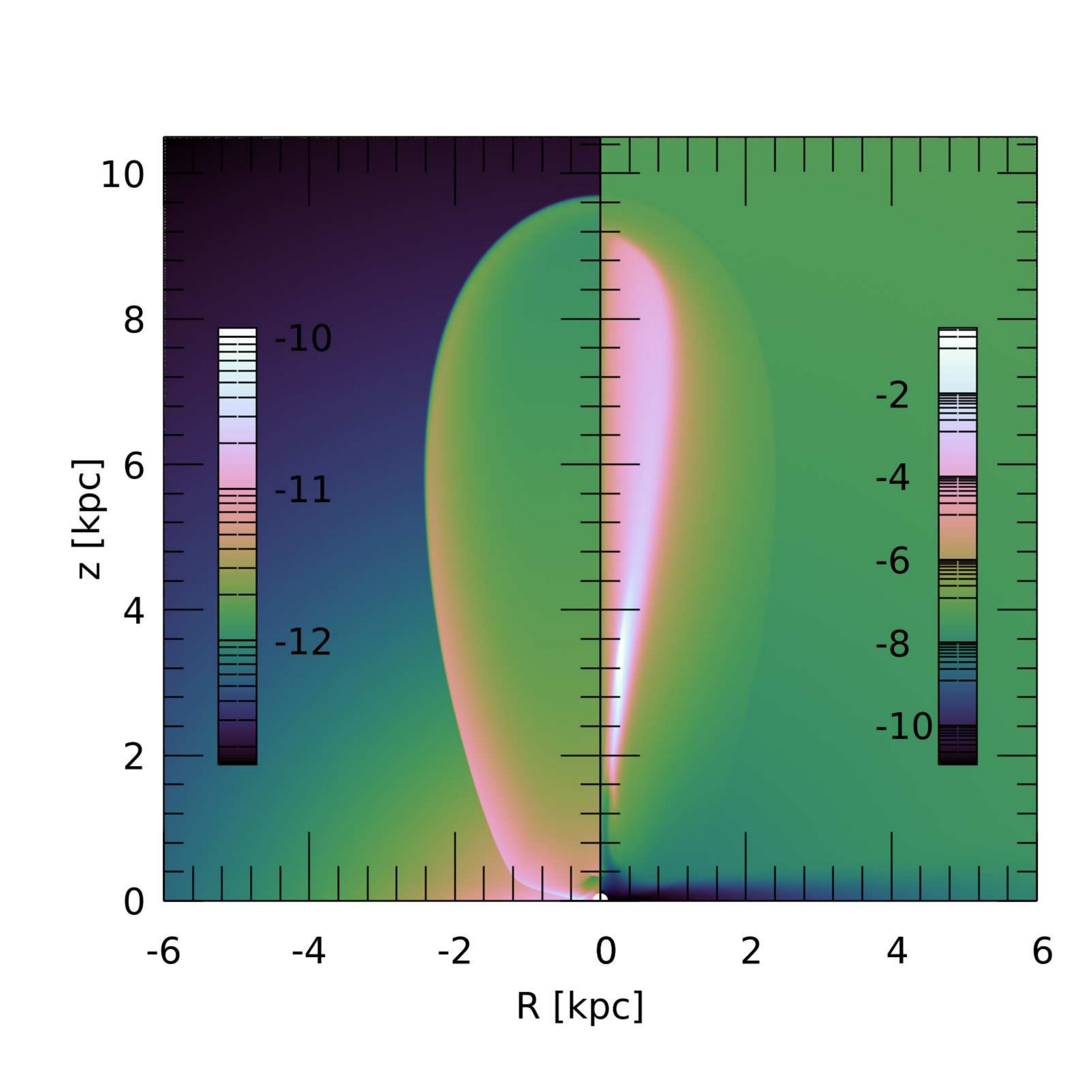}
\includegraphics[height=4.5truecm]{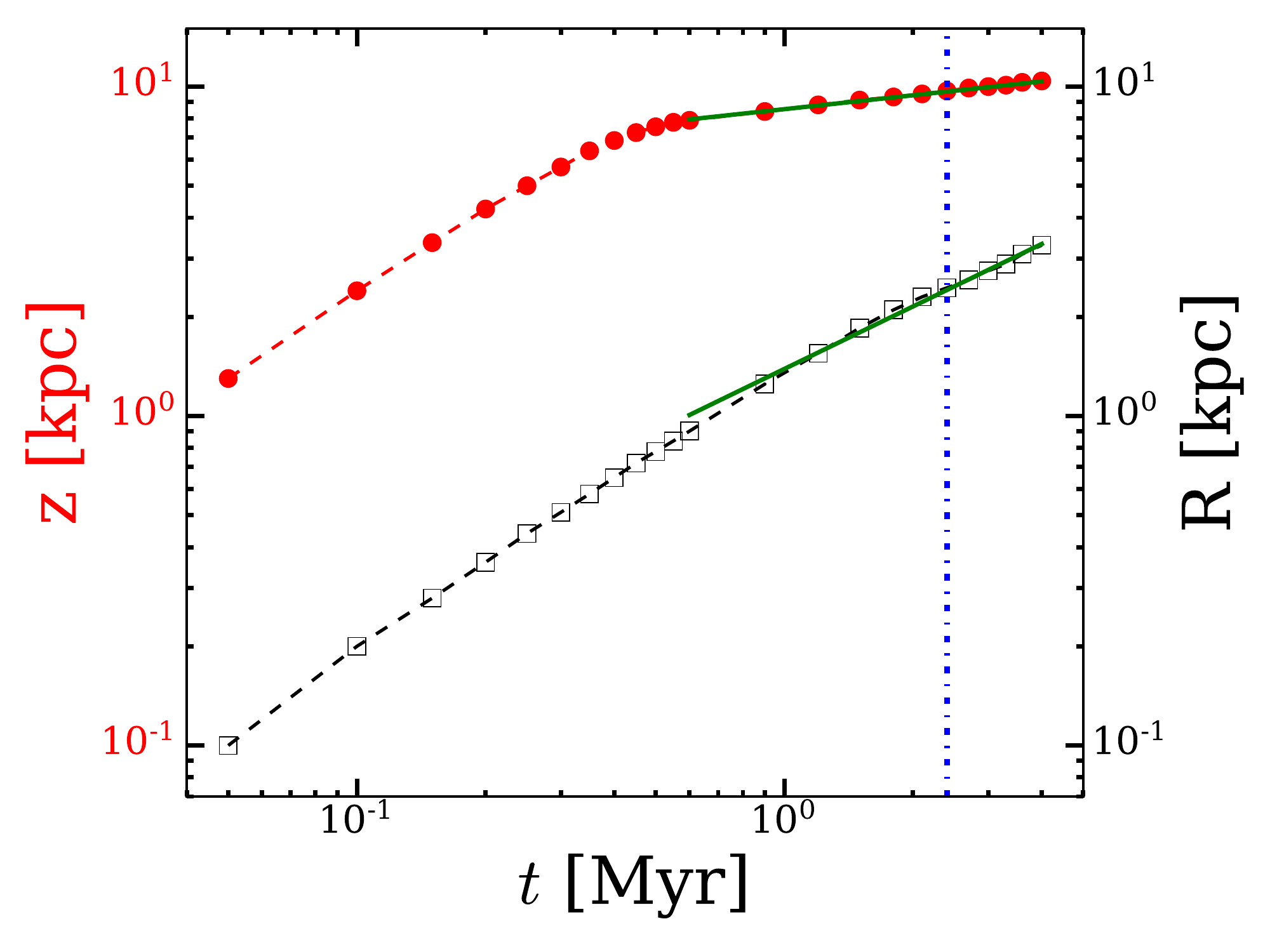}
\includegraphics[height=4.5truecm]{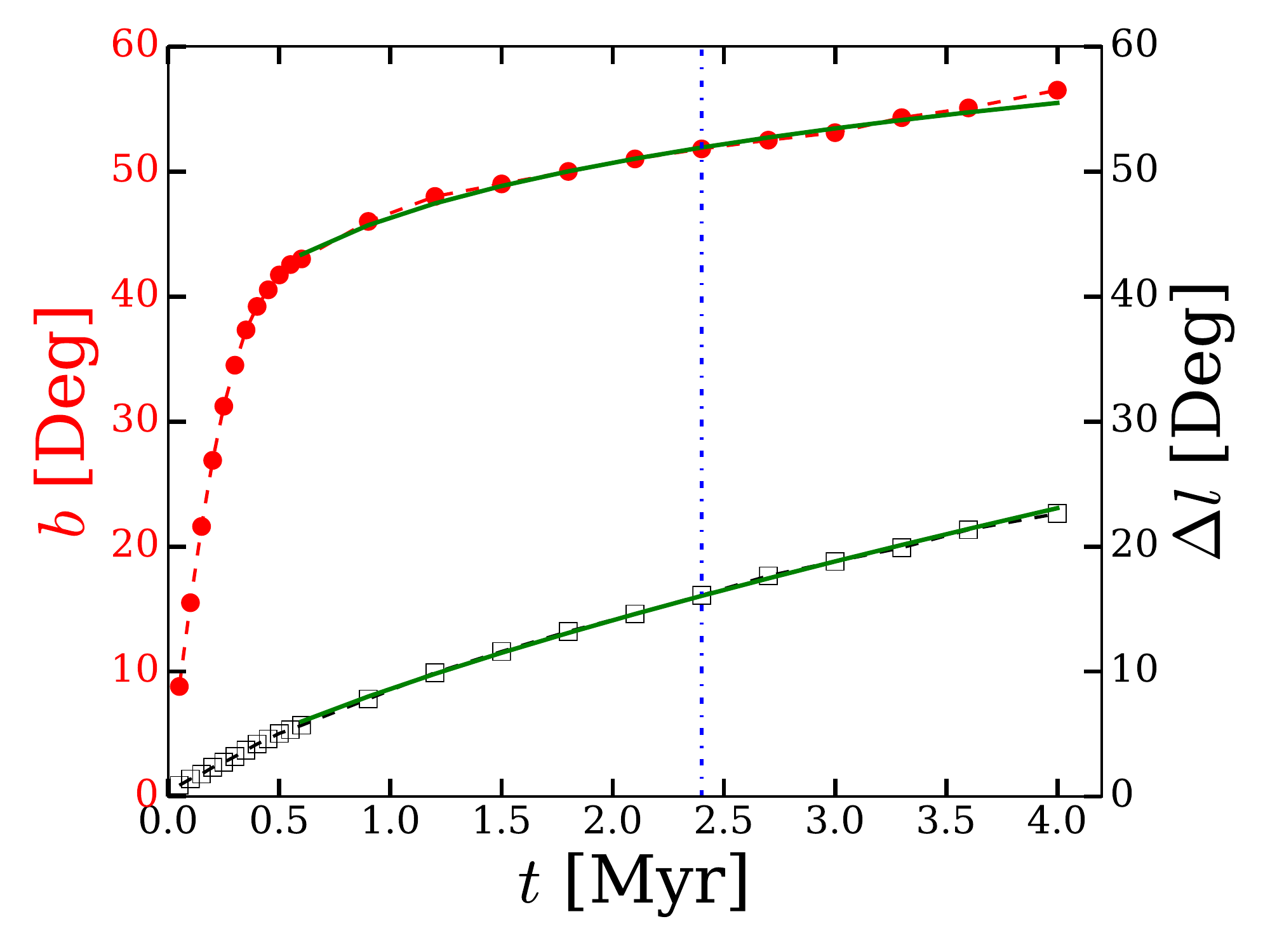}
}
}
\caption{
Same as \autoref{fig:ballisticCombo}, but for our fiducial slowdown bubble (setup J2).
\label{fig:NonBallisticCombo}}
\end{figure*}

The contact discontinuity trailing all parts of the slowing-down shock is evident in the figure, as well as in the corresponding pressure and entropy distributions presented in \autoref{fig:NonBallisticCombo} (left panel).
Unlike in the ballistic bubble,
this figure does not show any evidence of an inner shock or a low-pressure region trailing the bubble head,
due to an earlier thermalisation of the low momentum ($2 E_j/ c\:\beta_j$) jet for a given energy, as explained in section \ref{jettedresults}.
The pressure is fairly uniform in the inner parts of the bubble, but pressure variations behind the shock span more than an order of magnitude across the surface of the bubble.

We find that in the fast injection domain, the initial jet emerging by the injection time $t_{\inj}$ is substantially reshaped by the strong interactions with the ambient gas.
Unlike the slow jets studied for ballistic bubbles, which are initially sustained as cones of opening angle $\theta_{\modj}\simeq \theta_{\inj}$, the present
fast jets
substantially narrow at their top, presenting as top-pinched bubbles.
The upper structure of these initial jets are subsequently shaped by the interaction with the ambient gas,
rather than by the initial jet
parameters.
Indeed, we find that the simulation results in the high velocity regime are not sensitive to the initial setup parameters $3^\circ\lesssim\theta_{\inj}\lesssim5^\circ$ and $0.05<r_0/\mbox{kpc}<0.5$. Instead, a dependence on viscosity and injection time emerges, as anticipated in \autoref{subsec:BallisticOrSlowDown}.

Figure \ref{fig:NonBallisticCombo} also shows (middle and right panels) the temporal evolution of the bubble.
At early, $t\lesssim 0.5\Myr$ times, the bubble grows approximately linearly both in height and in width, as expected
in the ballistic phase.
For the fast jets in this regime, the
injection stage
has a much stronger effect in slowing and reshaping the jet.
From Eqs.~(\ref{eq:BallisticHeight}) and (\ref{eq:BallisticRmax}), we estimate that
$v_{\modj}\simeq 0.6v_{\inj}$ and $\theta_{\modj}\simeq 0^{\circ}.3$ (independent of $\theta_{\inj}$).
At late times ($t\gtrsim 0.5\Myr$), the bubble height grows noticeably slower, while its widening is only slightly diminished, as expected in the slowdown phase.
The ballistic-to-slowdown transition can therefore be identified at $t_{\tr}\simeq 0.5\Myr$ and $z_{\tr}\simeq 8\kpc$.

In the slowdown phase, the slow growth in height is well-fit by a power law, $z_H(t>0.6\Myr) \simeq 8.5t_1^{0.14}\kpc$, consistent with Eq.~(\ref{eq:zHvsT}) provided that $\Myfq\simeq 1/5$.
The widening of the bubble is not precisely a power law in this stage, slowing from $R_{\max}(t>0.6\Myr) \simeq 1.4t_1^{0.6}\kpc$  to $R_{\max}(t>1.8\Myr)\simeq 1.5t_1^{0.5}$.
The latter nicely matches (both in normalization and in power-law) the expected late-time behavior (\ref{eq:NonBallisticRmaxOft}) for the above parameters.
The projected dimensions of the bubble are best fit for $t>0.6\Myr$ by $b \simeq 46^\circ.3 t_1^{0.13}$ and $\Delta l \sim 8^\circ.6 t_1^{0.7}$.

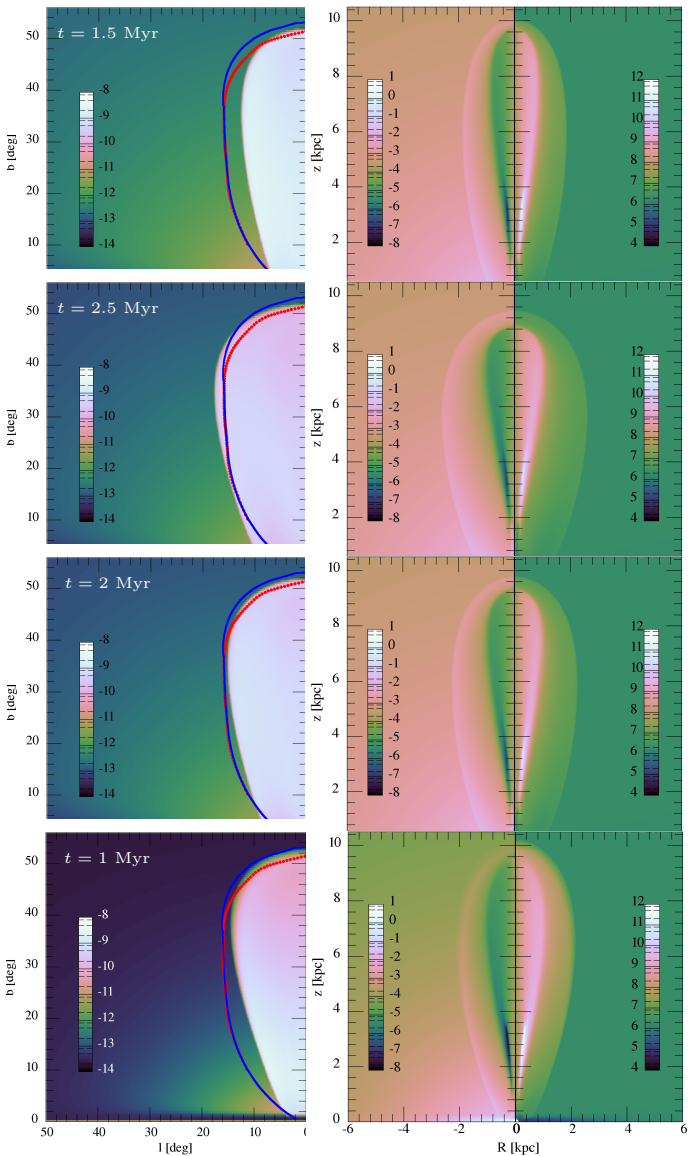
\begin{figure}
\DrawFigs{
\centering{
\hspace{-0.7cm}
\begin{tikzpicture}
\draw (0, 0) node[inner sep=0] {\raisebox{0.15cm}{\includegraphics[height=4.5truecm,trim={4.8cm 2.2cm 4.0cm 2.8cm}, clip]{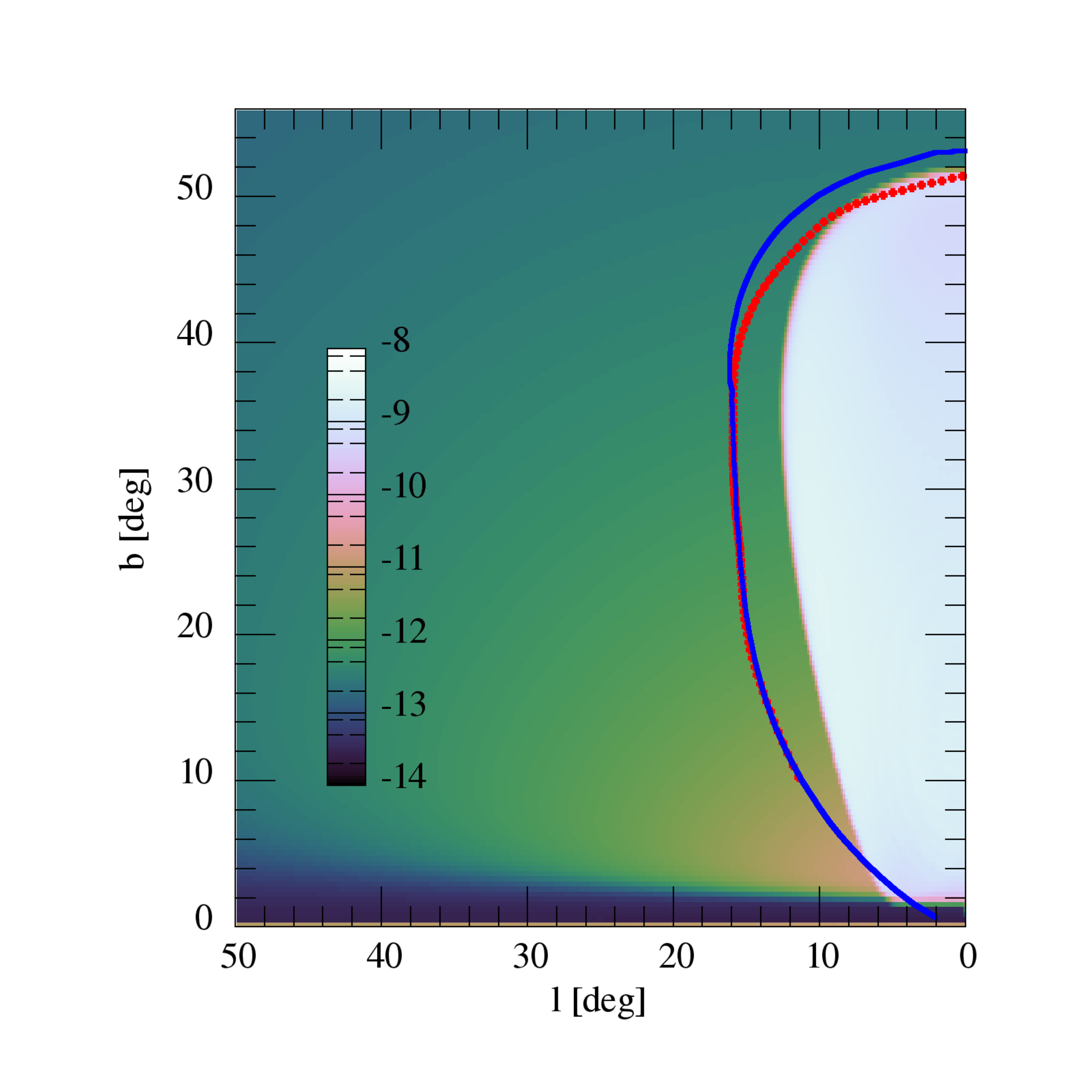}}};
\draw (-0.7, 1.8) node[text=white] {\scriptsize $t=1.5\Myr$};
\end{tikzpicture}
\hspace{-0.3cm}
\includegraphics[height=4.5truecm,trim={2.1cm 2.5cm 1.5cm 5.2cm}, clip]{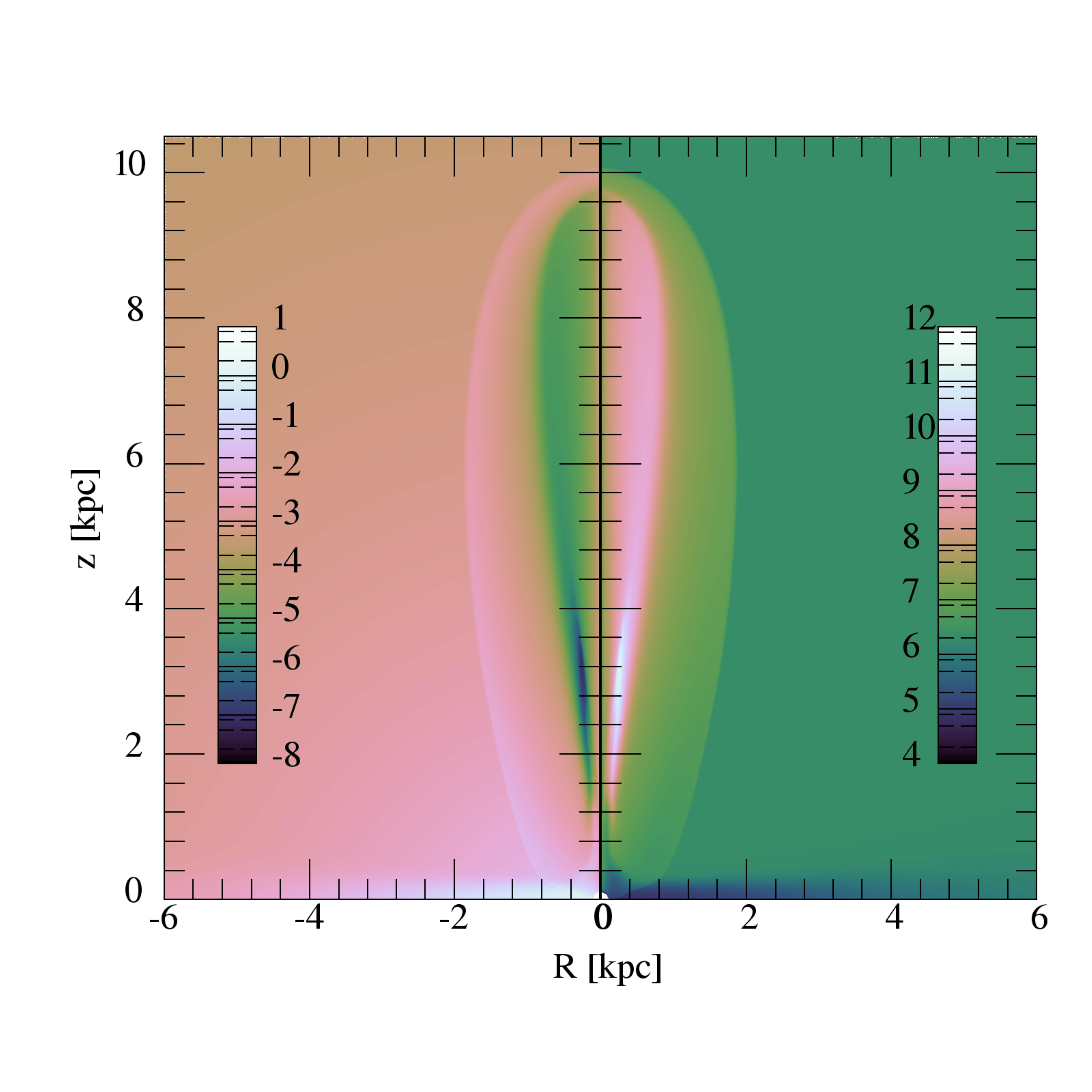}
\hspace{-0.5cm}
}\\
\vspace{-1.04cm}
\centering{
\hspace{-0.7cm}
\begin{tikzpicture}
\draw (0, 0) node[inner sep=0] {\raisebox{0.15cm}{\includegraphics[height=4.5truecm,trim={4.8cm 2.2cm 4.0cm 2.8cm}, clip]{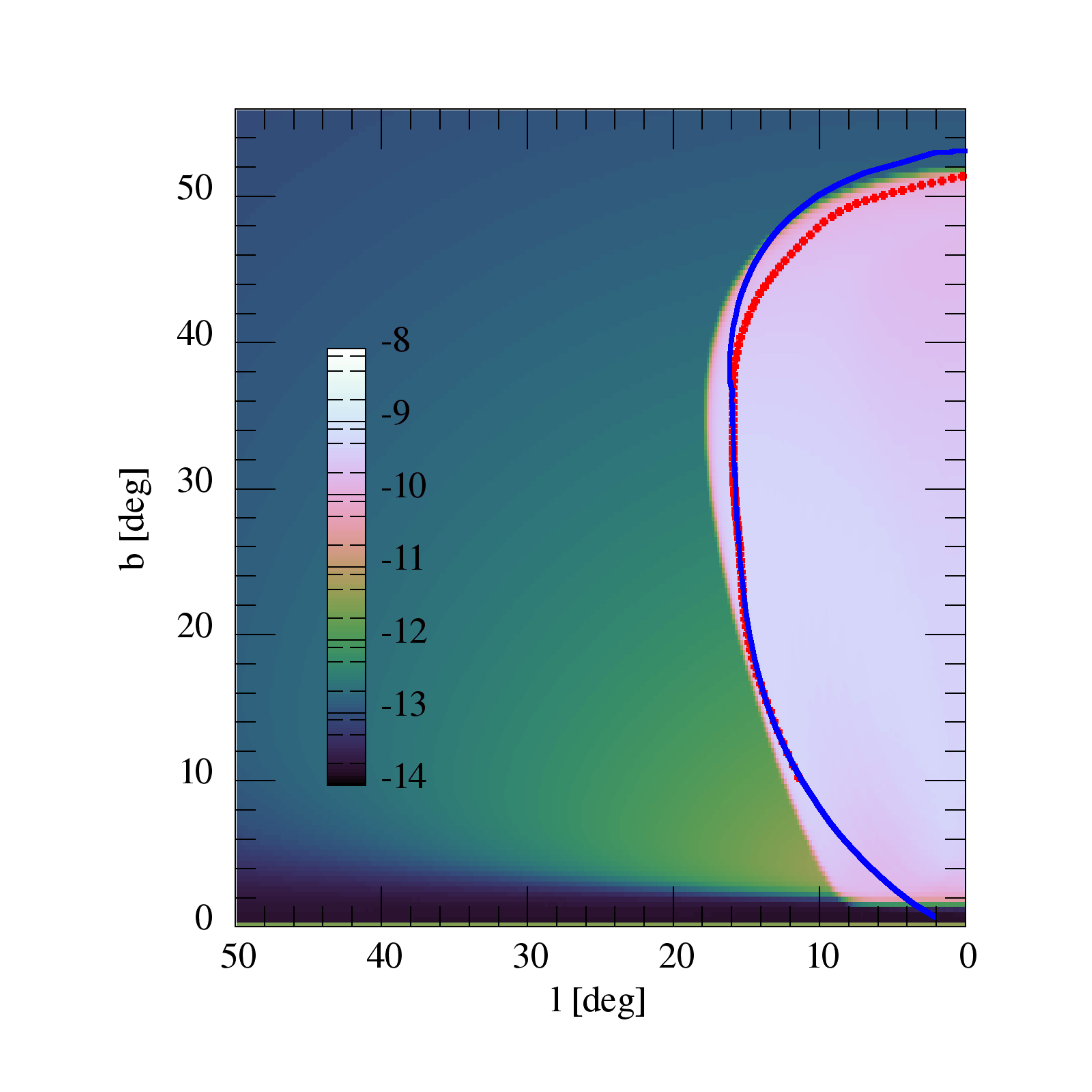}}};
\draw (-0.7, 1.8) node[text=white] {\scriptsize $t=2.5\Myr$};
\end{tikzpicture}
\hspace{-0.3cm}
\includegraphics[height=4.5truecm,trim={2.1cm 2.5cm 1.5cm 5.2cm}, clip]{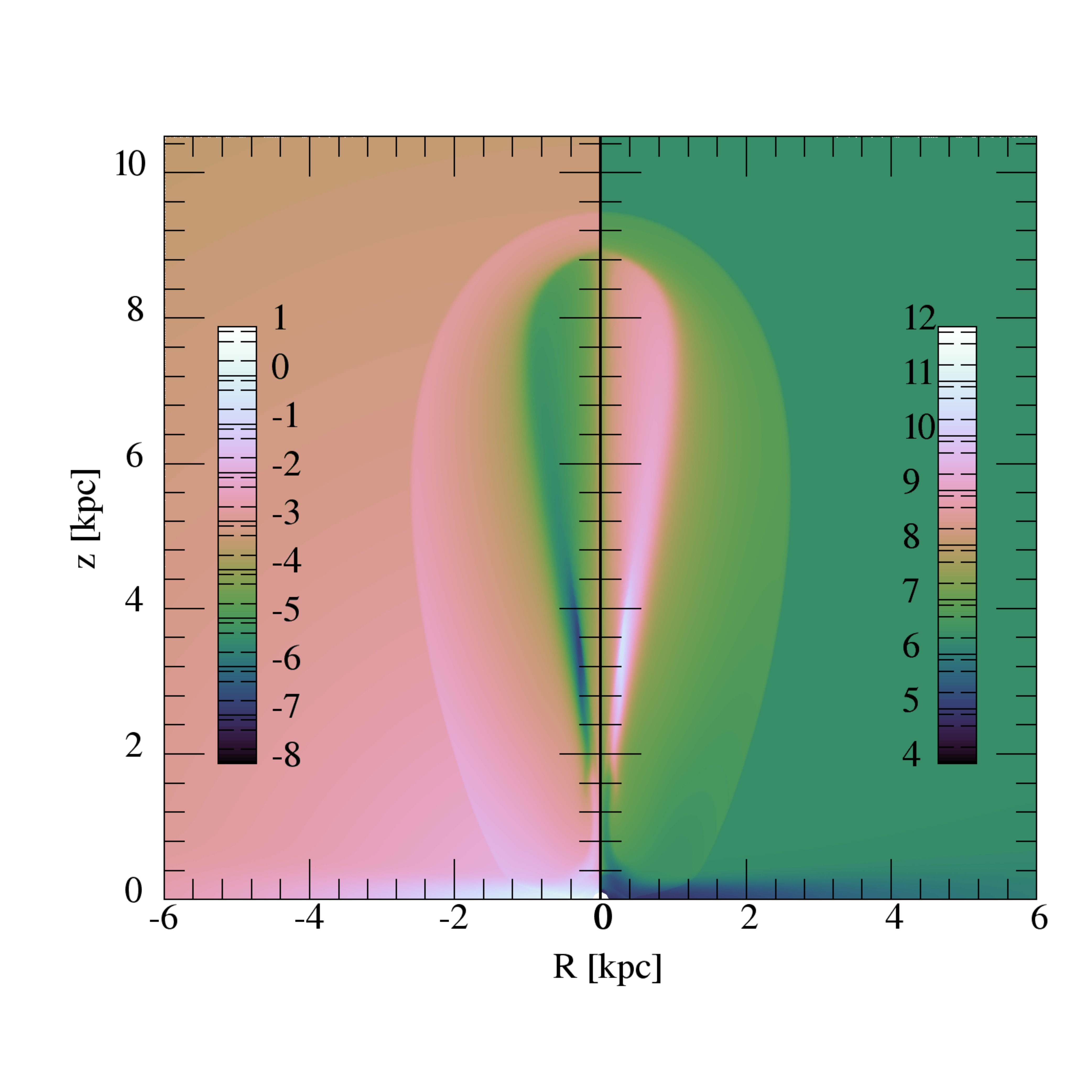}
\hspace{-0.5cm}
}\\
\vspace{-1.04cm}
\centering{
\hspace{-0.7cm}
\begin{tikzpicture}
\draw (0, 0) node[inner sep=0] {\raisebox{0.15cm}{\includegraphics[height=4.5truecm,trim={4.8cm 2.2cm 4.0cm 2.8cm}, clip]{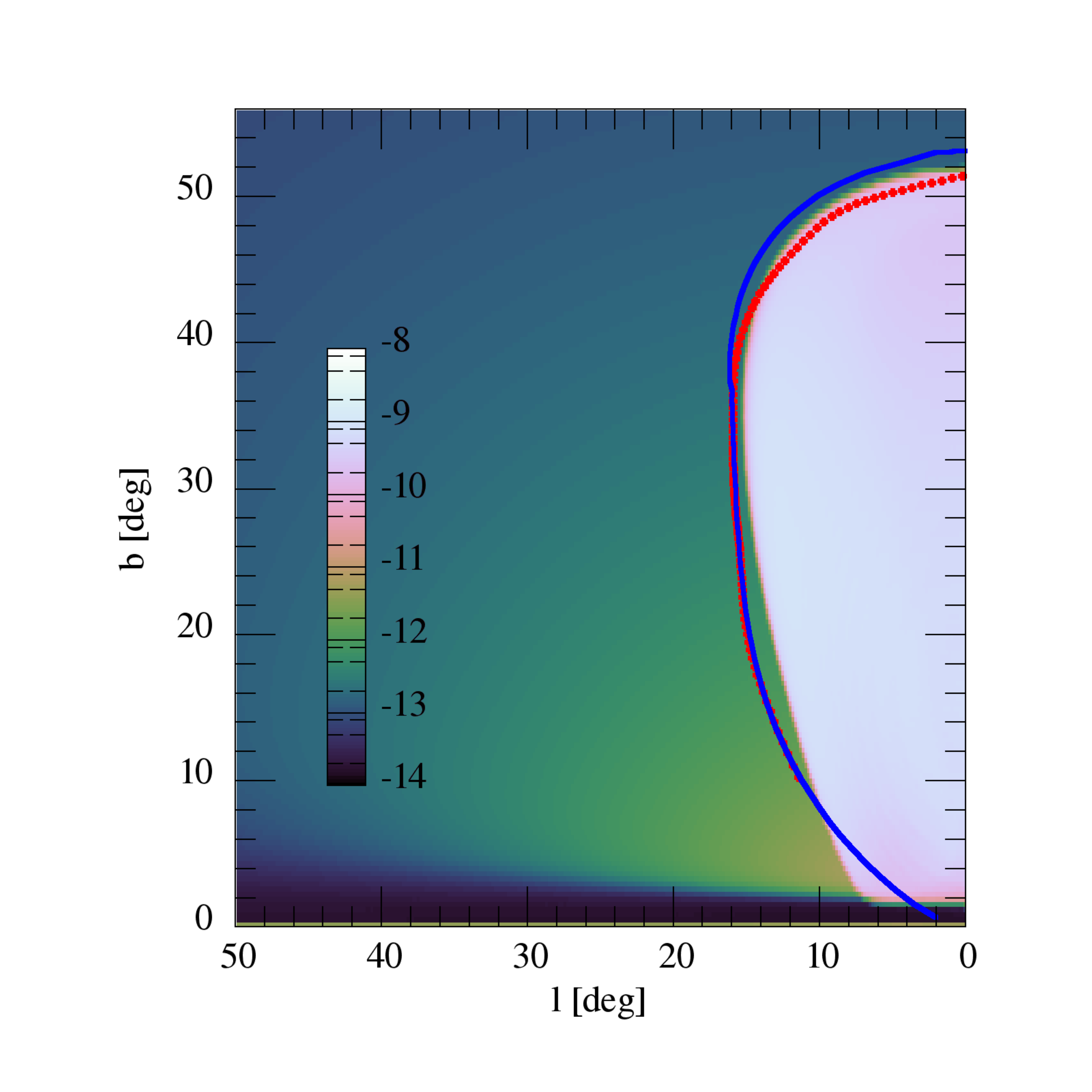}}};
\draw (-0.7, 1.8) node[text=white] {\scriptsize $t=2\Myr$};
\end{tikzpicture}
\hspace{-0.3cm}
\includegraphics[height=4.5truecm,trim={2.1cm 2.5cm 1.5cm 5.2cm}, clip]{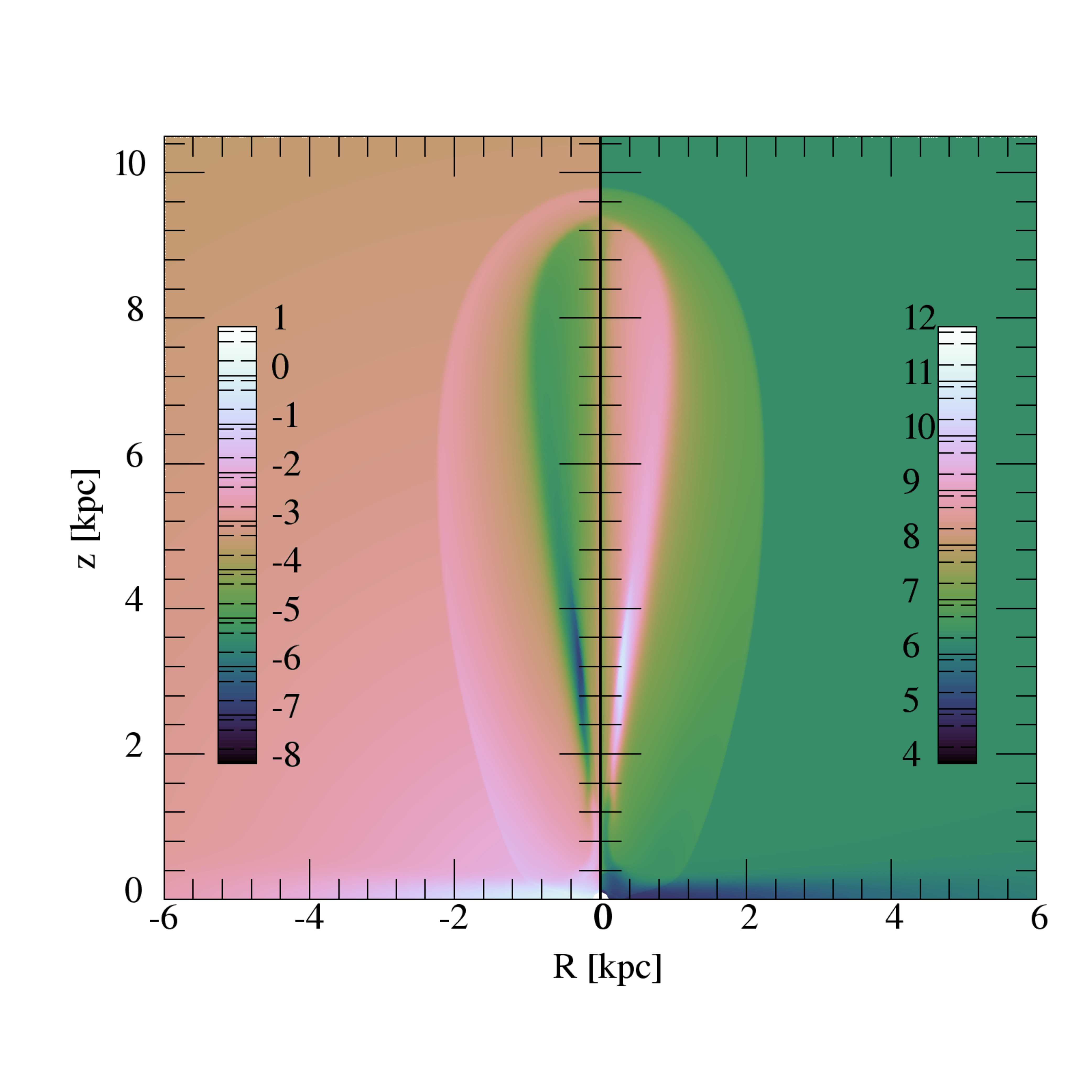}
\hspace{-0.5cm}
}\\
\vspace{-1.04cm}
\centering{
\hspace{-0.8cm}
\begin{tikzpicture}
\draw (0, 0) node[inner sep=0] {\raisebox{0.15cm}{\includegraphics[height=4.5truecm,trim={4.8cm 2.2cm 4.0cm 2.8cm}, clip]{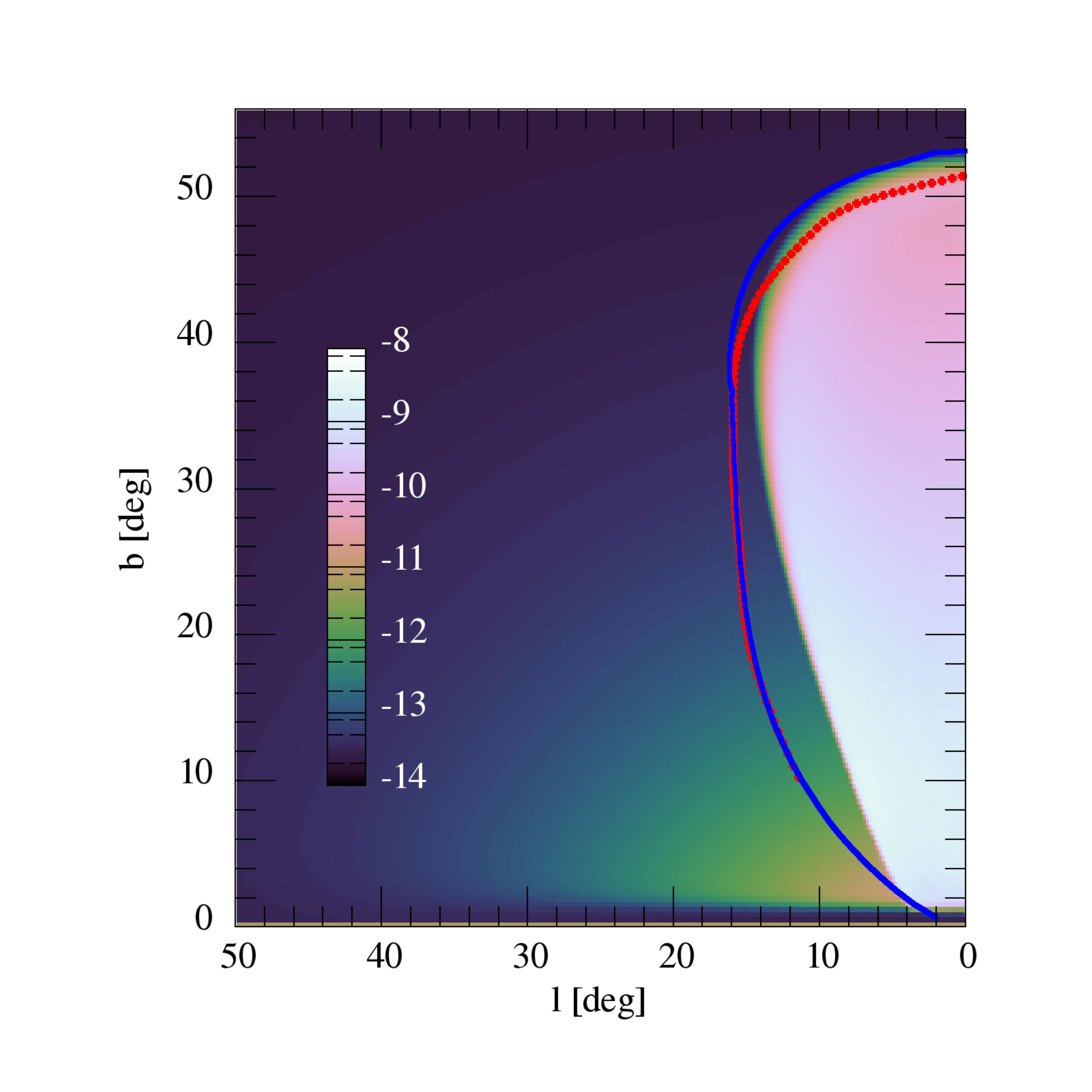}}};
\draw (-0.7, 1.8) node[text=white] {\scriptsize $t=1\Myr$};
\end{tikzpicture}
\hspace{-0.3cm}
\includegraphics[height=4.5truecm,trim={2.1cm 2.5cm 1.5cm 5.2cm}, clip]{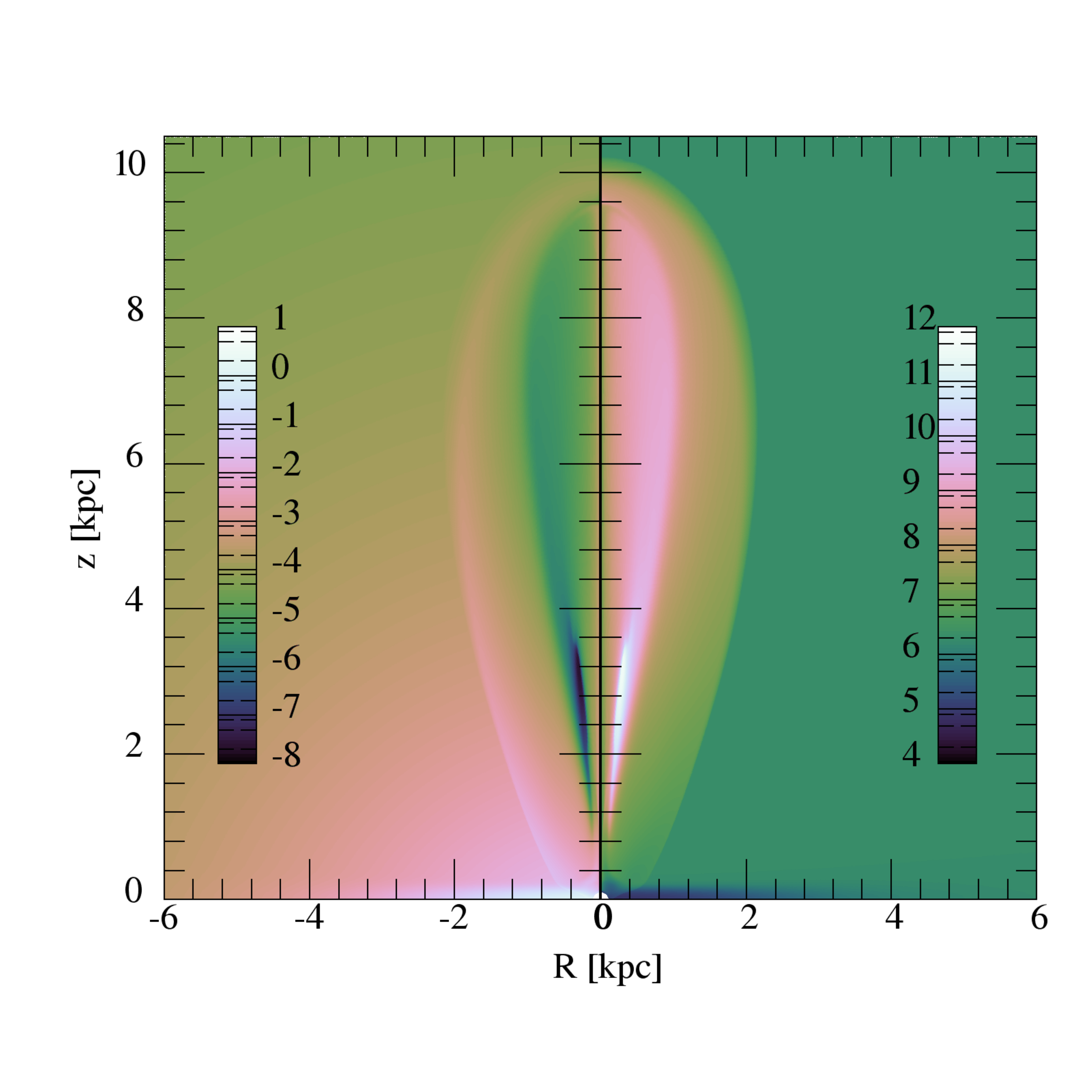}
\hspace{-0.4cm}
}
}
\caption{Variations on the slowing-down bubble setup J2a (top row), which itself is identical to J2 but with $E_{\inj}=3\times 10^{55}\erg$.
Single-parameter variations include $\beta_{\inj}=0.12$ (J2b, second row), $t_{\inj}=0.06\Myr$ (J2c, third row), and $M_{\rm d}=1.2\times 10^{11} M_\odot$ (J2d, fourth row).
Notations are the same as in \autoref{fig:base_GC_IL}.
\label{fig:nonballistic_variants}
}
\end{figure}

We carry out a suite of simulations that generate \emph{Fermi}-like bubbles in their slowing-down phase.
A representative sample of such simulations, including our fiducial setup J2 and its variants J2a through J2d, are summarised in the right columns of \autoref{Tab:Jetted}, and are depicted in \autoref{fig:nonballistic_variants}.
The observed height and width of the FBs strongly constrain the initial conditions of the simulations, such that large deviations from the J1 or J2 parameters generally yield simulated bubbles inconsistent with observations.

\begin{figure}
\centering{
\raisebox{-0.cm}{\includegraphics[height=4.1truecm]{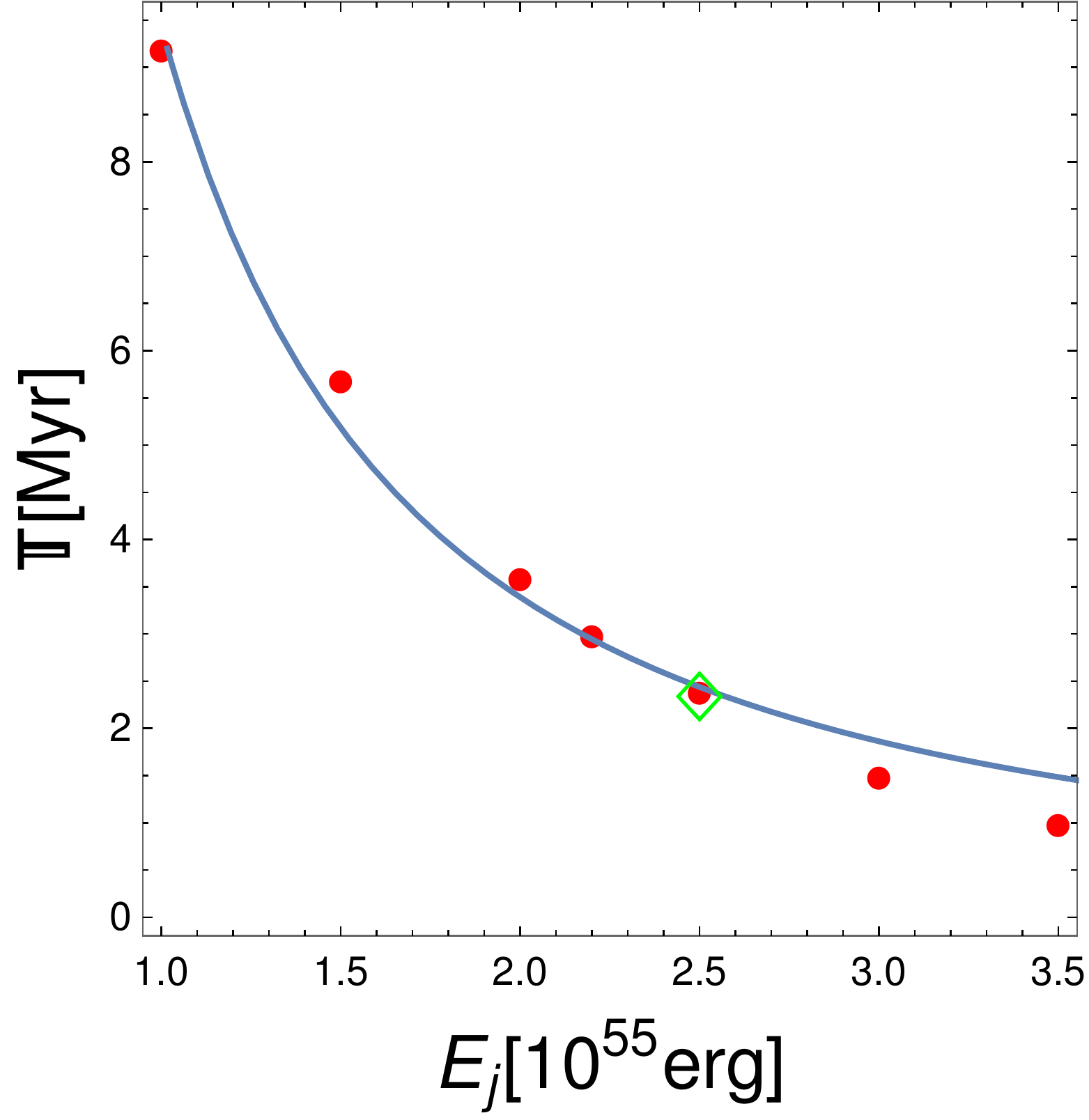}}
\raisebox{-0.cm}{\includegraphics[height=4.1truecm]{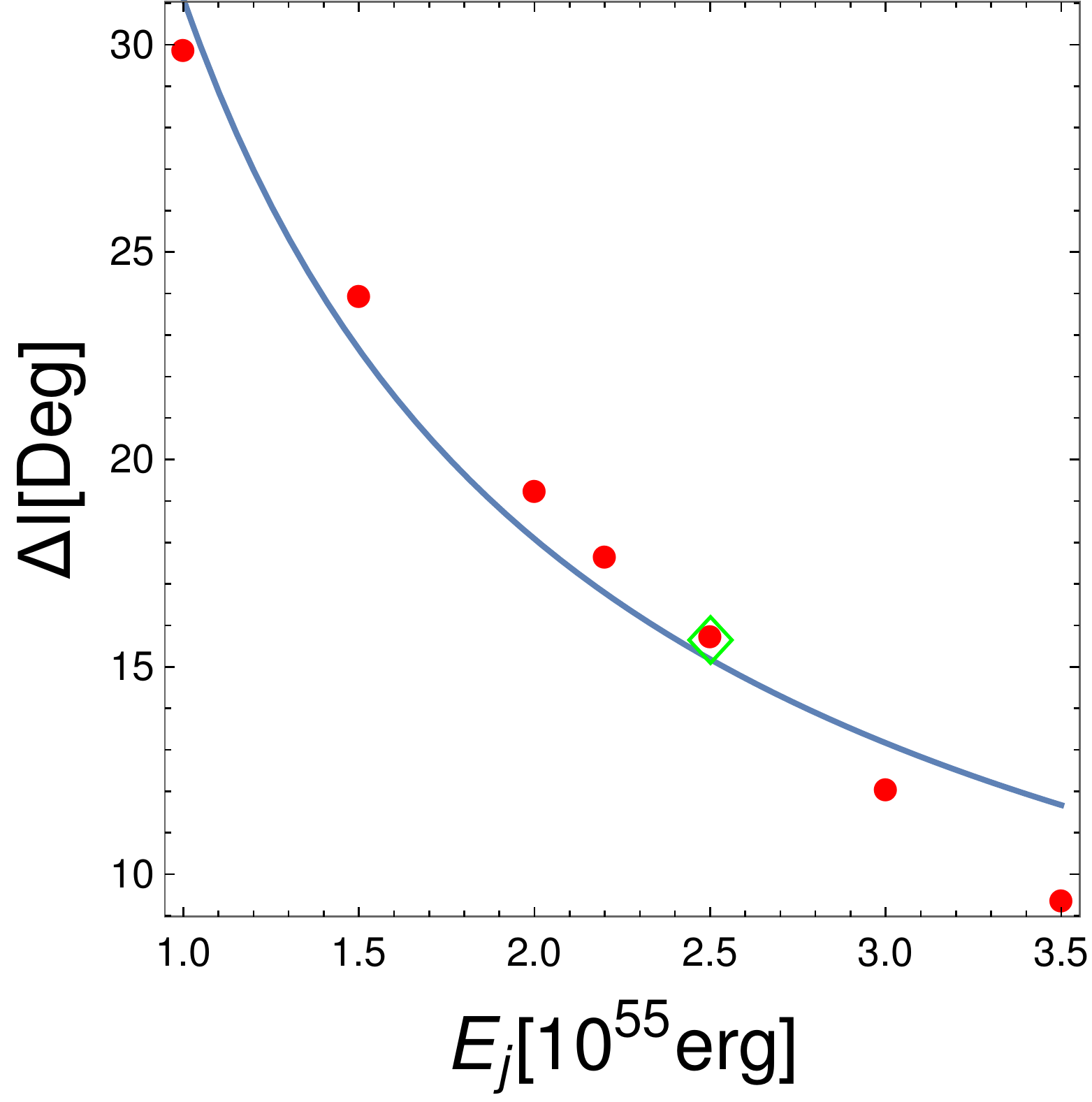}}
}\\
\centering{
\raisebox{-0.cm}{\includegraphics[height=4.1truecm]{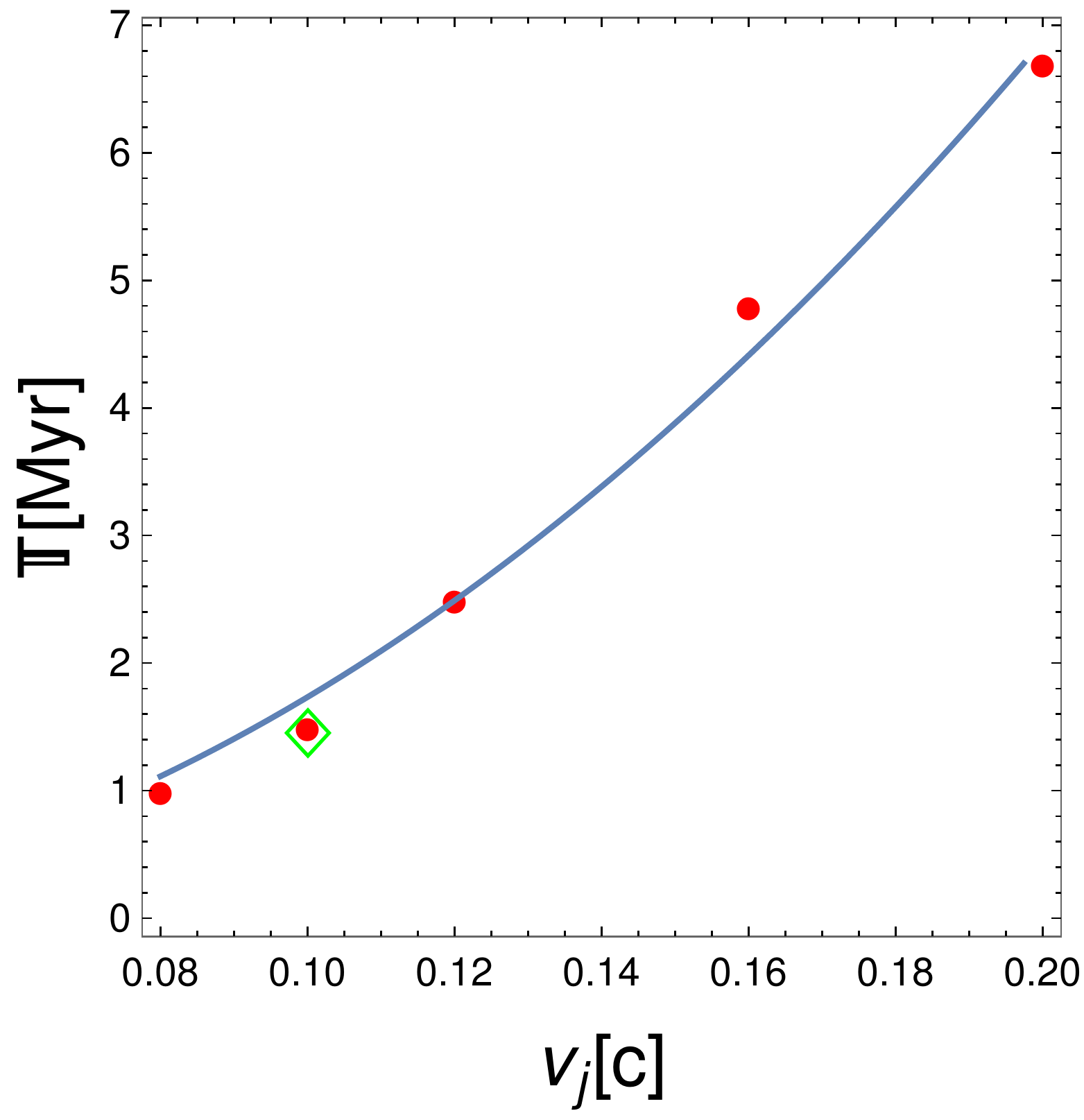}}
\raisebox{-0.cm}{\includegraphics[height=4.1truecm]{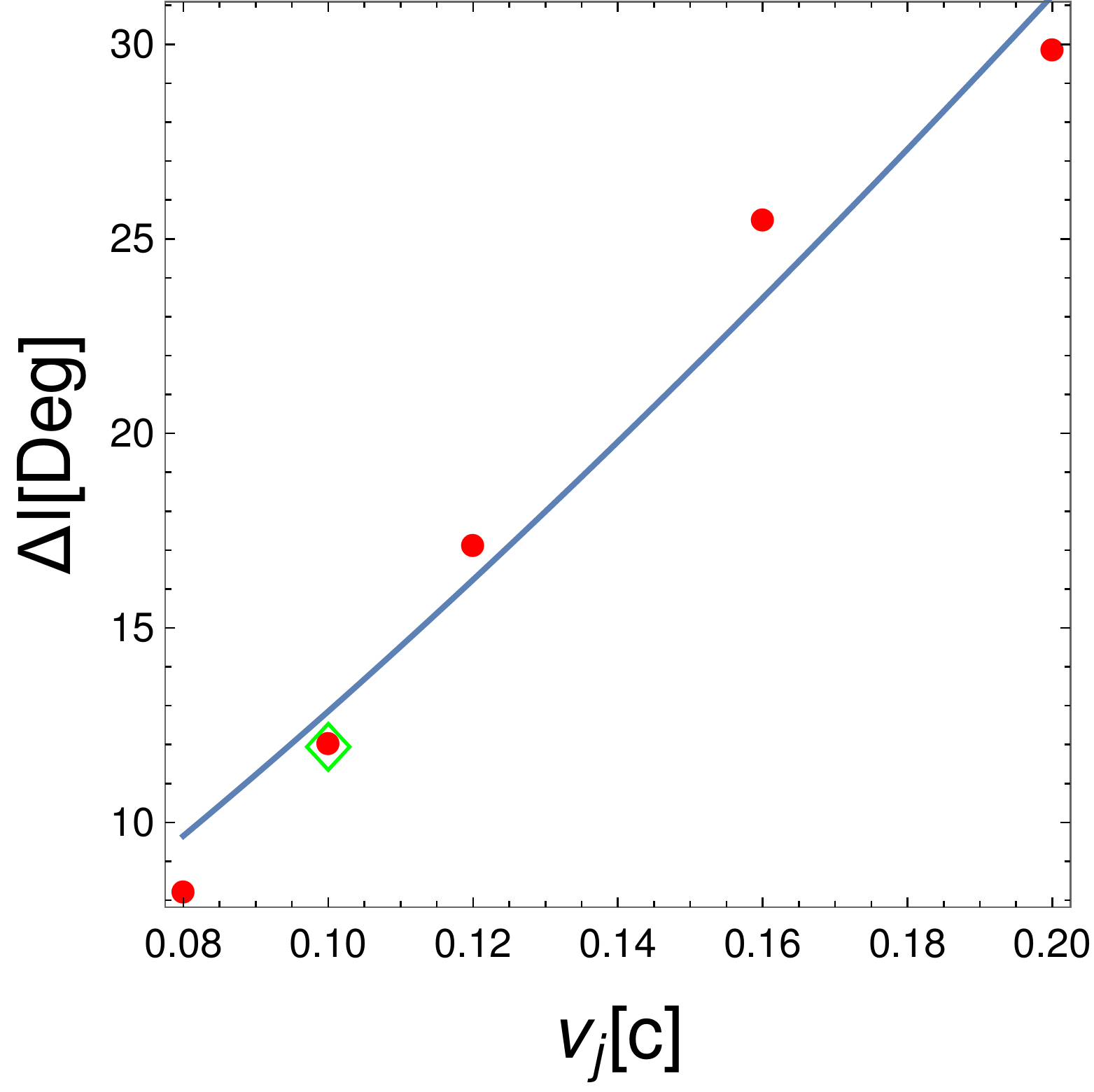}}
}\\
\centering{
\raisebox{-0.cm}{\includegraphics[height=4.1truecm]{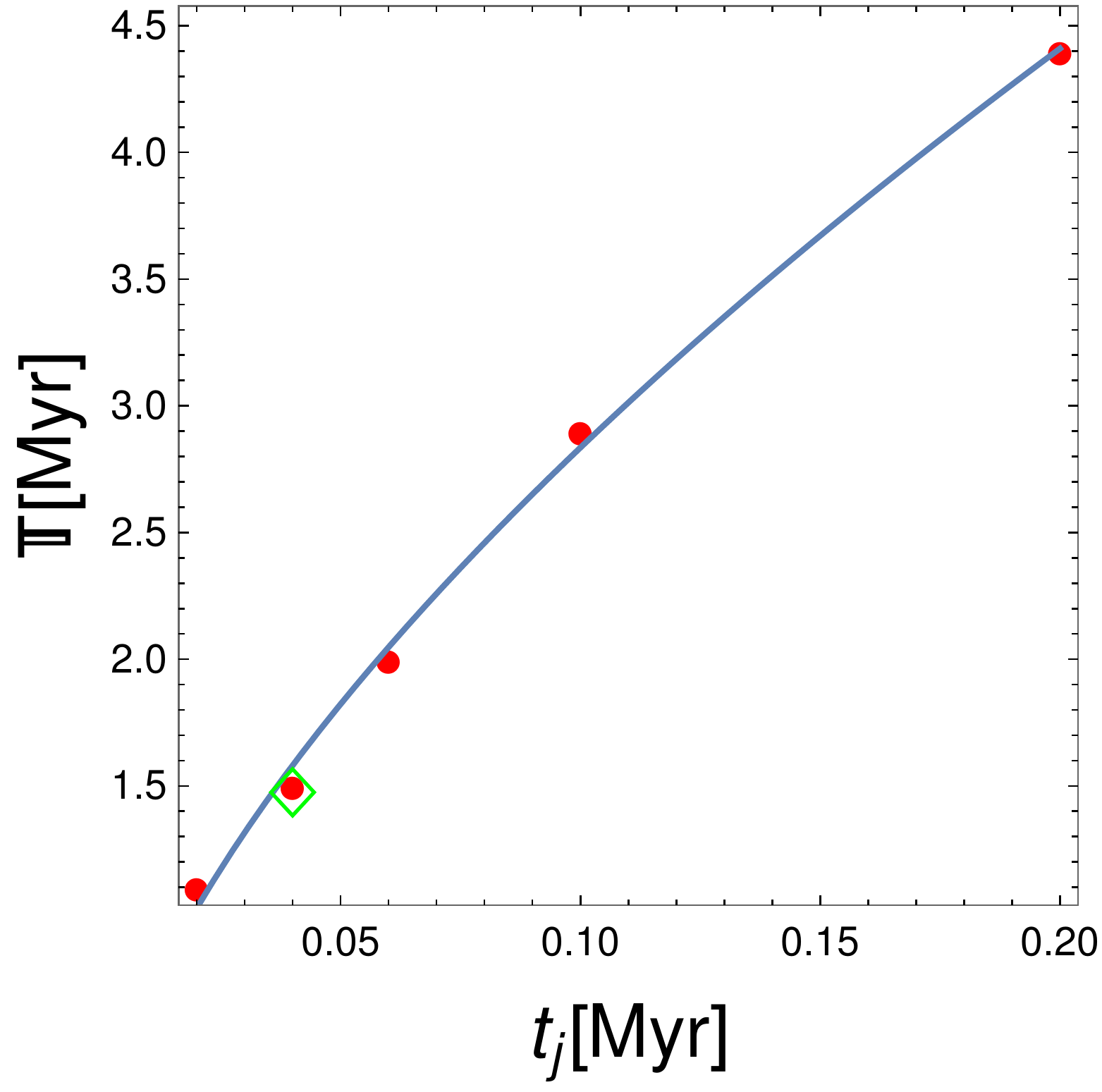}}
\raisebox{-0.cm}{\includegraphics[height=4.1truecm]{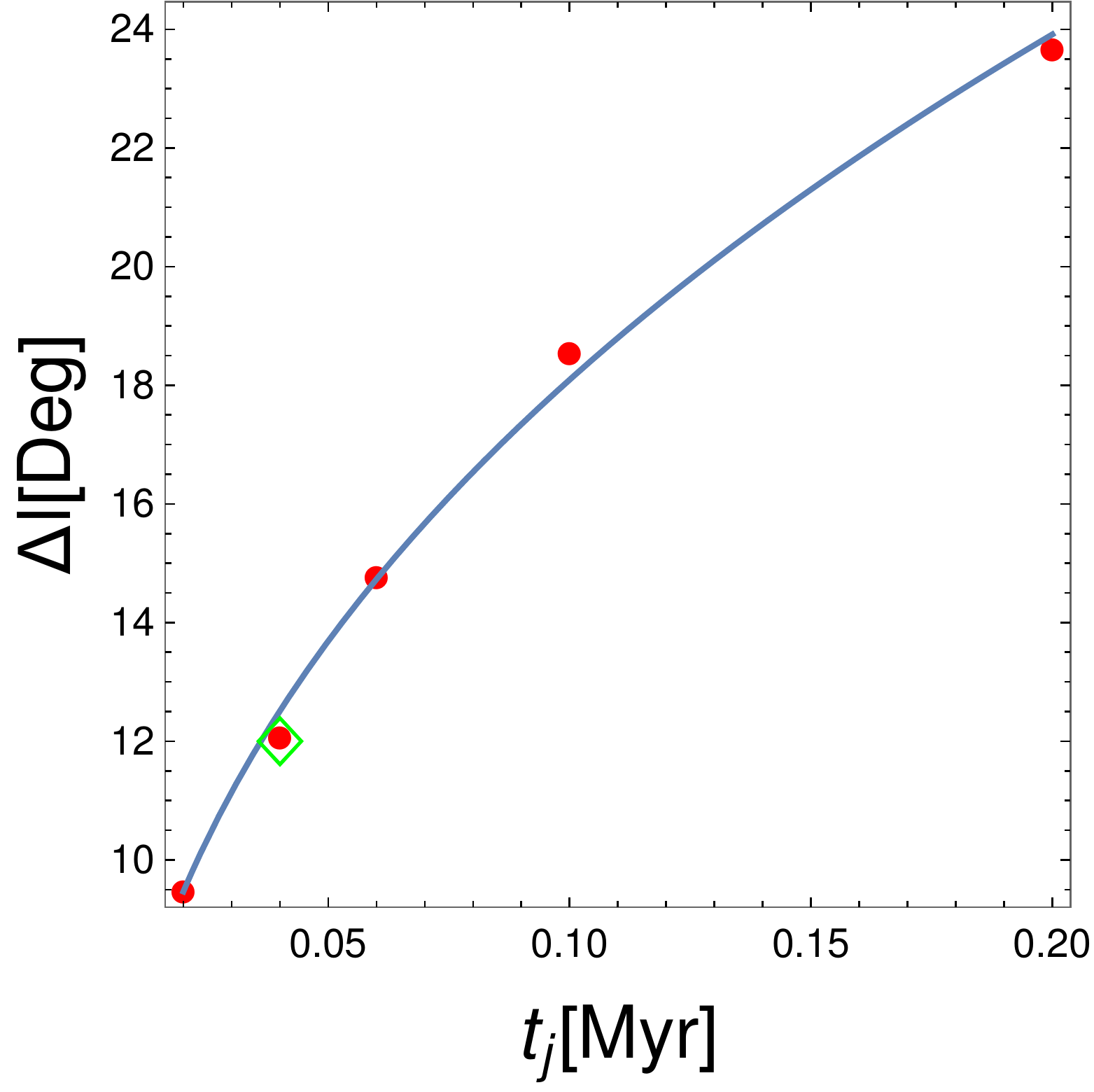}}
}\\
\centering{
\raisebox{-0.cm}{\includegraphics[height=4.1truecm]{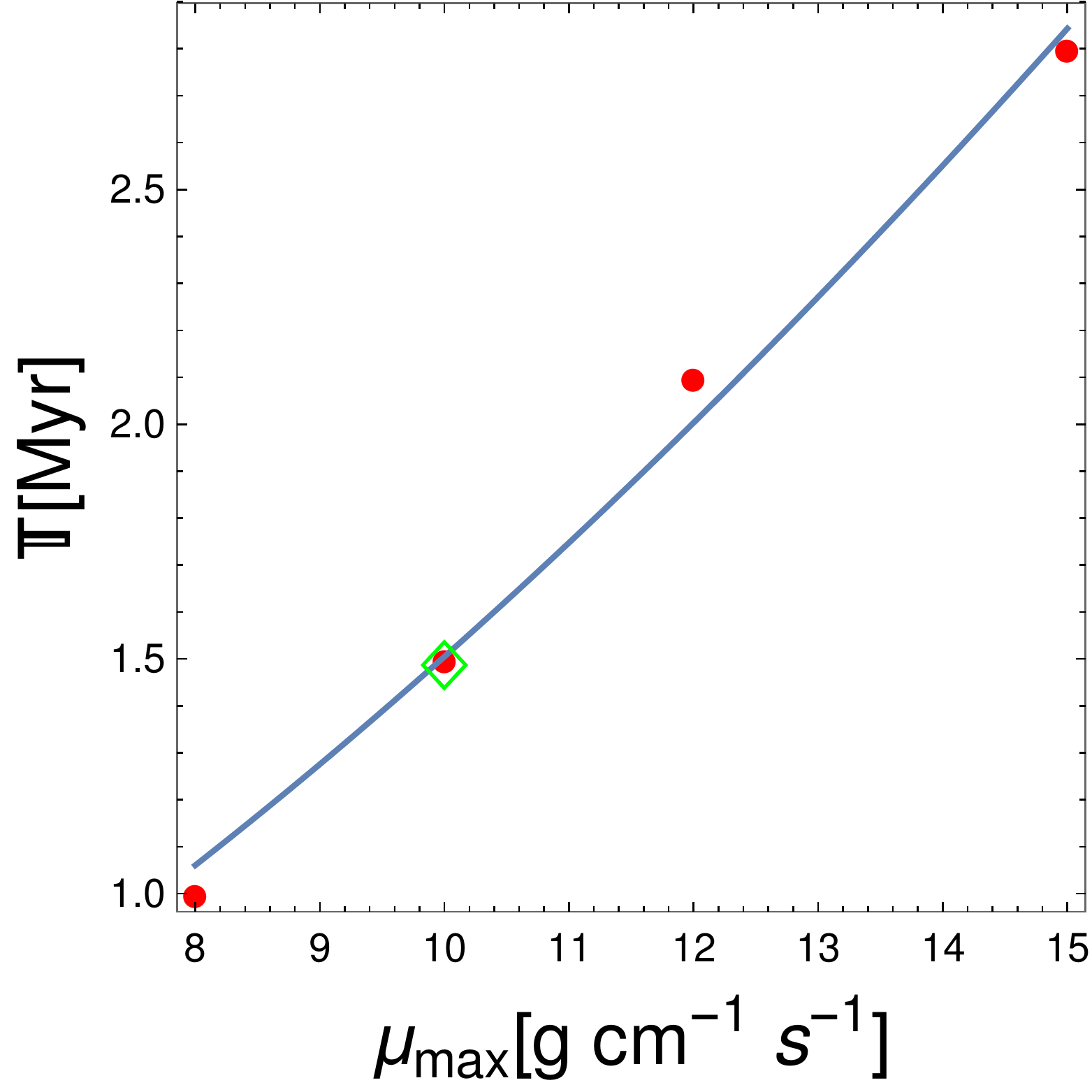}}
\raisebox{-0.cm}{\includegraphics[height=4.1truecm]{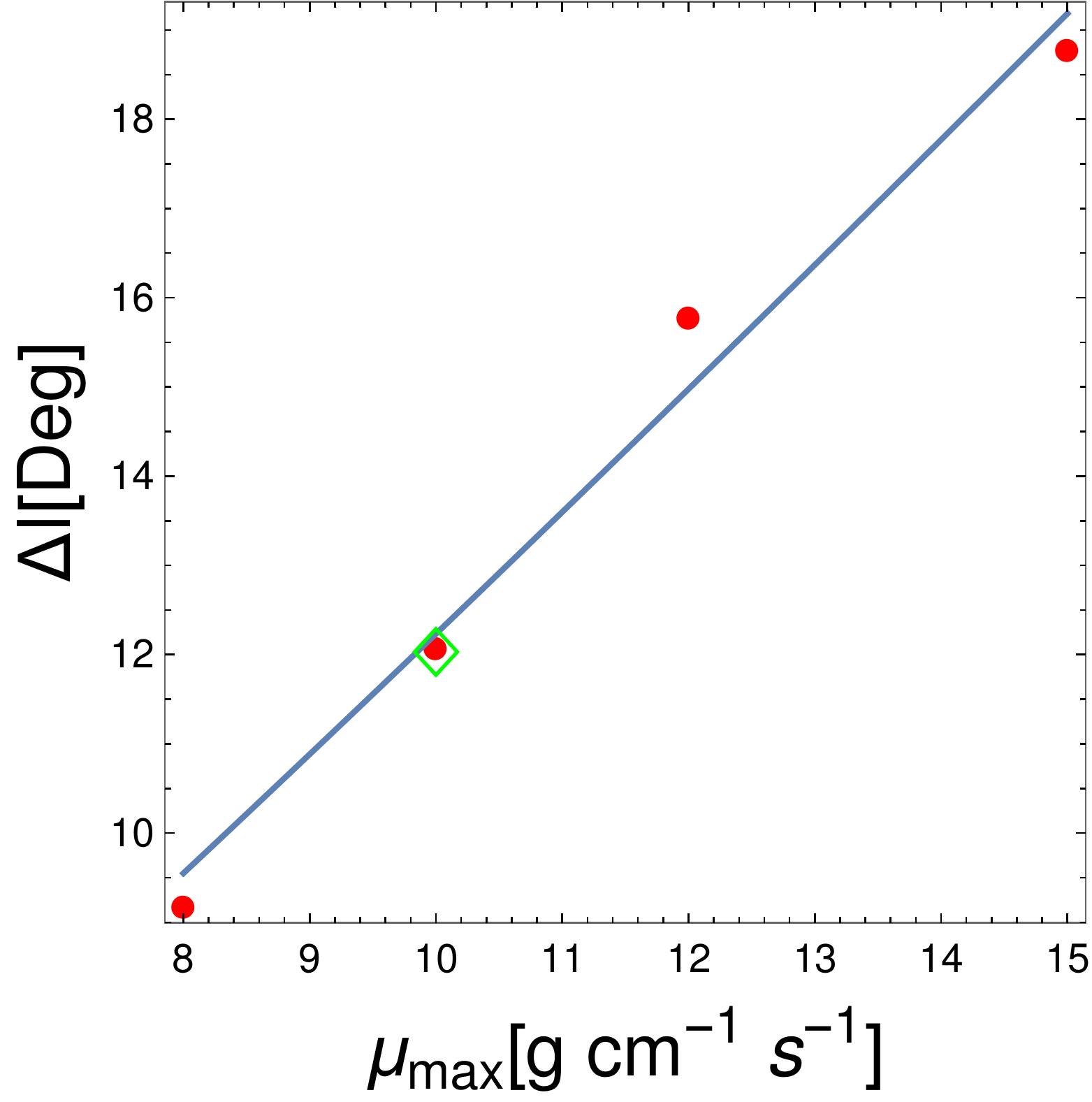}}
}
\caption{
The profiles of $\tage$ (right panel) and $\Delta l$ (left panel) for a range of $E_{\inj}$ (first row), v$_{\inj}$ (second row), $t_{\inj}$ (third row), and $\mu_{\max}$ (fourth row).
The red filled circles are the simulated results, fitted with power-law functions (blue solid lines).
For $\tage$ ($E_{\inj}$) dependence, $\beta_{\inj}$ and  $t_{\inj}$ are kept set at 0.1 and 0.04 Myr. During other parameters dependence test we kept $E_{\inj}$ set at 3$\times10^{55}\erg$. The $\theta_{\inj}$ parameter was set to $4^\circ$ for all dependencies test. The green diamonds show the best fitted parameters for the setup J2.
\label{fig:nonballistic_plots}}
\end{figure}

By separately varying each of the J2a setup parameters, one at a time, we derive the functional dependencies of the bubble age in the close vicinity of the J2a setup,
\begin{align}\label{eq:NonBallisticAgeScaling}
   \tage \simeq & 1.81\left(\frac{E_{\inj}}{3\times10^{55}\erg}\right)^{-1.51\pm0.06} \beta_{-1}^{1.92\pm0.10} \mu_{10}^{1.58\pm0.06} \nonumber \\
    &  \times t_{0.04}^{0.55\pm0.05} \Myr \coma
\end{align}
maximal half-width,
\begin{align} \label{eq:NonBallisticWidthScaling}
    R_{\max} \simeq & 1.95\left(\frac{E_{\inj}}{3\times10^{55}\erg} \right)^{-0.77\pm0.05} \beta_{-1}^{1.13\pm0.07} \mu_{10}^{1.2\pm0.1} \nonumber \\
    &  \times t_{0.04}^{0.37\pm0.03} \kpc \coma
\end{align}
maximal longitude,
\begin{align}\label{eq:NonBallisticLongScaling}
    \Delta l \simeq & 13^\circ.1\left(\frac{E_{\inj}}{3\times10^{55}\erg} \right)^{-0.79\pm0.05} \beta_{-1}^{1.26\pm0.08} \mu_{10}^{1.13\pm0.07} \nonumber \\
    & \times t_{0.04}^{0.37\pm0.04} \coma
\end{align}
and Mach number of the head,
\begin{align}\label{eq:NonBallisticMachScaling}
\Upsilon \simeq & 3.73\left(\frac{E_{\inj}}{3\times10^{55}\erg}\right)^{1.27\pm0.13} \beta_{-1}^{-1.14\pm0.14} \mu_{10}^{-1.0\pm0.1} \nonumber \\
& \times t_{0.04}^{-0.40\pm0.05} \fin
\end{align}
These individual parametric scans are illustrated in \autoref{fig:nonballistic_plots}.
Notice that the energy-dependencies in the top row are not well-fit by a power law.

These results are generally consistent with our model, considering the limited dynamical range explored.
For $z_{\tr}\simeq 8\kpc$ and $\tz=0.14$, the age of the bubble in Eq.~(\ref{eq:NonBallisticAge}) scales approximately as $\tage\propto v_{\modj}^{-1}z_{\tr}^{-2.2}$.
Combining this with Eqs.~(\ref{eq:zTransition}) and (\ref{eq:thetaEff}) yields $\tage\propto E_{\inj}^{-1.5}\beta_{\inj}^{1.9}\mu_{\max}^{1.5}$, in qualitative agreement with the fit (\ref{eq:NonBallisticAgeScaling}).
Equation (\ref{eq:NonBallisticRmaxB}) then gives $R_{\max}\propto E_{\inj}^{-0.7}\beta_{\inj}^{1.5}\mu_{\max}^{1.0}$, in fairly good agreement with the fit (\ref{eq:NonBallisticWidthScaling}).
Taking the time derivative of Eq.~(\ref{eq:NonBalZH}) yields, for late-times when $z_H(t)$ is a power-law, $\Upsilon\simeq (\tz z_H)/ (\tage c_s) \propto \tage^{-1}$, in qualitative agreement with the fit (\ref{eq:NonBallisticMachScaling}).
Taking into account corrections to $z_H(t)$ gives $\Upsilon\simeq (v_{\modj}/c_s)(z_H/z_\tr)^{1-\tz^{-1}} \propto v_{\modj}\tage^{-1.2}\propto E_{\inj}^{1.8}\beta_{\inj}^{-1.3}\mu_{\max}^{-1.8}$, partly consistent with (\ref{eq:NonBallisticMachScaling}).
Some positive dependence of $\theta_{\eff}$ upon the injection time is expected in (\ref{eq:thetaEff}); the $t_{\inj}$ dependence of the fits (\ref{eq:NonBallisticAgeScaling})--(\ref{eq:NonBallisticMachScaling}) would match the model if $\theta_{\modj}\propto t_{\inj}^{0.2}$.

The slowing-down bubbles are more sensitive to the Galactic model than their ballistic counterparts.
Increasing the disc mass significantly lowers both the age and the width of the slowing-down bubble, as it leads to a steeper, $\ah>2$ CGM density profile.  This effect, along with the slower sideways expansion inside the {\disk}, lead to younger bubbles more pinched at their base, as seen in the bottom panel of \autoref{fig:nonballistic_variants}.
Other parameters have a more modest effect. Introducing extreme CGM rotation, for example, renders the bubbles slightly thinner, but does not substantially modify their age, as shown in the bottom panel of the figure.

\section{Summary and Discussion} \label{sec:discussion}

We model the FBs under the assumption that their edges are strong forward shocks, as based on recent evidence (\autoref{sec:Intro}). We simulate a bubble in one Galactic hemisphere as either an undirected release of energy near or above the GC (\autoref{nonjettedresults}) or as a collimated injection of both energy and momentum as a jet emanating from CBH on $\sim 100\pc$ scales (\autoref{jettedresults}). The evolved FBs, computed for various injection and CGM parameters, are compared to observations in order to constrain both the FB engine and the Galactic model. The study combines analytic modeling with converged (\autoref{appendix:B}) axisymmetric hydrodynamic PLUTO-v4.0 simulations (\autoref{sec:Method}), which implement various hydrostatic models of the Galaxy (\autoref{appendix:A}) including halo, disc, and bulge components.

A putative non-directed FB engine, injecting energy but not momentum, leads to a fairly spherical bubble due to the fast sideways expansion after the bubble emerges from the {\disk}. The evolution is characterised by a $\lesssim 1\Myr$ linear, ballistic stage followed by an approximately self-similar, sublinear, spherical evolution. By the time the top of the bubble reaches the $|b|\simeq 52^\circ$ latitude of the FBs, it is too spherical to agree with observations, whether the engine is placed in the Galactic centre or is effectively offset to intermediate, order $z\sim1\kpc$ heights (\autoref{fig:base_GC_IL}).
This conclusion remains valid for a wide range of Galactic models (\autoref{fig:GC_setups} and \autoref{fig:IL_setups}), including extreme variations in Galactic {\disk} and bulge masses, fast and even maximal CGM halo rotation, and combining the above effects; see \autoref{Tab:NonDirected}. We conclude that a non-directed engine fails to generate FBs with forward shock edges consistent with observations for any plausible Galactic model.

In contrast, an outflow collimated into a thin jet, launched near the GC approximately perpendicular to the {\disk}, can reproduce the observed morphology of the FB edge as a strong forward shock.
The evolution can be crudely understood using a stratified model (\autoref{sec:ToyModels}), in which the gas behind the head of the jet is approximated as expanding sideways at a fixed $z$, as verified by numerical simulations (\autoref{subsec:JettedSims}).
The bubble initially evolves ballistically, $z_H\propto t$, until it accumulates sufficient mass and starts slowing down, with momentum conservation leading to a very slow, $z_H\propto t^{(0.1\mbox{\scriptsize--}0.2)}$ subsequent growth (\autoref{eq:zHvsT}).
As the transition height $z_{\tr}\gtrsim 5\kpc$ (\autoref{fig:FBTlim}) can be above or below the present-day $z_H\simeq 10\kpc$ tip of the bubble, the FBs can be either in the ballistic phase (large $E/\beta_j$; \autoref{eq:BallisticRegime}) or in the slowdown phase (small $E/\beta_j$; \autoref{eq:NonBallisticRegime}).
In both regimes, we identify plausible injection parameters (\autoref{Tab:Jetted}) that yield bubbles (modeled: left panel of \autoref{fig:FBModels}; simulated: \autoref{fig:NominalJetted}) consistent with the observed edges \citepalias{Keshetgurwich17} and X-ray shell \citepalias{Keshetgurwich18} of the FBs.

If the injection is sufficiently energetic or the jet is slow, $E_{55}\gtrsim 3(\beta_{-2}\theta_5)^2$, then the FBs are still ballistic.
A high Mach number $\Upsilon$ at the top of the bubble then requires $\beta_{-2}\simeq 0.35\Upsilon_5$, and the thickness of the bubbles requires $\theta_{\modj}\simeq 4^{\circ}$.
The injected energy (both hemispheres) is then $E_{55}\gtrsim 2\beta_{-2}^2\gtrsim 0.2\Upsilon_5^2$ and the age of the bubbles is $\tage\simeq 3.3\beta_{-2}^{-1}\Myr\lesssim 8\Upsilon_5^{-1}\Myr$.
Our simulations show bubbles broadly consistent with the model in terms of structure (\autoref{fig:FBModels} and the top panel of \autoref{fig:NominalJetted}), evolution (\autoref{fig:ballisticCombo}), and parameter dependence (\autoref{fig:ballistic_setups} and \autoref{fig:ballistic_plots}).
We identify a pronounced low-pressure region behind the head of the simulated ballistic bubble, consisting of unperturbed jet material.
The ballistic nature of the bubble renders it nearly independent of the Galactic model.

In contrast, if the injection is sufficiently fast or of low energy, $E/\beta_k$ is small (\autoref{eq:NonBallisticRegime}) and the observed FBs are already in their slowdown phase.
The injected energy is then $E_{55}\lesssim 3(\beta_{-2}\theta_5)^2$, in which case $0.05\Upsilon_5^2\theta_5^2\lesssim E_{55}\lesssim 2\beta_{-2}^2$ and $\tage\simeq 1.4(\beta_{-2}\theta_5)^{-1}\Myr\lesssim 4(\Upsilon_5\theta_5)^{-1}\Myr$.
Such slowing-down FBs can still be energetic, provided that $\beta_{-2}$ is large, but their age would then be of order a Myr only if $\theta_5$ is small.
Our simulations again show bubbles broadly consistent with the model in terms of structure (bottom panel of \autoref{fig:NominalJetted}), evolution (\autoref{fig:NonBallisticCombo}), and parameter dependence (\autoref{fig:nonballistic_variants} and \autoref{fig:nonballistic_plots}). These bubbles have a more regular inner structure, without the aforementioned low-pressure region. The edges of the bubbles in the slowdown phase are somewhat dependent on the {\disk} mass.

While the FB edges and X-ray shell can be reproduced equally well if observed in the ballistic or slowdown phases, the evolution of the inner structure of the bubble can distinguish between the phases.
The ballistic FB shows an inner cylindrical shock surrounded at its top by an irregular contact discontinuity surface, whereas the slowing-down FB shows a bubble-like contact discontinuity trailing the shock.
The high, $\Upsilon\gtrsim5$ Mach number of the forward shock, inferred from observations \citepalias{Keshetgurwich17, Keshetgurwich18}, is consistent with both ballistic and early slowdown phases, and is one of several indications that $z_{\tr}\gtrsim 5\kpc$.

The temperature behind the shock is a useful diagnostic of the FB energetics,
being directly related to the shock velocity and insensitive to high-energy processes.
The $(1\mbox{--}2)\keV$ temperatures in our nominal simulations
are slightly higher than the $0.4-0.5$ keV electron temperatures estimated based on the \ion{O}{VII} and \ion{O}{VIII} line ratios \citep{MillerBregman2016} and on the X-ray shell \citepalias{Keshetgurwich18}.
The difference is consistent with a partial equilibration of electrons at the high-Mach shock, as the Coulomb equilibration
timescale \citep{Spitzer1956},
\begin{equation}\label{eq:teq}
    t_{eq}\simeq 7.9\left(\frac{T}{10^7\K}\right)^{3/2}\left(\frac{n}{10^{-3}\cm^{-3}}\right)^{-1}\Myr\, ,
\end{equation}
is longer than the $\sim 2\mbox{--}3\Myr$ age of our simulated FBs.
A time-dependent calculation of the energy transfer between electrons and protons \citep[see appendix B of][]{Sarkar2017} indicates that by $\sim3$ Myr, the electron temperature reaches $\sim 4\times 10^6$ K, consistent with observations.

\begin{figure}
\centering{
\hspace{-2.5cm}
\raisebox{0.0cm}{\includegraphics[height=4.5truecm,trim={4.8cm 2.2cm 4.0cm 2.8cm}, clip]{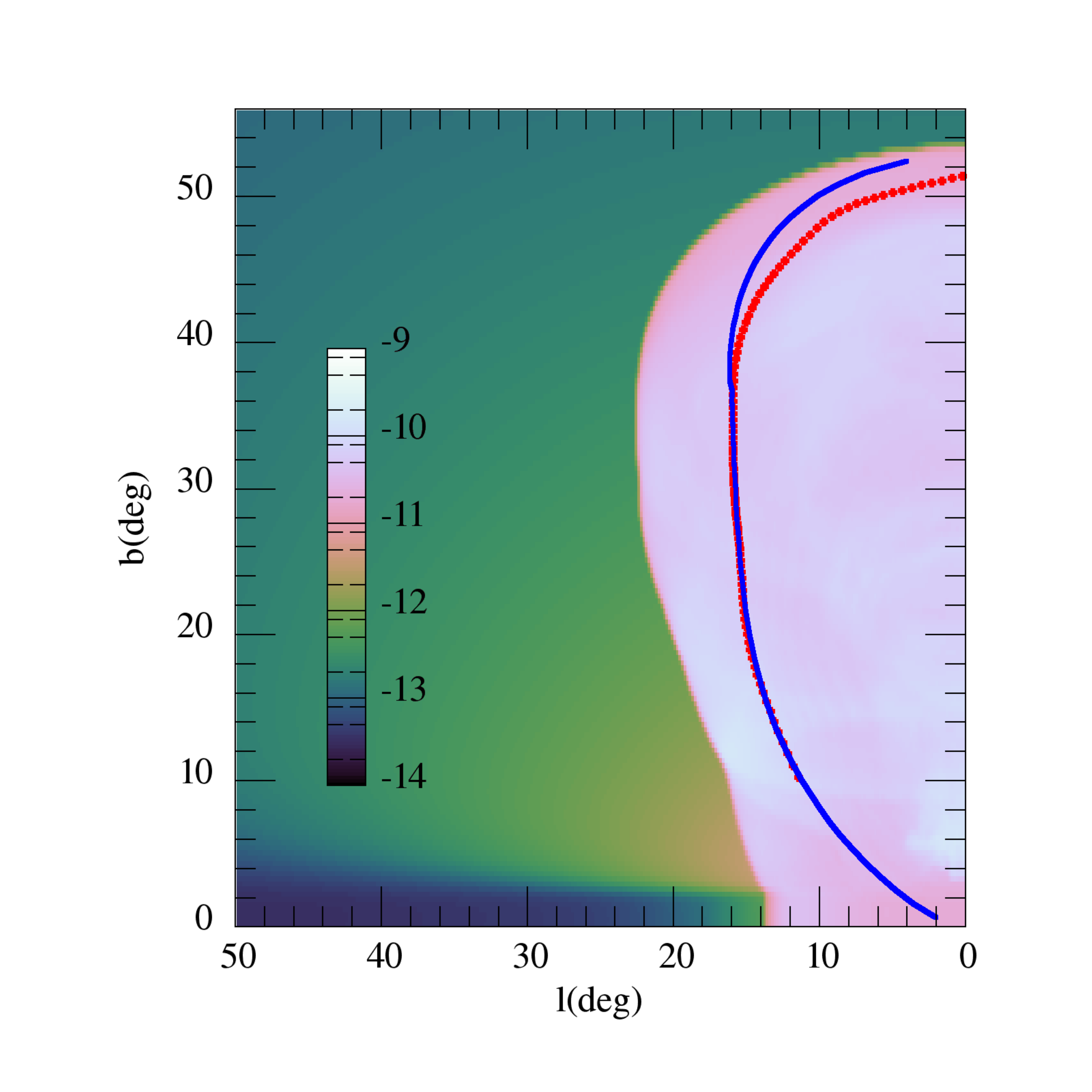}}
\hspace{-0.2cm}
\includegraphics[height=4.5truecm,trim={4.8cm 2.2cm 4.0cm 2.8cm}, clip]{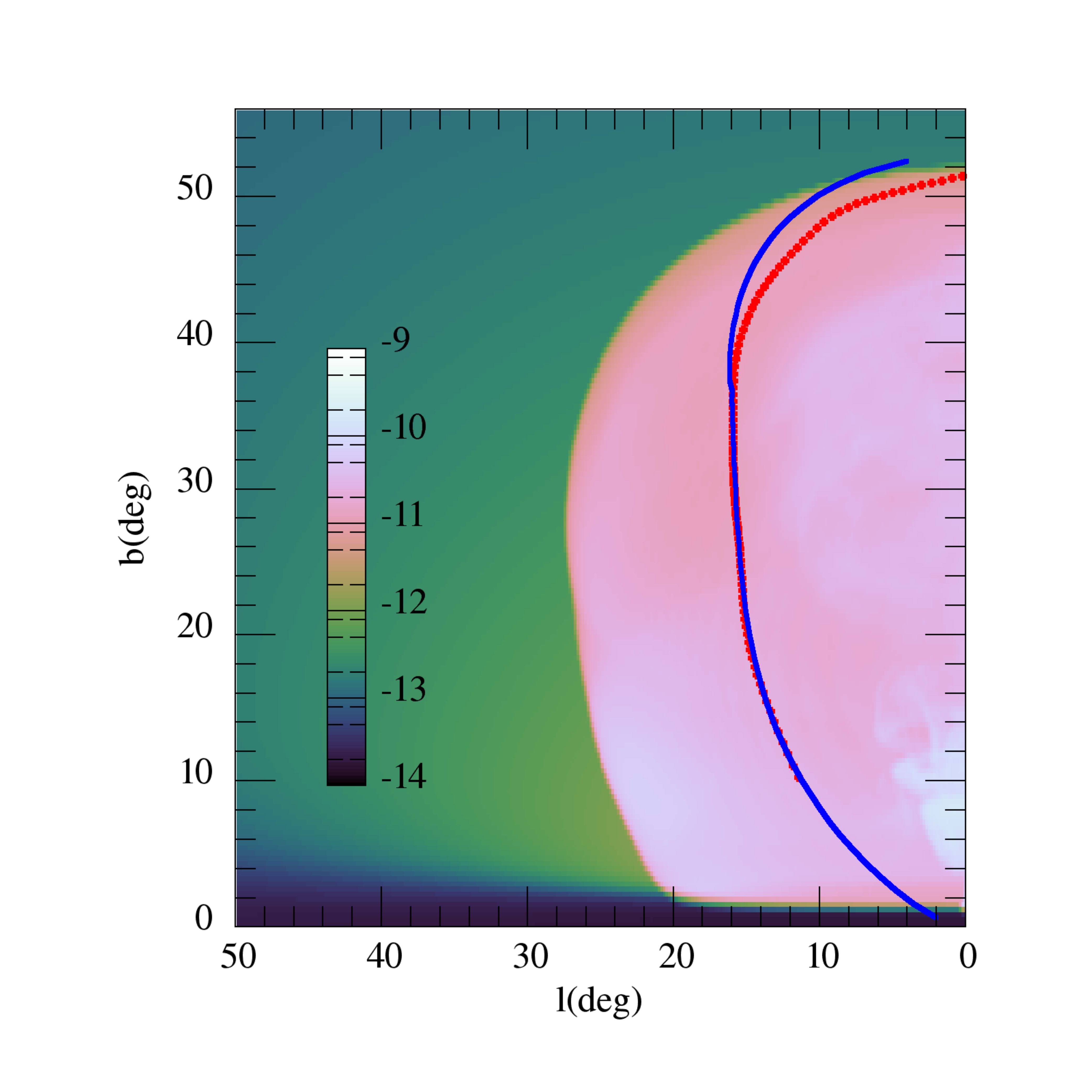}
\hspace{-2.5cm}
}
\caption{Simulated FBs generated using the setup of  \citetalias{ZhangGuo2020} with their injection above the {\disk} ($r_0=0.35\kpc$; left panel; at $\tage=5.0\Myr$) and for the same parameters but injection inside the {\disk} ($r_0=0.05\kpc$; right panel; at $\tage=7.5\Myr$). Notations are the same as in \autoref{fig:base_GC_IL}. In both cases, the FBs are too wide and the shocks too weak to match observations.
\label{fig:comparison}}
\end{figure}

During the course of this work, \citetalias{ZhangGuo2020} presented simulated FBs with forward shock edges from jet injection, considering a much narrower parameter range than studied here.
The preferred (case A) \citetalias{ZhangGuo2020} scenario is roughly comparable to our directed injection with slowing-down FBs, based on its parameters ($E\simeq 2.2\times 10^{55}\erg$ and $\beta_{\inj}\simeq0.08$) and declining shock Mach number.
However, the resulting FBs are not fully consistent with observations, as the simulated FBs are too wide (reaching $R\simeq 3.5\kpc$) and the shocks are too weak (with $\Upsilon\simeq 2.2$ at $z_H$ and $\Upsilon<1.5$ below $z_H/2$); indeed, the combination of thin FBs and a high Mach number imposes strong constraints on the injection setup.
Moreover, the \citetalias{ZhangGuo2020} injection appears unrealistically long,
with $t_{\inj}=1.0\Myr$, and is deposited at a height $r_0\simeq 350\pc$, avoiding the propagation of the jet through the {\disk}.
\autoref{fig:comparison} shows that while the \citetalias{ZhangGuo2020} setup already leads to FBs too wide, incorporating the {\disk} component exacerbates the discrepancy.

We have shown that FB observations identifying the edges as forward shocks imply a collimated injection, which we model as $\theta\lesssim4^\circ$ jets along the Galactic $z$ axis as they emerge on $\sim 100 \pc$ scales.
Sufficiently small inclinations with respect to the $z$ axis, of order $\theta\lesssim30^\circ$, say, may still provide a good match to the FBs, but we could not explain forward-shock FB edges with jets of large inclinations.
It is generally believed that highly collimated jets are launched by a rapidly spinning CBH along the spin axis \citep[\eg][]{McKinneyetal2013}.
While a jet perpendicular to the Galactic plane does not seem consistent with the presently inferred spin orientation \citep[\eg][]{Gravity2018}, rapid changes in spin direction may be natural for such a CBH \citep[\eg][]{DottiEtAl2012}.
Therefore, our results impose a challenge on modeling the launching of the jet and its collimation on small scales.

\section*{Acknowledgements}
We thank P. Sharma, C. M. Irwin, B. Nath, and T. Piran for useful discussions.
This research has received funding from the IAEC-UPBC joint research foundation (grant No. 300/18), the GIF (Grant No. I-1362-303.7/2016), and the Israel Science Foundation (ISF grant No. 1769/15), and was supported by the Ministry of Science, Technology \& Space, Israel.
KCS was supported by the Israeli Centers of Excellence
(I-CORE) program (center No. 1829/12) and the Israeli Science
Foundation (ISF grant No. 2190/20).

\bibliographystyle{mnras}
\bibliography{FermiBubbles}

\appendix

\section{Galactic model}
\label{appendix:A}

The galactic model implemented in the simulations
contains gas components which are in a steady state equilibrium with a static gravitational potential. The potential consists of a cylindrically symmetric stellar {\disk} and spherically symmetric stellar bulge and dark matter components.
For the stellar {\disk}, we use a \citet{Miyamotonagai75} potential in a cylindrical form,
\begin{equation}
\Phi_{d} =-\frac{GM_{d}}{\sqrt{R^2+\left(a_d+\sqrt{z^2+b_d^2}\right)^2}}
\coma
\end{equation}
where $a_d\geq0$ and $b_d\geq0$ represent the scale length and scale height of the stellar {\disk} of mass $M_d$. For the dark matter, we use a modified form of the NFW potential \citep{Navarroetal96} to include a core that produces a finite dark matter (DM) density at the centre,
\begin{equation}
\Phi_{\DM} =-\frac{G\Mvir}{f(c_{\text vir})}\frac{\ln\left(1+\sqrt{\frac{r^2+d^2}{r_s^2}}\right)}{\sqrt{r^2+d^2}} \nonumber
\coma
\end{equation}
where
$\cvir = \rvir/r_s$ is the DM concentration parameter,
$f(c_{\rm vir}) = \log(1+c_{\rm vir})-c_{\rm vir}/(1+c_{\rm vir})$,
$\rvir$ and $r_s$ are respectively the virial radius and scale radius for a DM halo
of mass $\Mvir$,
and $d$ is the core radius of the DM distribution.
The bulge potential is considered to have the form
\begin{equation}
\Phi_{b} =-\frac{GM_{b}}{\sqrt{r^2+a_b^2}}\,,
\end{equation}
where $a_b$ is the scale radius and $M_b$ is the mass of the stellar bulge. Thus the total Galactic potential becomes
\begin{equation}
\Phi(R,z) = \Phi_d(R,z) + \Phi_b(R,z) + \Phi_{\DM}(R,z) \fin
\end{equation}

For the gaseous components, we consider a circumgalactic medium (halo), an ISM {\disk}, and a central molecular zone (CMZ) component. Each of these components follows an independent steady-state equilibrium with the total gravitational potential $\Phi(R,z)$.
For example, the density distribution of the ISM {\disk} component, in equilibrium, can be written as \citep[for a derivation, see][]{Sarkaretal15a}
\begin{equation}
    \frac{\rho_d(R,z)}{\rho_{d0}} =
    \exp\left\{-\frac{\Phi(R,z)-\Phi_0-[\Phi(R,0)-\Phi_0]f_d^2}{c_{\sd}^2}\right\} \coma
\label{eq:rho_d}
\end{equation}
where the potential $\Phi_0\equiv\Phi(0,0)$ and ISM {\disk} mass-density $\rho_{d0}\equiv\rho_d(0,0)$ are evaluated at the Galactic centre, $c_{\rm sd} \equiv c_{\rm sd}(T_{\rm sd}) = (k_B T_{\rm sd}/\mu_m m_p)^{1/2}$ is the isothermal sound speed (including turbulence) of the disc gas, and $f_d$ is the rotation velocity ratio between the {\disk} gas $v_{\phi,d}(R)$ and stellar component $v_{\phi,G}(R) = (-R\:\partial \Phi/\partial R|_{z=0})^{1/2}$ at any given cylindrical radius $R$.  We assume this ratio and the sound speed to be independent of $R$ and $z$, in order to obtain an analytical expression for the density distribution.

We obtain the density distributions of the other components in a similar way. For the halo and the CMZ, the density distributions are, respectively,
\begin{equation}
    \frac{\rho_h(R,z)}{\rho_{h0}}=
    \exp\left\{-\frac{\Phi(R,z)-\Phi_0-[\Phi(R,0)-\Phi_0]f_h^2}{c_{\sh}^2}\right\} \coma
    \label{eq:rho_hot}
\end{equation}
and
\begin{equation}
   \frac{\rho_{\cmz}(R,z)}{\rho_{\cmz0}}=
    \exp\left\{-\frac{\Phi(R,z)-\Phi_0-[\Phi(R,0)-\Phi_0]f_{\cmz}^2}{c_{\scmz}^2}\right\} \coma
    \label{eq:rho_cmz}
\end{equation}
where, $\rho_{h0}\equiv\rho_h(0,0)$, $\rho_{\cmz0}\equiv\rho_{\cmz}(0,0)$, and rotation factors $f$ are defined analogously to $f_d$ in Eq.~(\ref{eq:rho_d}).

The total density at any computational cell is therefore given as
$\rho_{\tot}(R,z) = \rho_d(R,z)+\rho_h(R,z)+\rho_{\cmz}(R,z)$. Since each of these components may have a different rotation speed, the effective rotation speed $v_{\phi,\eff}$ at any computational cell is given by
\begin{equation}
  \frac{v_{\phi,\eff}(R,z)}{v_{\phi,G}(R)}=\sqrt{\frac{f_d^2 \rho_d(R,z) + f_h^2 \rho_h(R,z) + f_{\cmz}^2 \rho_{\cmz}(R,z)}{\rho_{\tot}(R,z)}} \quad .
  \label{eq:vphi_eff}
\end{equation}
Notice that, although we include the CMZ component, it is set to zero in practice in the simulations shown.

\section{Convergence tests}
\label{appendix:B}

\begin{figure}
\centering{
\hspace{-2.5cm}
\raisebox{0.0cm}{\includegraphics[height=4.5truecm,trim={4.8cm 2.2cm 4.0cm 2.8cm}, clip]{Figures/solar-p5-GC20-40pc-ang90.pdf}}
\hspace{-0.2cm}
\includegraphics[height=4.5truecm,trim={4.8cm 2.2cm 4.0cm 2.8cm}, clip]{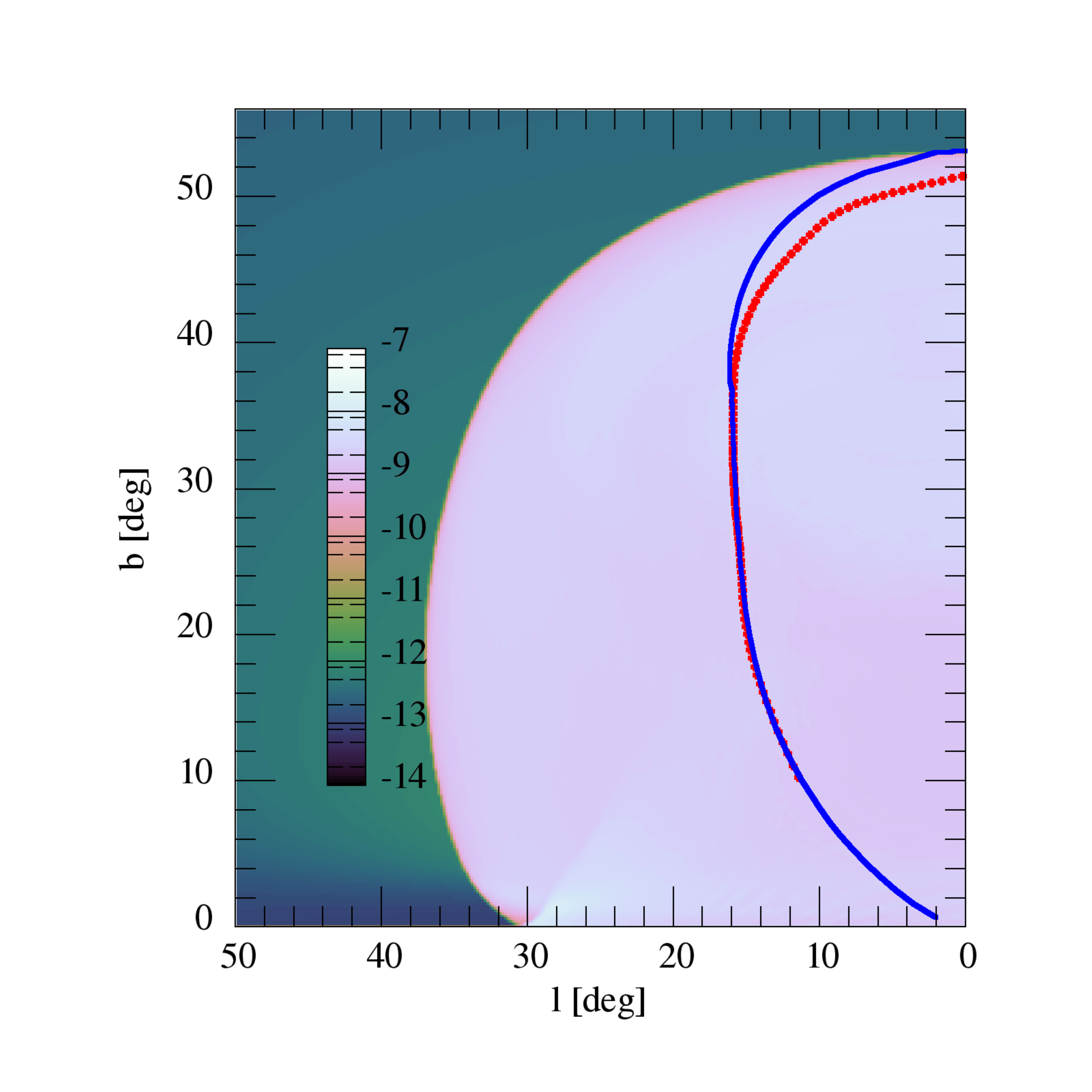}
\hspace{-2.5cm}
}
\caption{Same as top panel of \autoref{fig:base_GC_IL} (setup S1), but for simulation grids {1024$\times$256} (left) and {2048$\times$512} (right) in the $\{r, \theta\}$ directions.
\label{fig:resolution_test_GC}}
\end{figure}

\autoref{fig:resolution_test_GC} shows the contemporary projected FB images obtained for non-directed, GC injection (setup S1) for stretched $\{r,\theta\}$ grids of dimensions $\{1024\times256\}$ (left) and $\{2048\times512\}$ (right).
The two images are sufficiently similar
to warrant adopting $\{1024\times256\}$ as the base resolution for non-directed simulations.

\begin{figure}
\centering{
\hspace{-2.5cm}
\raisebox{0.0cm}{\includegraphics[height=4.5truecm,trim={4.8cm 2.2cm 4.0cm 2.8cm}, clip]{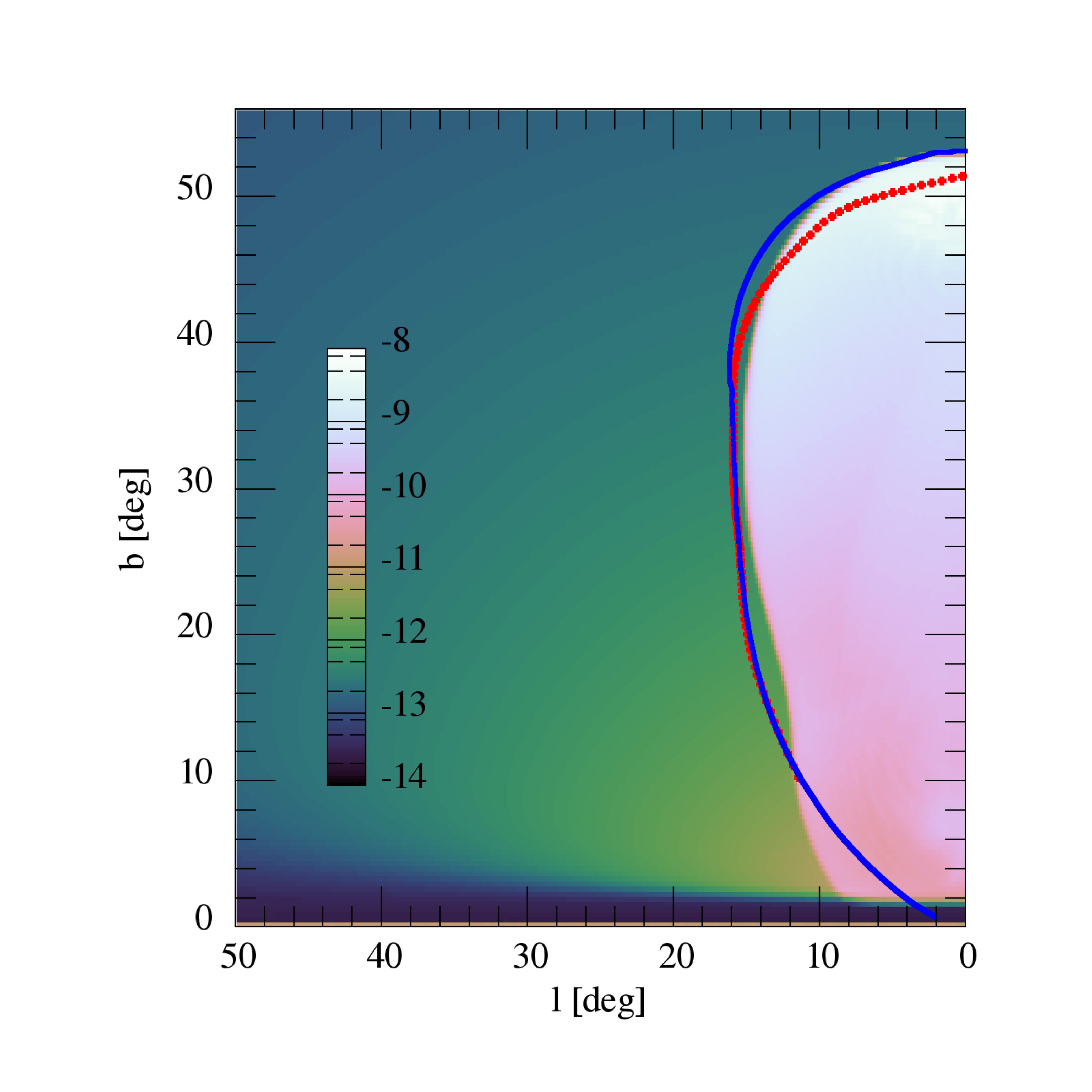}}
\hspace{-0.2cm}
\includegraphics[height=4.5truecm,trim={4.8cm 2.2cm 4.0cm 2.8cm}, clip]{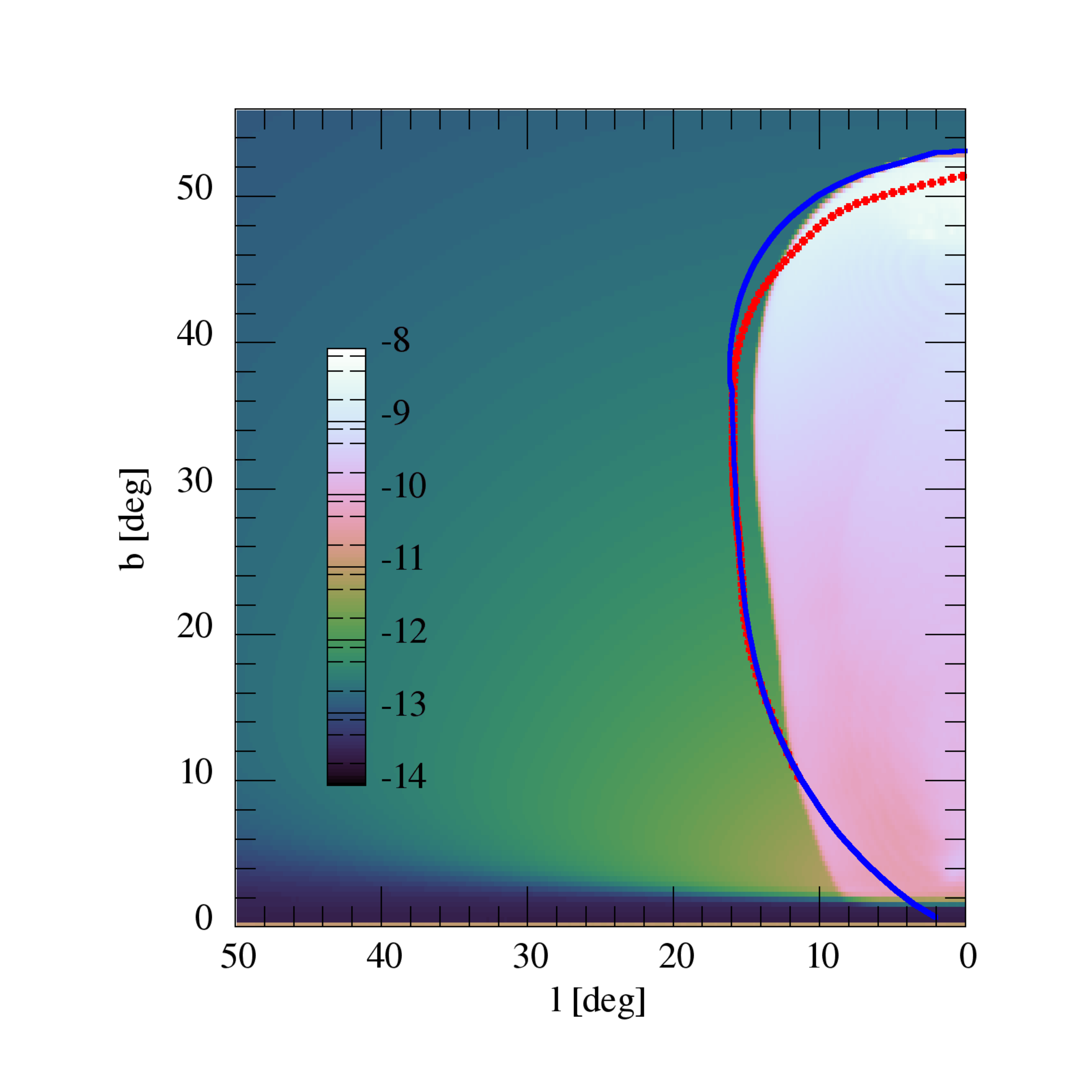}
\hspace{-2.5cm}
}
\caption{Same as top panel of \autoref{fig:NominalJetted} (setup J1), but for $\theta_{\inj}$=$4^\circ$, with simulation grids $\{1024\times512\}$ (left) and $\{2048\times1024\}$ (right).
\label{fig:resolution_test_bal}}
\end{figure}

\begin{figure}
\centering{
\hspace{-2.5cm}
\raisebox{0.0cm}{\includegraphics[height=4.5truecm,trim={4.8cm 2.2cm 4.0cm 2.8cm}, clip]{Figures/F-NBase-solar-1p5-E3.pdf}}
\hspace{-0.2cm}
\includegraphics[height=4.5truecm,trim={4.8cm 2.2cm 4.0cm 2.8cm}, clip]{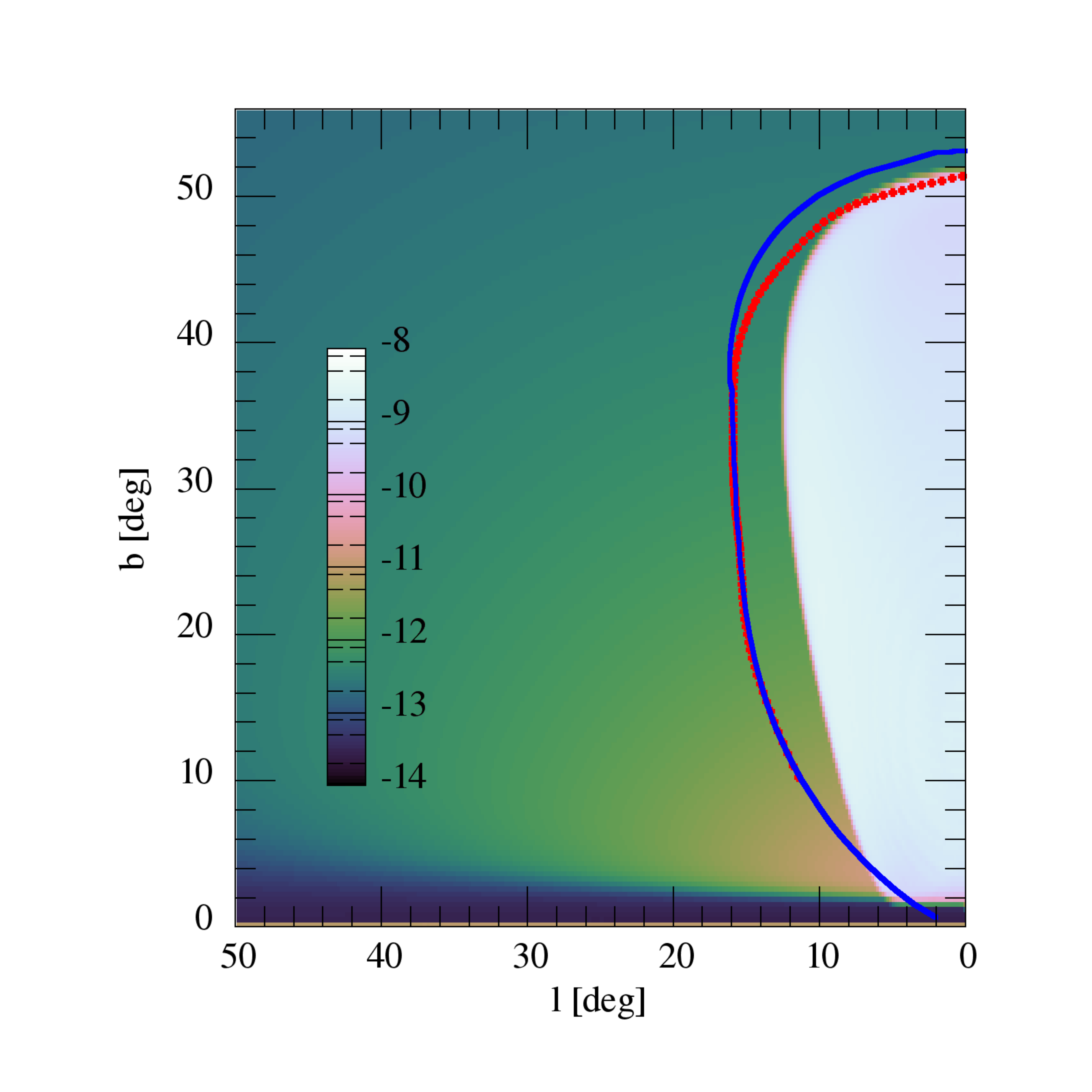}
\hspace{-2.5cm}
}
\caption{Same as top panel of \autoref{fig:nonballistic_variants}
(setup J2a) but for simulation grids $\{1536\times768\}$ (left) and $\{3072\times1536\}$ (right).
\label{fig:resolution_test_nonbal}}
\end{figure}

For directed injection, we find that a higher resolution is needed to guarantee convergence.
In the ballistic case, shown in \autoref{fig:resolution_test_bal}, we find that $\{1024\times512\}$ suffices as our base resolution.
In the slowdown regime, shown in \autoref{fig:resolution_test_nonbal}, a base resolution $\{1536\times768\}$ is sufficient.

\bsp
\label{lastpage}
\end{document}